\documentclass[12pt]{article}
\usepackage{setspace,amsmath,amsfonts,MnSymbol,amsthm,comment,paralist,graphicx,mathbbol,nicefrac,bm,mathtools,theoremref,xcolor,tikz,xurl,pgfplots}
\usetikzlibrary{patterns}
\usepackage{subcaption}
\newcommand{\RR}{\mathbb{R}}
\newcommand{\EE}{\mathbb{E}}
\newcommand{\PP}{\mathbb{P}}

\newcommand{\cD}{\mathcal{D}}

\newcommand{\ub}{\overline}
\newcommand{\lb}{\underline}
\definecolor{gray1}{gray}{0.8}
\definecolor{gray2}{gray}{0.6}
\definecolor{gray3}{gray}{0.4}
\definecolor{mixed}{rgb}{0.5,0,0.5}
\definecolor{mixyb}{rgb}{0.25,0.25,0.5}
\definecolor{mixgb}{rgb}{0,0.5,0.5}
\usetikzlibrary{patterns}
\usetikzlibrary{decorations.pathreplacing,calligraphy}

\tikzset{%
  move to start/.code=\tikz@arc@movetolineto\pgfpathmoveto,%
  line to start/.code=\tikz@arc@movetolineto\pgfpathlineto}
\makeatother
\makeatletter
\usepackage[left=2.5cm, bottom=3.5cm, right=2.5cm, top=3.5cm]{geometry}
\usepackage[normalem]{ulem}
\colorlet{linkequation}{blue}
\usepackage[authoryear,longnamesfirst,round]{natbib}
\setcitestyle{authoryear,open={(},close={)},aysep={},citesep={;}}
\colorlet{linkequation}{blue}
\usepackage[colorlinks=true,allcolors=blue]{hyperref}
\newcommand*{\SavedEqref}{}
\let\SavedEqref\eqref
\renewcommand*{\eqref}[1]{%
  \begingroup
    \hypersetup{
      linkcolor=linkequation,
      linkbordercolor=linkequation,
    }%
    \SavedEqref{#1}%
  \endgroup
}
\newcommand\bref[1]{{\hypersetup{linkcolor=blue}\autoref{#1}}}
\makeatletter
\renewcommand\@makefnmark{\hbox{\@textsuperscript{\normalfont\color{black}\@thefnmark}}}
\makeatother
\DeclareMathOperator*{\argmax}{\arg\!\max}

\newcommand*\supp{\mathrm{supp}}
\newcommand*\diff{\mathop{}\!\mathrm{d}}

\newtheorem{theorem}{Theorem}
\newtheorem*{theorem*}{Theorem}
\newtheorem{lemma}{Lemma}
\newtheorem{corollary}{Corollary}
\newtheorem{proposition}{Proposition}
\newtheorem{taggedlemmax}{Lemma}

\newenvironment{taggedlemma}[1]
 {\renewcommand\thetaggedlemmax{#1}\taggedlemmax}
 {\endtaggedlemmax}

\makeatletter
\renewenvironment{proof}[1][\proofname]{%
  \par\pushQED{\qed}\normalfont%
  \topsep6\p@\@plus6\p@\relax
  \trivlist\item[\hskip\labelsep\bfseries#1\@addpunct{.}]%
  \ignorespaces
}{%
  \popQED\endtrivlist\@endpefalse
}
\makeatother
\newtheorem{assumption}{Assumption}
\theoremstyle{definition}
\newtheorem{definition}{Definition}

\newtheorem{remark}{Remark}

\DeclareDocumentCommand\Pr{ m g g }{\ensuremath{
    { \IfNoValueTF {#3}
        {   \IfNoValueTF {#2}
          {\mathbb{P}\left[{#1}\right]}
          {\mathbb{P}\left[{#1}\mid{#2}\right]}
        }
        {\mathbb{P}_{#1}\left[{#2}\mid{#3}\right]}
    }
}}
\DeclareDocumentCommand\E{ m g g }{\ensuremath{
    { \IfNoValueTF {#3}
        { \IfNoValueTF {#2}
          {\mathbb{E}\left[{#1}\right]}
          {\mathbb{E}\left[{#1}\mid{#2}\right]}%
        }
        {\mathbb{E}_{#1}\left[{#2}\mid{#3}\right]}
    }
}}

\title{\textbf{Non-Discriminatory Personalized Pricing}\thanks{We thank Dirk Bergemann, Ben Brooks, Laura Doval, Yingni Guo, Elliot Lipnowski, Ellen Muir, Harry Pei, Alessandro Pavan, Jacopo Perego, Benjamin Polak, Phil Reny, Vasiliki Skreta, Alex Smolin, Rakesh Vohra, Jonas von Wangenheim, Frank Yang, Jidong Zhou, and various seminar audience, for their valuable comments and suggestions. We also thank Tsong-Hong Tenn for his research assistance. All errors are our own.}}
\author{Philipp Strack\thanks{Department of Economics, Yale University, Email: philipp.strack@yale.edu}
\and
Kai Hao Yang\thanks{School of Management, Yale University, Email: kaihao.yang@yale.edu} 
}
\date{\today}

\begin{document}

\maketitle

\begin{abstract}
A monopolist offers personalized prices to consumers with unit demand, heterogeneous values, and idiosyncratic costs, who differ in a protected characteristic, such as race or gender. The seller is subject to a non-discrimination constraint: consumers with the same cost, but different characteristics must face identical prices. Such constraints arise in regulated markets like credit or insurance. The setting reduces to an optimal transport, and we characterize the optimal pricing rule. Under this rule, consumers may retain surplus, and either group may benefit. Strengthening the constraint to cover transaction prices redistributes surplus, harming the low-value group and benefiting the high-value group.
\end{abstract}

\mbox{}


\newpage
\section{Introduction}\label{sec:intro}

\paragraph{Motivation} Advances in data collection have enabled firms to tailor prices to consumers based on a wide range of observable characteristics. In many markets, sellers now have access to rich datasets that allow for increasingly fine-grained segmentation, often approaching fully personalized pricing. At the same time, legal frameworks prohibit discrimination based on protected characteristics---such as gender, race, or age.\footnote{For instance, in the U.S., Title VII of the Civil Rights Act and the Equal Pay Act (EPA) protect workers from gender-based wage discrimination; the Fair Housing Act (FHA) prohibits housing discrimination based on protected characteristics such as race and gender; the Equal Credit Opportunity Act (ECOA) prevents lenders from offering different loan terms to borrowers based on protected characteristics; and recent legislation in California (AB1287) explicitly prohibits businesses from price discrimination based on gender.} This raises questions about how anti-discrimination laws should apply in data-rich environments, where pricing algorithms operate on rich data.

In such settings, legal prohibitions on \emph{disparate treatment}---that is, the explicit use of protected characteristics in decision-making---may have little effect: Because many non-protected variables can serve as proxies, firms may reproduce outcomes that disproportionately disadvantage protected groups.\footnote{As noted in \citet{whitehouse}:
``Big data naturally raises concerns among groups that have historically been victims of
discrimination. Given hundreds of variables to choose from, it is easy to imagine that statistical
models could be used to hide more explicit forms of discrimination by generating customer
segments that are closely correlated with race, gender, ethnicity, or religion [...], even if the profit
motive is different from, and in many cases fundamentally inconsistent with, the sort of
prejudice that our anti-discrimination laws seek to prohibit.'' }
Due to this concern, anti-discrimination regulations are often assessed based on the notion of \emph{disparate impact}, rather than disparate treatment. This means that, rather than simply prohibiting the use of protected characteristics as inputs for pricing decisions, consumers with different protected characteristics must face the same price distributions.\footnote{For example, according to \citet{whitehouse}:
``It is often straightforward to conduct statistical tests for disparate impact by
asking whether the prices generated by a particular algorithm are correlated with variables
such as race, gender or ethnicity.'' Likewise, in credit markets, according to 12 CFR Regulation B, ``The [ECO] Act and regulation may prohibit a creditor practice that is discriminatory in effect because it has a disproportionately negative impact on a prohibited basis, even though the creditor has no intent to discriminate and the practice appears neutral on its face.'' In the context of employment, title VII of the Civil Rights Act holds employers accountable for ``practice that causes a disparate impact on the basis of race, color, religion, sex, or national origin''. In the context of housing, the FHA and its regulations (c.f., 24 CFR) establishes ``liability [...] based on a practice’s discriminatory effect, even if not motivated by discriminatory intent.'' }

In this paper, we study how a seller optimally maximizes profit through personalized pricing while complying with anti-discrimination regulations that require no disparate impact across protected characteristics. We show that the seller’s problem reduces to a non-standard optimal transport formulation, in which consumers must be paired across groups to form segments with equal price distributions. We solve for the profit-maximizing non-discriminatory pricing rule and characterize how non-discrimination constraints shape price outcomes, surplus, and deadweight loss across different consumer groups.

Specifically, we consider a model where a monopolist faces a unit mass of consumers with unit demands, different values, and different costs of being served. Consumers differ along a binary protected characteristic: conditional on having the same cost, consumers in the ``$l$'' group have lower values and more elastic demand, while consumers in the ``$h$'' group have higher values and less elastic demand. The seller can charge consumers personalized prices, but the prices must satisfy a non-discrimination constraint---namely, among consumers who are equally costly to serve, the price distributions faced by the two protected groups must be identical.
    
\paragraph{Results} Our main result characterizes the profit-maximizing pricing strategies under the non-discrimination constraint. We show that finding an optimal non-discriminatory pricing rule is equivalent to solving a non-standard optimal transport problem: among all consumers who have the same cost, the seller chooses a matching scheme that matches the $l$-characteristic and $h$-characteristic consumers into pairs, where each pair faces the same price but has distinct values. The seller then selects the profit-maximizing price for each matched pair. This transport problem differs significantly from classical formulations: the objective is non-linear, non-convex and non-monotonic in consumer values, and it lacks properties such as translation invariance and supermodularity that are typically used to derive closed-form solutions. 
Using duality results, \bref{thm1} solves the optimal transport problem and explicitly constructs the profit-maximizing non-discriminatory pricing rule that is Pareto undominated. 
Under this pricing rule, consumers with intermediate values are matched assortatively and face a price equal the lower value of the matched pair; while consumers with high values are matched with consumers from the other protected group who have low values, and face a price equal the higher value of the matched pair. 

We then turn to the welfare consequences of non-discriminatory pricing. While consumers from both protected groups can retain positive surplus under optimal pricing rules, not all consumers are served. In particular, consumers with lower values are priced out of the market, generating deadweight loss (\bref{prop:surplus}). Meanwhile, consumers with high values have their surplus extracted, and thus only consumers with intermediate values retain positive surplus. The surplus distribution is shaped by the underlying value distributions and by the relative sizes of the groups. As one group becomes more prevalent, the seller gains greater incentives to tailor prices more finely to that group, thereby reducing its surplus---an effect analogous to diminishing information rents in screening models (\bref{prop:population-size}). As a result, although anti-discrimination regulations may strictly benefit consumers from both groups, they do not necessarily favor the disadvantaged group the regulations are designed to protect. In some cases, the advantaged group may benefit more, while some consumers in the disadvantaged group may be completely excluded from the market.

In addition to price-based fairness, we explore a stricter notion of non-discrimination that requires outcome distributions---not just price distributions---to be identical across groups. This stronger constraint ensures that, for a fixed cost, transaction probabilities and transaction prices are statistically independent of protected characteristics. In \bref{prop:compare}, we establish that the profit-maximizing policy under this notion differs from the optimal non-discriminatory pricing rule: it increases surplus for $h$-consumers while reducing surplus for $l$-consumers. These results highlight that the choice of fairness definition---whether it is based on inputs, distributions, or outcomes---can meaningfully influence both efficiency and equity in personalized pricing.

Lastly, we consider a number of extensions of our main model. First, we study a model that allows for imperfect price discrimination, where the seller observes only a noisy signal of consumer values. The pricing problem continues to admit an optimal transport representation and can sometimes be solved explicitly, leading to similar insights as in our main model (\bref{sec:4.1}). We also examine the set of implementable welfare outcomes and show that while some surplus-maximizing segmentations remain feasible under non-discrimination constraints, others do not (\bref{sec:4.2}). In \bref{sec:all} and \bref{sec:general}, we characterize all optimal pricing rules, including the undominated ones, and the ones where the seller can extract the full surplus despite non-discrimination requirements.

\paragraph{Related Literature}
The literature on price discrimination has studied the welfare effects of monopolistic price discrimination. In particular, they explore whether third degree price discrimination benefits consumers \citep*[see, e.g.,][]{V85,ACV10,C16}. \citet*{BBM15} show that any surplus division between the consumers and a monopolist can be achieved by some market segmentation.\footnote{See also: \citet*{HS22} and \citet*{HS23}, who further consider segmentations in environments that feature nonlinear pricing; and \citet{FHS25}, who characterize when does more information on consumers' characteristics lead to higher (lower) welfare.} In an environment where the seller only observes protected characteristics, \citet*{CEL22} introduce multiple non-discrimination constraints and characterize the optimal prices. Similarly, \citet{KZ21} introduces several notions of non-discriminatory pricing, and characterize the optimal prices in a linear demand model.  
In contrast, this paper characterizes the optimal non-discriminatory pricing rules when the seller can engage in personalized pricing. 

The personalized pricing model, where sellers are able to offer each consumer a price that depends on their values, has also been widely adopted in oligopoly models: \citet{TV88} show that in a  Hotelling duopoly model consumer surplus can be higher under personalized pricing compared to uniform pricing. This framework is adopted by various papers that further investigate the effects of brand name \citep{SZ02}, advertisement \citep{CI02}, or data sharing \citep{MSV19}. \citet{RZ24} provide a comprehensive welfare analysis in a general oligopoly setting.    

\citet{SY24} and  \citet{he2021private}, characterize signals that do not reveal certain information, which are referred to as privacy-preserving signals. A non-discriminatory pricing rule is mathematically equivalent to a privacy-preserving signal where the privacy sets are defined by the protected characteristics. Section~5.3 in \citet{SY24} illustrates the relation of privacy and non-discriminatory pricing through an example. 
The notion of non-discriminatory pricing is also related to the notion of statistical parity in the algorithmic fairness literature \citep[see, e.g.,][]{D71,CV10,HPS16}.\footnote{Two other commonly adopted criteria are \emph{separation} and \emph{sufficiency}. It is well-known that none of any pairs of these three common fairness criteria can be satisfied at the same time (see \citet{BHN19} and \citet{CW23} for a comprehensive review of these criteria).}  These papers study the optimal fair algorithms for specific decision problems, typically with a binary state or a binary action.\footnote{In economics, \citet{LLM23} and \citet{doval2023persuasion} further characterize the entire Pareto frontier in terms of the payoffs of each protected group in a general setting.}

The rest of the paper is organized as follows: \bref{sec:model} introduces the model, \bref{sec:optimal} solves for the profit-maximizing non-discriminatory pricing rules, and discusses welfare implications and comparative studies. \bref{sec:extensions} presents extensions. \bref{sec:con} concludes.

\section{Model}\label{sec:model}

A monopolist sells a good or service to a continuum of consumers, each of whom demands one unit.
We normalize the total mass of consumers to one.

\paragraph{Consumer Types}
Each consumer is described by their value for the good $v  \in V \subseteq \RR_+$, the cost $c \in C \subseteq \RR_+$ of being served, a \emph{protected characteristic} $\theta \in \Theta:=\{l,h\}$, and an auxiliary index $r \in [0,1]$.
We denote by $\omega = (v,c,\theta,r) \in \Omega := V \times C \times \Theta \times [0,1]$ a consumer's type.

The value $v$ is the willingness-to-pay of the consumer and $c$ is the (potentially consumer-specific) cost the seller incurs for supplying the good or service. For example, in an insurance market, the cost $c$ could capture the expected damages; in a credit market, it could capture the expected cost of default; and for a physical good, it could simply be the production cost.
The protected characteristic $\theta$ could indicate whether the consumer is male or female, or black or white, which might be correlated with both a consumer's value $v$ and cost $c$. The index $r$ serves as a randomization device that allows the seller to charge different prices to consumers with the same $v$, $c$, and $\theta$.\footnote{All our results remain unchanged without $r$, as the optimal pricing rules we obtain turn out to be non-random.}

\paragraph{Distribution of Consumer Types}
Let $\PP$ be the product of a probability distribution on $V \times C \times \Theta$ and the Lebesgue measure on $[0,1]$. 

We denote by $G(\cdot)=\Pr{c \leq \cdot}$ the distribution of cost, by $\alpha_c = \Pr{ \theta = h }{c}$ the fraction of consumers with characteristic $h$ conditional on having cost $c$,\footnote{When there is no risk of confusion, we slightly abuse the notation and use $v,c,\theta,r$ to denote the random variable as well as a realization.} and by $F_{c,\theta}(\cdot):=\PP[v \leq \cdot \mid c,\theta]$ the distribution of values $v$ of consumers of characteristic $\theta$ and cost $c$. 
We assume that $F_{c,\theta}$ admits a density $f_{c,\theta}$, has full support on an interval $[\lb{v}_c,\ub{v}_c]$ for some $0 \leq \lb{v}_c <\ub{v}_c \leq \infty$,\footnote{In particular, $F_{c,l}$ and $F_{c,h}$ have common supports. This assumption is for the ease of exposition, and the result can be readily extended to distributions with different (interval) supports.} and $h$-consumers have higher values for the product in the likelihood ratio order. That is, $f_{c,h}(v)/f_{c,l}(v)$ is increasing in $v$ on $[\lb{v}_c,\ub{v}_c]$ for all $c$. This assumption implies that, conditional on having the same cost, consumers with protected characteristic $l$ have lower values (in first-order stochastic dominance) and react more strongly to price changes (i.e., are more elastic).

One natural case captured by the above assumption is that of a normal good when consumers with $\theta = h$ are richer. Alternatively, consumers of with $\theta =h$ could be the group with worse outside options. In particular, depending on the context, $h$-consumers could be either the advantaged group (e.g., rich consumers) or the disadvantaged group (e.g., those who have worse outside options).

\paragraph{Pricing Rules} 
A \emph{pricing rule} $p:\Omega \to \RR_+$ is a random variable, where $p(\omega) \in \RR_+$ is the price faced by consumers with type $\omega \in \Omega$. In particular, a pricing rule $p$ allows prices to be personalized, as different consumers could face different prices.   
The seller's profit under pricing rule $p$ equals
\[
    \Pi(p):=\EE[(p(\omega)-c)\mathbf{1}\{v \geq p(\omega)\}]\,.
\]

For a pricing rule to be non-discriminatory, the distribution of prices consumers face can depend on the cost of serving them, but not on their protected characteristic (even if it correlates with their values).

\begin{definition}
A pricing rule $p$ is \emph{non-discriminatory} if for all $c\in C$ and $M \subseteq \RR_+$,
\[
    \Pr{p \in M }{c, \theta = l} = \Pr{ p \in M }{c, \theta = h} \,. 
\]

\end{definition}

Let $\cD$ be the set of all non-discriminatory pricing rules. Non-discriminatory pricing rules exist, since charging a constant price to all consumers is always non-discriminatory.

As an example, U.S. fair lending laws require that in a loan market, black and white consumers with the same expected cost of default must be offered the same interest rates. This regulation is enforced: The Consumer Financial Protection Bureau (CFPB) launched 32 fair lending probes in 2022.
For example, the CFPB investigated Wells Fargo  for ``statistically significant disparities'' in the rates at which the bank offered pricing exemptions (which correspond to $0.25\% - 0.75\%$ interest reductions relative to the rate calculated based on credit risk) to female and black loan applicants \citep{cnbc_wells_fargo_2023}.

\begin{remark}[Pricing Rules and Market Segmentations]\label{rem:segmentation}
A pricing rule is closely related to \emph{market segmentation}, in the sense of \citet{BBM15}. A market segmentation $s:\Omega \to S$ is a random variable that maps consumers' types into some measurable space $S$. Each realization $s(\omega)$ corresponds to a \emph{market segment}, so that $s(\omega)=s(\omega')$ means consumers with type $\omega$ and $\omega'$ belong to the same segment. In this regard, a pricing rule $p$ itself is a market segmentation, where consumers who face the same price belong to the same segment. The converse is also true: given any market segmentation $s$, any pricing rule $p$ that is measurable with respect to $s$ can be interpreted as a rule that charges all consumers in the same segment the same price.  
\end{remark}

\paragraph{Consumer Surplus and Welfare Loss}
Finally, we denote by $$CS(c,\theta;p)= \E{ (v-p)^+ }{ c,\theta }$$ the average consumer surplus, and by $$WL(c,\theta;p)= \E{ \mathbf{1}\{p>v\} (v-c)^+ }{ c,\theta }$$ the welfare loss, of a $\theta$-consumer with cost $c$ under pricing rule $p$.

\section{Optimal Non-Discriminatory Pricing}\label{sec:optimal}
We now maximize the seller's profit over non-discriminatory pricing rules. That is, we solve
\begin{equation}\label{eq:optimal-pricing}
    \Pi^\star:=\sup_{p \in \cD} \Pi (p) \,.
\end{equation}
A pricing rule $p \in \cD$ is undominated if there does not exist another pricing rule $p' \in \cD$ such that both $h$-consumers and $l$-consumers have a higher average surplus, and the seller has a higher profit, with at least one of them being strictly higher. We focus on the undominated pricing rules among all profit-maximizing pricing rules.\footnote{We will further characterize the welfare outcomes of all profit-maximizing pricing rules later in \bref{sec:all}.}

\subsection{Optimal Pricing as an Optimal Transport}
We begin the analysis by establishing that the pricing problem \eqref{eq:optimal-pricing} is equivalent to an optimal transport problem. Fix a non-discriminatory pricing rule $p \in \cD$. For all cost $\tilde{c} \in C$, define a probability measure $\rho_{\tilde{c}} \in \Delta( V^2 )$ on pairs of values $(v_l,v_h)$:\footnote{Note that $\rho_c$ is indeed a probability measure, as it is a mixture of product measures.} For all measurable sets $V_l,V_h \subseteq V$,

\begin{equation}\label{eq:match}
    \rho_{\tilde{c}} ( V_l \times V_h ) :=  \E{ \Pr{v \in V_l }{ p, c = \tilde{c}, \theta=l } \times \Pr{v \in V_h }{p, c = \tilde{c}, \theta= h } }{c = \tilde{c}} \,.
\end{equation}

That is, among those consumers who face the same price and have a cost $c$, $\rho_c$ randomly matches the values of $l$-consumers to values of $h$-consumers into pairs.\footnote{For example, if a constant price $p \in \RR_+$ is charged to all consumers with a given cost $c$, the resulting distribution $\rho_{c}$ is the product distribution generated by $F_{c,l}$ and $F_{c,h}$.} Since $p$ is non-discriminatory, it follows that the marginals of $\rho_c$ equal $F_{c,l}$ and $F_{c,h}$, respectively.

\begin{lemma}\label{lem:lem1}
If $p \in \cD$, then $\rho_c$ has marginal distributions $(F_{c,l},F_{c,h})$ for all $c \in C$.
\end{lemma}

Given such matching schemes $(\rho_c)_{c \in C}$, an upper bound on the expected profit of the seller is thus given by setting the price optimally for each matched pair $(v_l,v_h)$ and each cost $c$:
\[
 \Pi(p) \leq   \int_C\left(\int_{V^2} \pi_c(v_l,v_h) \diff \rho_c\right) G(\diff c)\,,
\]
where $\pi_c(v_l,v_h)$ is the optimal profit when selling to a pair of consumers with values $(v_l,v_h)$ and cost $c$:
\begin{equation}\label{eq:expost-pricing}
    \pi_c(v_l,v_h) := \max_{\tilde{p}\geq 0} (\tilde{p}-c)\left[ (1-\alpha_c) \mathbf{1}\{v_l \geq \tilde{p}\} + \alpha_c \mathbf{1}\{v_h \geq \tilde{p}\} \right] \,.
\end{equation}
Clearly, the optimal price when trade occurs must be either $v_l$ or $v_h$, and thus
\[
    \pi_c(v_l,v_h) = \max \big\{ \min\{v_l,v_h\}-c, \,\,\alpha_c (v_h-c)^+, \,\, (1-\alpha_c)(v_l-c)^+ \big\} \,.
\]
Denote by $\mathcal{R}_c \subset \Delta(V^2)$ the set of all probability measures on $V^2$ with marginals $F_{c,l},F_{c,h}$. The above arguments imply that the seller's optimal profit $\Pi^\star$ is bounded from above by choosing a joint distribution $\rho_c \in \mathcal{R}_c$ to maximize $\pi_c$ for all $c$. Moreover, given any matching schemes $(\rho_c)_{c \in C}$ with $\rho_c \in \mathcal{R}_c$ for all $c$, the pricing rule induced by charging an optimal price that solves $\eqref{eq:expost-pricing}$ for each realized matched pair $(v_l,v_h)$ must be non-discriminatory. Together, we have the following representation of the seller's problem \eqref{eq:optimal-pricing}:

\begin{proposition}[Optimal Transport Representation]\label{prop:opt-transport}
Let $\pi^\star$ be the value of the optimal transport problem: 
\begin{equation}\label{eq:transport}
    \pi^\star:=\int_C \left(\max_{\rho_c \in \mathcal{R}_c} \int_{ V^2} \pi_c(v_l,v_h) \diff \rho_c\right) \diff G( \diff c)\,.
\end{equation}
Then $\pi^\star=\Pi^\star$. Moreover, any solution of \eqref{eq:transport} induces a solution of \eqref{eq:optimal-pricing}; while any solution of \eqref{eq:optimal-pricing} corresponds to a solution of \eqref{eq:transport}, via \eqref{eq:match}. 
\end{proposition}

Intuitively, while the non-discrimination constraint prohibits the seller from tailoring prices to each individual consumer, the seller will optimally tailor to \emph{pairs} of consumers, according to \bref{prop:opt-transport}.  

\paragraph{Relation to Other Optimal Transport Problems} The optimal transport problem given by \eqref{eq:transport} is a non-standard problem along several dimensions. To illustrate, suppose that $\alpha_c = 1/2$ and $c=0$.
In this case, maximizing the profit function $\pi_c$ is equivalent to minimizing $\tilde{\pi}_c(v_l,v_h) := \min \{ | v_h - v_l|, v_h, v_l  \}$.\footnote{To see this, note that $\max \{ \min\{v_l,v_h\}, 0.5 v_l, 0.5 v_h  \} = \max \{ 0.5\min\{v_l - v_h, v_h-v_l\}, - 0.5 v_h, -0.5 v_l\} + 0.5(v_h+v_l) = -0.5 \min \{ |v_h-v_l|, v_l, v_h \}\,,$ where the last equality follows as the marginals of $v_l,v_h$ are fixed.}
In comparison, the objective function in classical optimal transport problems take form of $\hat{\pi}(v_l,v_h) = d(|v_h-v_l|)$, where $d:\RR_+ \to \RR_+$ is a convex function. It is well-known that the assortative matching is optimal \citep[see, e.g.,][]{V09} for these problems.

\begin{figure}[t]
\begin{center}
\begin{tikzpicture}[scale=1.0]
    \begin{axis}[
        domain=0:10, 
        samples=200, 
        axis lines=middle,
        xlabel={$v_h$}, ylabel={$\tilde{\pi}_c(3,v_h)$},
        xtick={0, 3, 10},
        ytick={0, 3},
        ymin=0, ymax=4,
        xmin=0, xmax=10,
        width=10cm, height=4.5cm,
        grid=both,
        major grid style={line width=.2pt,draw=gray!50},
        minor grid style={line width=.1pt,draw=gray!50},
        enlargelimits=false,
        thick
    ]
    
    \addplot[
        domain=0:10,
        blue, thick,
    ] {min(3, x, abs(x - 3))};  
    \end{axis}
\end{tikzpicture}
\end{center}
\caption{\label{fig:cost}The cost function $\min\{v_l,v_h, |v_h - v_l| \} $ for $\alpha=0.5$, $c=0$, and $v_l=3$.}
\end{figure}
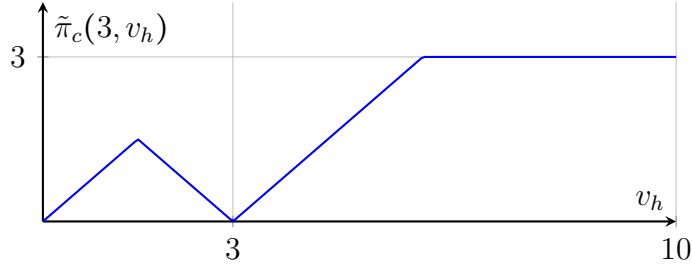

More broadly, the objective function $\tilde{\pi}_c$ does not satisfy the common properties studied in the optimal transport literature (see \bref{fig:cost} for an illustration of $\tilde{\pi}_c$):
\begin{compactenum}[(i)]
\item The profit function is not supermodular or submodular. 
\item The profit function is not translation invariant, i.e. $\tilde{\pi}_c(v_l,v_h) \neq \tilde{\pi}_c(v_l+\epsilon,v_h+\epsilon)$.
\item The profit function is non-monotone, i.e. $|v_h - v_l| > |v_h' - v_l| \nRightarrow \tilde{\pi}_c(v_l,v_h) > \tilde{\pi}_c(v_l,v_h')$.
\item The profit function is non-convex/concave, i.e. $v_h \mapsto \tilde{\pi}_c(v_h,v_l)$ is neither (quasi) convex, nor (quasi) concave.
\end{compactenum}
Due to these differences, the solution to our problem will be quite different from the typical solutions in the optimal transport literature.\footnote{To our knowledge, the only other paper that establishes explicit properties of the solution for a concrete optimal transport problem without imposing these assumption is \cite{boerma2025sorting}, who study the function $|v_l-v_h|^\beta$ for $\beta \in (0,1)$ and thus relax condition (i) while keeping (ii)-(iv).}

\subsection{Profit-Maximizing Pricing Rules}
By \bref{prop:opt-transport}, the seller's problem \eqref{eq:optimal-pricing} is equivalent to a family of optimal transport problems indexed by $c$. For the ease of exposition, we first impose the following assumption on $F_{c,l}$ and $F_{c,h}$, which allows us to focus on the more economically interesting cases, and defer the characterization of optimal pricing rules for general distributions to \bref{sec:general}.
\begin{assumption}\label{ass:no-full-surplus-extraction}
    $\Pr{v \leq c }{c,\theta=l} < \| F_{c,l} - F_{c,h}\|$ for almost all $c \in C$.
\end{assumption}
Here, $\|\cdot\|$ denotes the total variation distance.\footnote{Formally, $\|G-H\| = \sup_{A \subseteq \RR_+} \left| \int_A \diff G - \int_A \diff H\right|$ for all CDFs $G$,$H$.}
\bref{ass:no-full-surplus-extraction} thus requires that, conditional on each cost, the distance between the value distribution of $h$-consumers and that of $l$-consumers is always greater than the share of $l$-consumers whose cost exceed their value.
As we show in \bref{cor:full-surplus-extraction}, the seller can in fact fully extract all gains from trade conditional on some $c \in C$ if and only if \bref{ass:no-full-surplus-extraction} does not hold. 
Note that when providing the good is always costless (i.e., $c=0$), \bref{ass:no-full-surplus-extraction} is trivially satisfied, and the only pricing rule that achieves full surplus extraction is to charge each consumer their value $p(v,c,\theta,r)=v$, which is discriminatory if $F_{c,l} \neq F_{c,h}$.

To make the pricing rule non-discriminatory, in the spirit of \bref{prop:opt-transport}, the seller could first match consumers into pairs and charge each matched pair the same price. As demonstrated by the following examples:  

\begin{figure}[t!]
\centering
\begin{subfigure}[b]{0.4\linewidth}
\begin{tikzpicture}[scale=5]
\draw [gray1, very thick] plot [smooth, tension=0.8] coordinates { (0,0) (0.25,0.04) (0.75,0.13) (1,0)};
\draw [gray1, very thick] plot [smooth, tension=0.8] coordinates { (0,-0.3) (0.25,-0.17) (0.75,-0.26) (1,-0.3)};
\draw (0.95,0.07) node [above=2pt] {  $f_{c,h}$};
\draw (0.05,-0.23) node [above=2pt] {  $f_{c,l}$};
\draw (0,0) node [left=2pt] { $\theta=h$};
\draw (0,-0.3) node [left=2pt] { $\theta=l$};
\draw [->, very thick] (0.6,-0.01)--(0.4,-0.29);
\draw [ultra thick, red] (0,0)--(1,0);
\draw [ultra thick, red] (0,-0.3)--(1,-0.3);
\draw [very thick] (0.2,-0.02)--(0.2,0.02);
\draw [very thick] (0.2,-0.32)--(0.2,-0.28);
\draw [ultra thick] (0,0)--(1,0);
\draw [ultra thick] (0,-0.02)--(0,0.02);
\draw [ultra thick] (1,-0.02)--(1,0.02); 
\draw [ultra thick] (0,-0.3)--(1,-0.3);
\draw [ultra thick] (0,-0.32)--(0,-0.28);
\draw [ultra thick] (1,-0.32)--(1,-0.28);
\draw [ultra thick, blue] (0,0)--(1,0);
\draw [ultra thick, blue] (0,-0.3)--(1,-0.3);
\draw (0.2,0) node [below=2pt] {  $c$};
\draw (0.2,-0.3) node [below=2pt] {  $c$};
\draw (0.6,0.075) node [above=2pt]{  $\color{white}F_{c,h}^{-1}(1-q_c)$};
\draw (0.4,-0.3) node [below=2pt] {  $\color{white}F_{c,l}^{-1}(q_c)$};
\end{tikzpicture}
\caption{Assortative.}
\label{fig:example1}
\end{subfigure}
\begin{subfigure}[b]{0.4\linewidth}
\begin{tikzpicture}[scale=5]
\draw [gray1, very thick] plot [smooth, tension=0.8] coordinates { (0,0) (0.25,0.04) (0.75,0.13) (1,0)};
\draw [gray1, very thick] plot [smooth, tension=0.8] coordinates { (0,-0.3) (0.25,-0.17) (0.75,-0.26) (1,-0.3)};
\draw (0.95,0.07) node [above=2pt] {  $f_{c,h}$};
\draw (0.05,-0.23) node [above=2pt] {  $f_{c,l}$};
\draw (0,0) node [left=2pt] { $\theta=h$};
\draw (0,-0.3) node [left=2pt] { $\theta=l$};
\draw [very thick] (0.2,-0.02)--(0.2,0.02);
\draw [very thick] (0.2,-0.32)--(0.2,-0.28);
\draw [very thick] (0.4,-0.32)--(0.4,-0.28);
\draw [very thick] (0.6,-0.02)--(0.6,0.02);
\draw [->, very thick] (0.3,-0.29)--(0.7,-0.01);
\draw [->, very thick] (0.7,-0.29)--(0.3,-0.01);
\draw [ultra thick] (0,0)--(1,0);
\draw [ultra thick] (0,-0.02)--(0,0.02);
\draw [ultra thick] (1,-0.02)--(1,0.02); 
\draw [ultra thick] (0,-0.3)--(1,-0.3);
\draw [ultra thick] (0,-0.32)--(0,-0.28);
\draw [ultra thick] (1,-0.32)--(1,-0.28);
\draw [ultra thick, red] (0.4,-0.3)--(1,-0.3);
\draw [ultra thick, red] (0,0)--(0.6,0);
\draw [ultra thick, blue] (0.6,0)--(1,0);
\draw [ultra thick, blue] (0,-0.3)--(0.4,-0.3); 
\draw (0.2,0.01) node [above=2pt] {  $c$};
\draw (0.2,-0.3) node [below=2pt] {  $c$};
\draw (0.6,0.075) node [above=2pt]{  $F_{c,h}^{-1}(1-q_c)$};
\draw (0.4,-0.3) node [below=2pt] {  $F_{c,l}^{-1}(q_c)$};
\end{tikzpicture}
\caption{Partly Anti-Assortative.}
\label{fig:example2}
\end{subfigure}
\caption{Assortative and Partly Anti-Assortative Pricing Rules. Regions of values matched are of the same color. The arrows illustrate a generic pair of values that are matched together, and the direction indicate the price each matched pair faces, conditional on being above $c$.}
\label{fig:examples}
\end{figure}
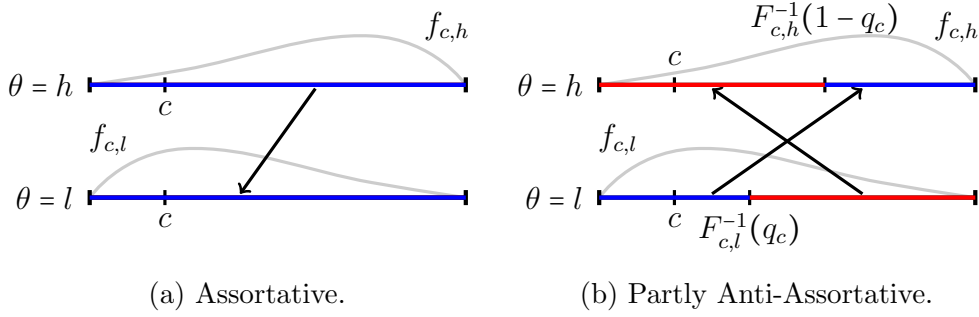

\begin{definition}[Assortative Matching]
The assortative pricing rule $p^{ass} \colon \Omega \to V$ is defined by matching consumers into pairs assortatively conditional on $c$, and charging each pair the maximum of the lower value of the pair and $c$. That is: 
\begin{equation}\label{eq:assortative-matching}
    p^{ass}(v,c,\theta,r ) = \begin{cases} \max\{(F_{c,l}^{-1} \circ F_{c,h})(v),c\} &\text{ if } \theta = h \\
     \max\{v,c\} &\text{ if } \theta = l\\
    \end{cases}
\end{equation}
\end{definition}

The pricing rule $p^{ass}$ yields a profit equals the gains from trade of $l$-consumers: $\Pi(p^{ass})=\E{ (v-c)^+}{\theta = l}$. Alternatively, the seller could charge higher prices on average, by matching some high-value $l$-consumers with low-value $h$-consumers and charge these pairs the value of the $h$-consumer, while matching the rest of the low-value $l$-consumers with the remaining high-value $h$-consumers and charge these pairs the higher value of the pair.

\begin{definition}[Partly Anti-Assortative Matching]
A partly anti-assortative pricing rule $p^{anti} : \Omega \to V$ is defined by
\begin{equation}\label{eq:partly}
    p^{anti}(v,c,\theta,r) = \begin{cases}
    \max\{F_{c,h}^{-1} (F_{c,l}(v)-q_c ),c\} &\text{ if } \theta = l \text{ and } v > F_{c,l}^{-1}(q_c) \\
    \max\{F_{c,h}^{-1} (F_{c,l}(v)+1-q_c ),c\} &\text{ if } \theta = l \text{ and } v \leq F_{c,l}^{-1}(q_c) \\
     \max\{v,c\} &\text{ if } \theta = h
     \end{cases} \,,
\end{equation}
for some quantiles $(q_c)_{c \in C}$. 
\end{definition}
Regardless of the quantiles $(q_c)_{c \in C}$, $h$-consumers always have their surplus extracted under $p^{anti}$, and thus the seller's profit is at least $\E{\alpha_c (v-c)^+}{\theta=h}$. How many more $l$-consumers purchase, on the other hand, depends on the choice of quantiles. For instance, if $q_c=1$, then no $l$-consumers would purchase. More specifically, all $l$-consumers with values below $F_{c,l}^{-1}(q_c)$ would not purchase, while $l$-consumers with values above $F_{c,l}^{-1}(q_c)$ may or may not purchase, and fewer of these consumers would purchase as $q_c$ becomes smaller.\footnote{As we show in the Appendix, the smallest $q_c$ such that all these consumers would purchase is given by $q_c=q^\star_c:=\max_{x \in V}(F_{c,l}(x)-F_{c,h}(x))$.} 

In essence, $p^{anti}$ charges higher prices to $h$-consumer at the cost of excluding some $l$-consumers and thus might obtain a higher or lower profit than $p^{ass}$, which sells to more consumers at lower prices. The trade-off between efficiency and profit that results from the non-discrimination constraint resembles that of standard screening concerns, even though buyers hold \emph{no private information} here.

\bref{fig:examples} illustrates the assortative pricing rule and the partly anti-assortative pricing rule. Although both of these pricing rules are simple and non-discriminatory, it turns out that neither is optimal and the optimal non-discriminatory pricing rule takes a more intricate form that balances the efficiency-profit trade-off.

\paragraph{An Optimal Pricing Rule}
We now describe an optimal pricing rule, which we denote by $p^\star$. For all $c \in C$, let $\Delta_c(v):=F_{c,l}(v)-F_{c,h}(v)$ and let $\overline{\Delta}_c^{-1},\underline{\Delta}_c^{-1} : [0,1] \to V$ be the larger and smaller inverses of $\Delta_c$, and let $v^\star_c$ be the unique solution to $f_{c,l}(v^\star_c)=f_{c,h}(v^\star_c)$.\footnote{Formally, since $F_{c,h}$ dominates $F_{c,l}$ in the likelihood ratio order, $\Delta_c$ is quasi-concave and is maximized at $v^\star_c$ Thus, for any $q \in [0,\Delta_c(v^\star_c)]$, there are exists a unique pair $(\lb{\Delta}_c^{-1}(q),\ub{\Delta}_c^{-1}(q)) \in V^2$ such that $\lb{\Delta}_c^{-1}(q) \leq v^\star_c \leq \ub{\Delta}_c^{-1}(q)$ and $\Delta_c(\lb{\Delta}_c^{-1}(q))=q=\Delta_c(\ub{\Delta}_c^{-1}(q))$.}
The following lemma identifies some critical cutoffs.    
\begin{lemma}\label{lem:system}
For all $c \in C$, there exists a unique increasing vector $\kappa_c \in \RR^5$ with $\kappa_c^4<v^\star_c<\kappa_c^5$ 
such that 
\begin{equation}\label{eq:system}
\begin{aligned}
    &\kappa^2_c=F_{c,l}^{-1}(\Delta_c(\kappa_c^3)+F_{c,h}(\kappa^1_c))=F_{c,l}^{-1}(\Delta_c(\kappa_c^4))=F_{c,l}^{-1}(\Delta_c(\kappa_c^5))\\
    &\kappa^1_c-c=(1-\alpha_c)\cdot(\kappa^3_c-c)=\alpha_c\cdot(\kappa^5_c-\kappa^4_c)\,.
\end{aligned}
\end{equation}
\end{lemma}
\noindent Henceforth, we will call $\kappa_c$ the unique solution to \eqref{eq:system} and define the pricing rule $p^\star$ as: 
\begin{equation}\label{eq:p-star}
\begin{aligned}
    p^\star(v,c,l,r) & :=\begin{cases}
     \overline{\Delta}_c^{-1}(\Delta_c(\kappa^5_c) - F_{c,l}(v)),&\mbox{if } v < \kappa_c^2\\
    F_{c,h}^{-1}(F_{c,l}(v)-F_{c,l}(\kappa_c^2)+F_{c,h}(\kappa_c^1)),&\mbox{if } v \in [\kappa_c^2,\kappa_c^3)\\
    v,&\mbox{if } v \geq \kappa_c^3
    \end{cases}\,;\\
 p^\star(v,c,h,r) & :=\begin{cases}
    \lb{\Delta}_{c}^{-1}\left(F_{c,h}(v)+\Delta_c(\kappa_c^3)\right),\hspace{1.75cm}&\mbox{if } v < \kappa_c^1\\
    F_{c,l}^{-1}(F_{c,h}(v)+\Delta_c(\kappa_c^4)),&\mbox{if } v \in [\kappa_c^4,\kappa_c^5)\\
    v,&\mbox{if } v \in [\kappa_c^5,\infty) \cup (\kappa_c^1,\kappa_c^4)
    \end{cases}\,.
\end{aligned}
\end{equation}

\begin{theorem}[Optimal Pricing]\label{thm1}$ $
\begin{compactenum}[(i)]
\item $p^\star$ is a profit-maximizing non-discriminatory pricing rule. That is, $p^\star$ solves \eqref{eq:optimal-pricing}. 
\item Every undominated profit-maximizing non-discriminatory pricing rule $p$ induces the same average surplus for consumer of each protected characteristic and cost. That is, $CS(c,\theta;p)=CS(c,\theta;p^\star)$ for all $c \in C$ and $\theta \in \{l,h\}$.
\end{compactenum}
\end{theorem}

\begin{figure}[h!]
\centering
\tikzset{
solid node/.style={circle,draw,inner sep=1.25,fill=black},
hollow node/.style={circle,draw,inner sep=1.25}
}
\begin{subfigure}[b]{0.4\linewidth}
\hspace*{-2.6cm}
\begin{tikzpicture}[scale=4]
\draw [<->, very thick] (-1.1,1.05) node (yaxis) [above] {  $p^\star$}
        |- (-0.05,0) node (xaxis) [right] { $v$};
\draw [dashed, thick] (-1.1,0)--(-0.1,1); 
\draw (-1.1,0.7) node [left=2pt] {  $\kappa_c^5$};
\draw (-1.1,0.1) node [left=2pt] {  $\kappa_c^1$};
\draw [dotted, thick] (-1.1,0.1)--(-0.95,0.1); 
\draw [dotted, thick] (-0.4,0.7)--(-1.1,0.7);
\draw [dotted, thick] (-0.4,0.7)--(-0.4,0); 
\draw (-0.4,0) node [below=2pt] {  $\kappa_c^5$};
\draw [dotted, thick] (-0.95,0)--(-0.95,1);
\draw [dotted, thick] (-0.8,0)--(-0.8,0.3);
\draw [dotted, thick] (-0.8,0.3)--(-1.1,0.3);
\draw (-1.1,0.3) node [left=2pt] {  $\kappa_c^3$};
\draw (-0.95,0.01) node [below=2pt] {  $\kappa_c^2$};
\draw (-0.8,-0.01) node [below=2pt] {  $\kappa_c^3$};
\draw [blue,ultra thick] (-0.8,0.3)--(-0.1,1);
\draw [blue,ultra thick] plot [smooth, tension=0.8] coordinates { (-1.1,0.7) (-1.05,0.85) (-1,0.95) (-0.95,1)};
\draw [blue,ultra thick] plot [smooth, tension=0.5] coordinates { (-0.95,0.1) (-0.9,0.12) (-0.85,0.18) (-0.8,0.3)};
\draw (-0.6,-0.12) node [below=2pt] {  $\theta=l$};
\draw [<->, very thick] (0.5,1.05) node (yaxis) [above] {  $p^\star$}
        |- (1.55,0) node (xaxis) [right] { $v$};
\draw [dashed, thick] (0.5,0)--(1.5,1); 
\draw (0.5,0.3) node [left=2pt] {  $\kappa_c^3$};
\draw (0.5,0.5) node [left=2pt] {  $\kappa_c^4$};
\draw [dotted, thick] (0.5,0.5)--(1,0.5);
\draw [dotted, thick] (0.6,0)--(0.6,0.5);
\draw [dotted, thick] (0.5,0.3)--(0.8,0.3);
\draw [dotted, thick] (0.8,0.3)--(0.8,0);
\draw (0.8,-0.01) node [below=2pt] {  $\kappa_c^3$};
\draw (0.6,0.01) node [below=2pt] {  $\kappa_c^1$};
\draw (1,0) node [below=2pt] {  $\kappa_c^4$};
\draw (1.2,0) node [below=2pt] {  $\kappa_c^5$};
\draw [dotted, thick] (1,0)--(1,0.5);
\draw [dotted, thick] (1.2,0)--(1.2,0.7); 
\draw [blue, ultra thick] plot [smooth, tension=0.5] coordinates {(0.5,0.3) (0.53,0.4) (0.57,0.47)  (0.6,0.5)};
\draw [blue, ultra thick] plot [smooth, tension=0.8] coordinates {(1,0.5) (1.05,0.52) (1.15,0.6) (1.2,0.7)};
\draw [blue, ultra thick] (0.6,0.1)--(1,0.5);
\draw [blue, ultra thick] (1.2,0.7)--(1.5,1);
\draw (0.5,0.7) node [left=2pt] {  $\kappa_c^5$};
\draw [dotted, thick] (0.5,0.7)--(1.2,0.7);
\draw (1,-0.12) node [below=2pt] {  $\theta=h$};
\end{tikzpicture}
\label{figure1}
\end{subfigure}%
\caption{Optimal Pricing Rule $p^\star$.}
\label{fig:optimal-pricing}
\end{figure}
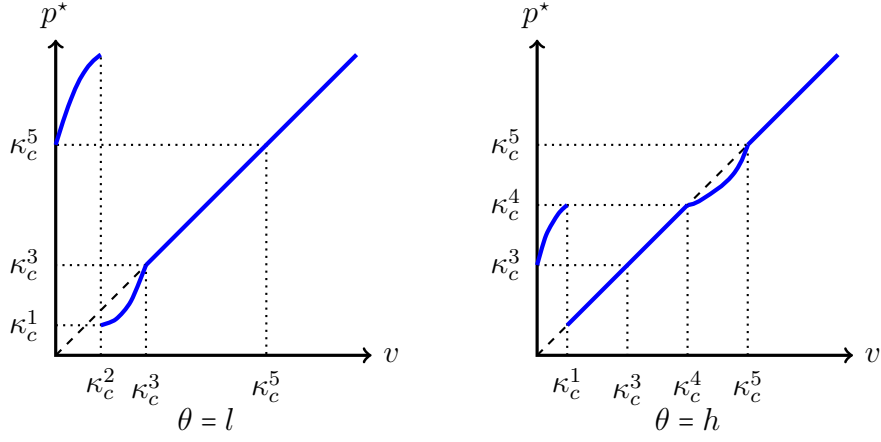

\bref{fig:optimal-pricing} plots the optimal pricing rule $p^\star$. Under $p^\star$, for $l$-consumers, those with values above the cutoff $\kappa^3_c$ face a price equal to their value; those with values in the interval $[\kappa^2_c, \kappa_c^3)$ face a price less than their value; and those with values below $\kappa_c^2$ face a price that exceeds their value. The prices faced by $h$-consumers have the same feature, except that the cutoffs are different. Using \eqref{eq:system}, it can be verified that for all $c \in C$, the distributions of $p^\star$ conditional on $(c,l)$ and on $(c,h)$ are the same, and thus $p^\star$ is indeed non-discriminatory.  Notably, the pricing rule $p^\star$ is non-monotone in consumers' values given $c$ and $\theta$; and does not depend on the randomization device $r$. 

The optimality of $p^\star$ stems from delicately balancing the efficiency-profit trade-off imposed by the non-discrimination constraint. Under $p^\star$, low-value consumers do not purchase (i.e., $p^\star$ is above the 45-degree line for low values in \bref{fig:optimal-pricing}), which in turn allows the seller to carefully choose the price they face in order to be able to target the high-value consumers with a different protected characteristic (i.e., $p^\star$ coincides with the 45-degree line for high values) while maintaining the same price distributions for both groups. In the meantime, intermediate-value consumers all purchase, and some of them purchase at a price below their values (i.e., $p^\star$ is below the 45-degree line for some intermediate values). This allows the seller to sell to more consumers while leaving them as little surplus as possible.   

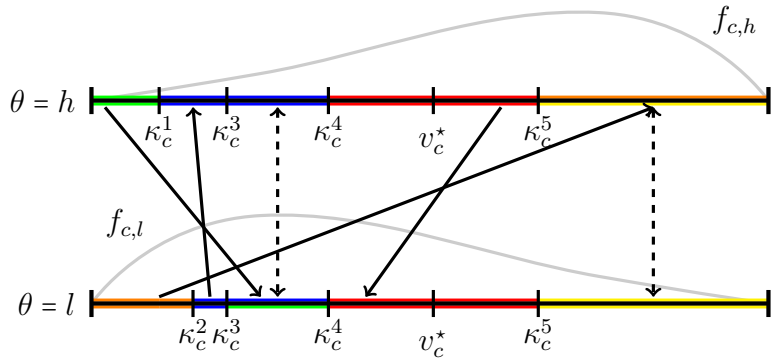
\begin{figure}[h!]
\centering
\begin{tikzpicture}[scale=9]
\draw [gray1, very thick] plot [smooth, tension=0.8] coordinates { (0,0) (0.25,0.04) (0.75,0.13) (1,0)};
\draw [gray1, very thick] plot [smooth, tension=0.8] coordinates { (0,-0.3) (0.25,-0.17) (0.75,-0.26) (1,-0.3)};
\draw (0.95,0.07) node [above=2pt] {  $f_{c,h}$};
\draw (0.05,-0.23) node [above=2pt] {  $f_{c,l}$};
\draw [very thick, green] (0,0.005)--(0.1,0.005);
\draw [very thick, green] (0,-0.005)--(0.1,-0.005);
\draw [very thick, orange] (0,-0.295)--(0.1,-0.295);
\draw [very thick, orange] (0,-0.305)--(0.1,-0.305);
\draw [very thick, orange] (0.1,-0.295)--(0.15,-0.295);
\draw [very thick, orange] (0.1,-0.305)--(0.15,-0.305);
\draw [very thick, blue] (0.15,-0.295)--(0.2,-0.295);
\draw [very thick, blue] (0.15,-0.305)--(0.2,-0.305);
\draw [very thick, blue] (0.1,-0.005)--(0.2,-0.005);
\draw [very thick, blue] (0.1,0.005)--(0.2,0.005);
\draw [very thick, blue] (0.2,0.005)--(0.35,0.005);
\draw [very thick, blue] (0.2,-0.005)--(0.35,-0.005);
\draw [very thick, blue] (0.2,-0.295)--(0.35,-0.295);
\draw [very thick, green] (0.2,-0.305)--(0.35,-0.305);
\draw [very thick, red] (0.35,0.005)--(0.66,0.005);
\draw [very thick, red] (0.35,-0.005)--(0.66,-0.005);
\draw [very thick, red] (0.35,-0.295)--(0.66,-0.295);
\draw [very thick, red] (0.35,-0.305)--(0.66,-0.305);
\draw [very thick, orange] (0.66,0.005)--(1,0.005);
\draw [very thick, yellow] (0.66,-0.005)--(1,-0.005);
\draw [very thick, yellow] (0.66,-0.295)--(1,-0.295);
\draw [very thick, yellow] (0.66,-0.305)--(1,-0.305);
\draw [->, very thick] (0.02,-0.01)--(0.25,-0.29); 
\draw [<-, very thick,] (0.83,-0.01)--(0.1,-0.29);
\draw [->, very thick] (0.605,-0.01)--(0.405,-0.29);
\draw [<-, very thick] (0.15,-0.01)--(0.175,-0.29);
\draw [<->, dashed, very thick] (0.275,-0.01)--(0.275,-0.29);
\draw [<->, dashed, very thick] (0.83,-0.01)--(0.83,-0.29);
\draw [very thick] (0.1,-0.02)--(0.1,0.02);
\draw [very thick] (0.15,-0.32)--(0.15,-0.28);
\draw [very thick] (0.2,-0.02)--(0.2,0.02);
\draw [very thick] (0.35,-0.02)--(0.35,0.02);
\draw [very thick] (0.66,-0.02)--(0.66,0.02);
\draw [very thick] (0.2,-0.32)--(0.2,-0.28);
\draw [very thick] (0.35,-0.32)--(0.35,-0.28);
\draw [very thick] (0.66,-0.32)--(0.66,-0.28);
\draw [very thick] (0.505,-0.02)--(0.505,0.02);
\draw [very thick] (0.505,-0.32)--(0.505,-0.28);
\draw [ultra thick] (0,0)--(1,0);
\draw [ultra thick] (0,-0.02)--(0,0.02);
\draw [ultra thick] (1,-0.02)--(1,0.02); 
\draw [ultra thick] (0,-0.3)--(1,-0.3);
\draw [ultra thick] (0,-0.32)--(0,-0.28);
\draw [ultra thick] (1,-0.32)--(1,-0.28);
\draw (0,0) node [left=2pt] { $\theta=h$};
\draw (0,-0.3) node [left=2pt] { $\theta=l$};
\draw (0.1,0) node [below=2pt] {  $\kappa_c^1$};
\draw (0.15,-0.3) node [below=2pt]{  $\kappa^2_c$};
\draw (0.2,0) node [below=2pt] {  $\kappa^3_c$};
\draw (0.2,-0.3) node [below=2pt] {  $\kappa^3_c$};
\draw (0.505,-0.015) node [below=2pt] {  $v^\star_c$};
\draw (0.505,-0.315) node [below=2pt] {  $v^\star_c$};
\draw (0.35,0) node [below=2pt] {  $\kappa^4_c$};
\draw (0.35,-0.3) node [below=2pt] {  $\kappa^4_c$};
\draw (0.66,0) node [below=2pt] {  $\kappa^5_c$};
\draw (0.66,-0.3) node [below=2pt] {  $\kappa^5_c$};
\end{tikzpicture}
\caption{Matching Scheme $\rho^\star_c$.}
\label{fig1}
\end{figure}

From \bref{prop:opt-transport}, the pricing rule $p^\star$ can alternatively be described by a family of matching schemes $\{\rho^\star_c\}_{c\in C}$ that solves \eqref{eq:transport}. \bref{fig1} plots the matching scheme $\rho_c^\star$ for a given $c$. In \bref{fig1}, the top interval depicts values of $h$-consumers, and the bottom interval depicts values of $l$-consumers. Subintervals with the same colors on each side are matched together: subintervals connected by solid arrows are matched positively assortatively, whereas subintervals connected by dashed arrows are matched by pairing consumers with the same values. The direction of the arrow indicates which value in a matched pair equals the price under $p^\star$. According to \bref{fig1}, $\rho^\star_c$ matches $h$-consumers who have values $v \leq \kappa^1_c$ with $l$-consumers who have values $v \in (\kappa^3_c,\kappa^4_c]$. The seller's optimal price, by \eqref{eq:system}, for each of these matched pairs, equals the high value of the pair. 
Meanwhile, $l$-consumers with $v \leq \kappa_c^2$ are matched with an equal mass of $h$-consumers with $v>\kappa^5_c$, and the the seller's optimal price for each of these matched pairs, by \eqref{eq:system}, equals the high value of the pair; $l$-consumers with $v \in (\kappa^2_c,\kappa^3_c]$ are matched assortatively with $h$-consumers with $v \in (\kappa^1_c,\kappa^3_c]$, and the seller's optimal price for each of these matched pairs, by \eqref{eq:system}, equals the low value of the pair; consumers with $v \in (\kappa^4_c,\kappa^5_c]$ are matched assortatively, and the seller's optimal price for each of these matched pairs, by \eqref{eq:system}, equals the low value of the pair. Lastly, all the remaining consumers are matched with those with the same values, and the seller's optimal price equals their values. By \eqref{eq:system}, each of these matching regions have equal mass of consumer values and thus the matching scheme $\rho_c^\star$ is well-defined. 

As some consumers face a price below their values under $p^\star$, \bref{thm1} implies that the seller cannot fully extract all gain from trade under the non-discrimination constraint. 

\begin{remark}
There are many other pricing rules beyond $p^\star$ that maximize the sellers profit.
For example, in the interval $[\kappa_c^4,\kappa_c^5]$ where the seller matches $l$ and $h$ consumers, any other matching that ensures every $l$-consumer is matched with an $h$-consumer of higher value yields the same expected profit of $\E{v}{c, \theta=l, v \in [\kappa_c^4,\kappa_c^5]}-c$.
As a result, for each individual consumer, their surplus might be different under different undominated profit-maximizing pricing rules. Nonetheless, \bref{thm1} ensures that all undominated optimal pricing rules lead to the same \emph{average} consumer surplus for each protected characteristic $\theta$ and cost $c$.
\end{remark}

\paragraph{An Example of Insurance Demand}

To illustrate \bref{thm1}, we next present a simple example in the context of insurance markets.
Suppose that the values of consumers with cost $c$ and protected characteristic $\theta$ are exponentially distributed with mean $\E{v}{c,\theta}=\lambda_\theta c$, for some $0<\lambda_l<\lambda_h$.\footnote{Formally, $F_{c,l}(x)=1-e^{-\nicefrac{x}{\lambda_l c}}$ and $F_{c,h}(x)=1-e^{-\nicefrac{x}{\lambda_h c}}$. Since $\lambda_h>\lambda_l$, $F_{c,h}$ dominates $F_{c,l}$ in the likelihood ratio order.
} 
In the context of insurance, this means that consumers who face greater risk (i.e., higher $c$) have on average higher value for insurance (i.e., higher $\EE[v\mid c]$). 

Defining $\gamma = \nicefrac{\lambda_h}{\lambda_l}$, we show in the appendix that \bref{ass:no-full-surplus-extraction} is satisfied if and only if
\[
    1 - e^{-\frac{1}{\lambda_l}} < \gamma ^{\frac{-\gamma}{\gamma-1}} \left( \gamma - 1\right) \,.
\]
Intuitively, this assumption is satisfied if either (i) the difference in the expected valuations for the product as measured by $\gamma$ is large or (ii) low type consumers value the product not too little relative to its production cost (i.e.,
$\lambda_l=\E{v/c}{c,\theta=l}$ is large).
For example, if $h$-consumers value the product on average twice as much than $l$-consumers, then \bref{ass:no-full-surplus-extraction} is satisfied whenever $l$-consumers' values are approximately three times has high as their costs on average. Furthermore, under this distribution, it follows that the cutoffs defined by \eqref{eq:system}, as well as the average consumer surplus, must be linear in $c$: $\kappa_c=c\cdot \kappa_1$ and $CS(c,\theta;p^\star)=c\cdot CS(1,\theta;p^\star)$ for all $c$ and $\theta$.\footnote{In fact, we show in the Appendix that for any distributions that take the form of $F_{c,\theta}(x)=F_{\theta}(\nicefrac{x}{c})$, and $\alpha_c=\alpha \in (0,1)$, for all $c$ and $\theta$, \bref{ass:no-full-surplus-extraction} holds if and only if $F_{l}(1)<\|F_l-F_h\|$, and the cutoffs $\kappa_c$ and consumer surplus $CS(c,\theta;p^\star)$ must be linear in $c$.  }

\begin{figure}[t]
    \centering
    \includegraphics{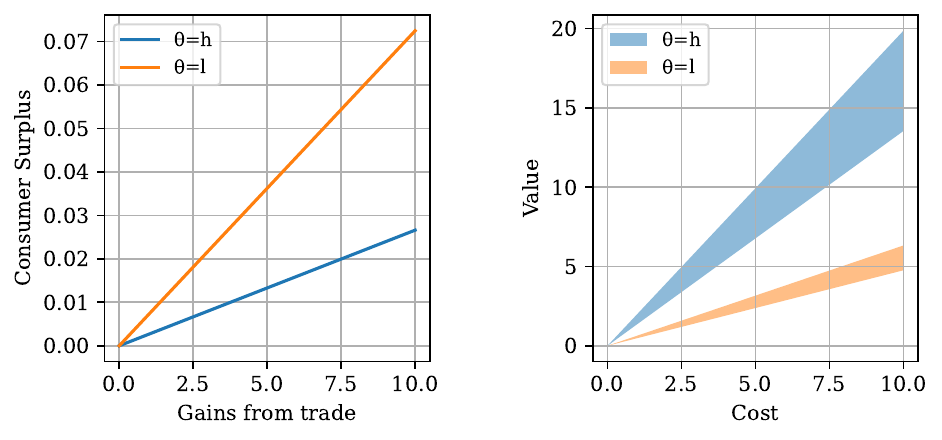}
    \caption{Consumer Surplus for $\alpha_c=\nicefrac{1}{2}$, $\lambda_l=1,\lambda_h=3$. The left panel plots the average surplus of $l$ and $h$ consumers as a function of gains from trade. The right panel plots the consumers who retain a positive surplus under $p^\star$.}
    \label{fig:surplus-example}
\end{figure}

\bref{fig:surplus-example} illustrates the consumer surplus under the optimal non-discriminatory pricing rule $p^\star$ for the case of $\alpha_c=\nicefrac{1}{2}$, $\lambda_l=1$, and $\lambda_h=3$. The left panel displays the average surplus of $l$ and $h$ consumers with each level of gains from trade $\E{v}{c}-c$. Meanwhile, the right panel displays all consumer types $(v,c,\theta,r)$ who receive strictly positive surplus. According to this panel, some  consumers receive positive surplus, and $l$-consumers who receive positive surplus always value the product less than $h$-consumers who receive positive surplus. 

\section{Welfare Implications}\label{sec:welfare}
In this section, we discuss the welfare implications of non-discrimination regulations using the characterization given by \bref{thm1}. 
\subsection{Consumer Surplus and Welfare Losses}
An immediate consequence of \bref{thm1} is that consumers generally retain a positive surplus under any optimal non-discriminatory pricing rule, as stated in \bref{prop:surplus} below.

\begin{proposition}\label{prop:surplus}
Under the optimal pricing rule $p^\star$, for all $c \in C$,
\begin{compactenum}[(i)]
    \item $CS(c,h;p^\star)>0$; while $WL(c,h;p^\star)>0$ whenever $\lb{v}_c \leq c$;
    \item $CS(c,l;p^\star)>0$ and $WL(c,l;p^\star)>0$ if and only if $\alpha_c \cdot (\ub{v}_c-c) > \lb{v}_c-c$.
\end{compactenum} 
\end{proposition}

According to \bref{prop:surplus}, $h$-consumers always retain a positive surplus under the optimal non-discriminatory pricing rule $p^\star$, and would have a positive deadweight loss whenever the lowest value does not have strictly positive gains from trade (e.g., when $\lb{v}_c=0$, as in the insurance example in \bref{sec:optimal} with exponential value distributions). In the meantime, $l$-consumers retain a positive surplus if and only if $\alpha_c(\ub{v}_c-c)>\lb{v}_c-c$ for some $c$. This condition means that when the highest-value consumer is matched with the lowest-value consumer, it would be more profitable for the seller to only sell to the high-value consumer by charging a high price, which is satisfied whenever the support $[\lb{v}_c,\ub{v}_c]$ of the value distribution conditional on cost is wide enough, and, in particular, whenever $\lb{v}_c \leq c$. Overall, \bref{prop:surplus} implies that consumers would typically retain a positive surplus under the optimal pricing rule $p^\star$, but at the expense of some consumers who are efficient to trade with being excluded.      

The fact that consumers generally retain a positive surplus and the deadweight loss is generally positive under the optimal non-discriminatory pricing rule $p^\star$ is reminiscent of the notion of information rents in screening problems. In standard monopolistic screening problems, agents typically retain some information rents because the principal does not observe the agent's private type, and thus has to pay the agent some rents to elicit this information. In the context of non-discriminatory personalized pricing, although the seller \emph{observes} the consumers' types and can propose personalized prices that depend on each consumer's type, the non-discrimination constraint effectively prohibits the seller from \emph{using} certain information conveyed by a consumer's type. Indeed, by requiring the price distribution to be the same for different protected characteristics, the non-discrimination constraint prohibits the seller from using any information---even though it is observable---conveyed by the protected characteristics $\theta$ when designing personalized prices. As a result, consumers would be able to keep some rents as a part of their types are \emph{effectively} private. 

However, the information rents are manifested differently under non-discriminatory personalized pricing. In standard screening problems with one-dimensional types and single-crossing preferences, information rents are enjoyed by high-type agents. However, under non-discriminatory personalized pricing, it is the consumers with \emph{intermediate} values (i.e., those with $v \in (\kappa_c^2,\kappa_c^3), \theta=l$ and $v \in (\kappa_c^4,\kappa_c^5), \theta=h$) who retain a positive surplus, while the high-value consumers have their surplus extracted and the low-value consumers are excluded. In other words, while both creating information rents, unobserved information and prohibited information would generally lead to different distribution of welfare among consumers. Under non-discriminatory pricing, intermediate-value consumers benefit from the regulation at the expense of high-value consumers being extracted and low-value consumers being excluded.

\subsection{Profit Loss Due to Non-Discrimination Constraints}
\begin{figure}[t]
    \centering
    \includegraphics{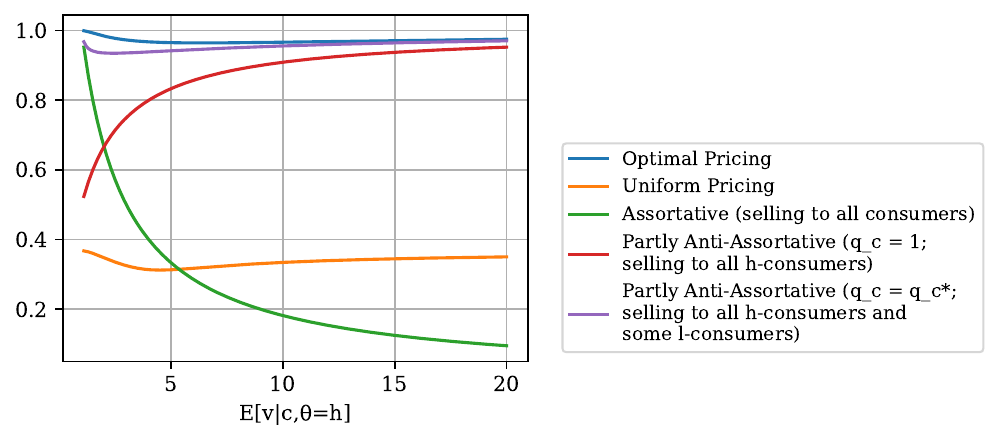}
    \caption{Seller profit divided by total surplus for different pricing rules with $c=0$, $\alpha_c=\nicefrac{1}{2}$, and $\E{v}{c,\theta=l}=1$.}
    \label{fig:profit-ratio}
\end{figure}
In this section, we briefly explore how much profit the seller loses due to the non-discrimination constraint. In the case where there is no cost $c=0$, $\alpha_c=\nicefrac{1}{2}$, and values are exponentially distributed with means $\E{v}{c,\theta=l}=1$ and $\E{v}{c,\theta=h} \geq 1$,\footnote{Note that \bref{ass:no-full-surplus-extraction} always holds here as $F_{c,l}(c)=0$.}  \bref{fig:profit-ratio} plots the share of the seller's profit relative to the total surplus, under various non-discriminatory pricing rules, including the optimal pricing rule $p^\star$, the assortative pricing rule $p^{ass}$, the anti-assortative pricing rule $p^{anti}$ with $q_c=1$ and $q_c=q_c^\star:=\Delta_c(v_c^\star)$, as well as the (optimal) uniform pricing rule. As illustrated by \bref{fig:profit-ratio}, even though the seller is prevented from full surplus extraction due to the non-discrimination constraint, the seller can still guarantee a significant share of the total gains from trade using non-discriminatory personalized pricing ($\geq 95\%$).
This contrasts with the optimal non-personalized pricing rule, which always yields the seller less than 40\% of the social surplus.

In what follows, we provide a general lower bound on the share of total surplus the seller can guarantee using a non-discriminatory pricing rule. To this end, for any cost $\tilde{c} \in C$, let $r_{\tilde{c}} + 1$ be the ratio of gains from trade of $l$ and $h$ types
\[
r_{\tilde{c}}:=\frac{\EE[(v-c)^+\mid c=\tilde{c}, \theta=h]}{\EE[(v-c)^+\mid c=\tilde{c}, \theta=l]}-1\,.
\]
The following proposition establishes a lower bound on the share of surplus the seller can extract from consumers under $p^\star$ conditional on $c$, which depends only on $r_c$ but not on other details of the distributions of values $F_{c,l}$ and $F_{c,h}$.

\begin{proposition}\label{prop:profit-bound}
For all $\tilde{c} \in C$, 
\[
    \frac{\E{(p^\star-c)\mathbf{1}\{v \geq c\}}{c=\tilde{c}}}{\E{(v-c)^+}{c=\tilde{c}}} \geq \frac{\max \{ 1, \alpha_{\tilde{c}} \, (r_{\tilde{c}}+1) \}}{ \alpha_{\tilde{c}} \, r_{\tilde{c}} + 1 } \geq \frac{r_{\tilde{c}} + 1 }{2 \, r_{\tilde{c}} + 1 } > \frac{1}{2} \,.
\]
In particular, 
\[
\Pi^\star=\Pi(p^\star) \geq \EE\left[\frac{r_c+1}{2r_c+1}\right]>\frac{1}{2}\,.
\]
\end{proposition}

According to \bref{prop:profit-bound}, the seller can always guarantee $\E{r_c/(2r_c+1)}$ share of total surplus under the non-discrimination constraint. 
For example, if $h$-consumers have 40\% higher gains from trade compared to $l$-consumers (i.e., $r_c = 0.4$), then the seller can extract at least 77\% of total gains from trade under the optimal non-discriminatory personalized pricing rule.
We note that optimal non-discriminatory personalized pricing always guarantees the seller strictly more than half of the surplus, which exceeds the best guarantee uniform pricing can give.\footnote{Uniform pricing can guarantee half the surplus when the seller's profit---as a function of the uniform price---is concave, but not otherwise \citep{BCW22}. In the example of \bref{fig:profit-ratio}, the seller's profit function is not concave and thus uniform pricing only gives a profit that is less than 40\% of the total surplus.} The bound in \bref{prop:profit-bound} is not tight in general, and the seller could typically obtain an even higher surplus extraction rate under the optimal pricing rule $p^\star$, as illustrated by \bref{fig:profit-ratio}.

\subsection{Who Benefits More from Anti-Discrimination Regulation}
Next, we explore which protected characteristic benefits more from anti-discrimination regulations. The answer depends on the relative size of the population of different consumers and the underlying value distributions. As the next result establishes, the more consumers with the same protected characteristic there are in the market, the lower their surplus would be.

\begin{proposition}[Effects of Population Sizes]\label{prop:population-size} Fix the value distributions $F_{c,l},F_{c,h}$. Let $CS(c,\theta;p^\star, \alpha_c)$ denote the consumer surplus under pricing rule $p^\star$ when $\PP[\theta=h\mid c]=\alpha_c$. 
\begin{compactenum}[(i)]
\item The surplus  of $h$-consumers $CS(c,h;p^\star, \alpha_c)$ decreases in $\alpha_c$.
\item The surplus  of $l$-consumers $CS(c,l;p^\star, \alpha_c)$ increases  in $\alpha_c$. 
\end{compactenum}
Furthermore, $\lim_{\alpha_c \to 1 } CS(c,h;p^\star, \alpha_c) = \lim_{\alpha_c \to 0} CS(c,l;p^\star,\alpha_c) =0$. In particular, for all $c \in C$, there exists $\alpha_c,\alpha_c'$ such that $CS(c,h;p^\star,\alpha_c)>CS(c,l;p^\star,\alpha_c)$ and $CS(c,h;p^\star,\alpha_c')<CS(c,l;p^\star,\alpha_c')$.
\end{proposition}

Intuitively, if there are more consumers of the same characteristic, the seller has higher incentives to tailor prices finely to that consumer group, which reduces their surplus. Again, this is reminiscent of classical information rents stemming from private information in screening problems, where the agent's information rent decreases as their types become more similar. 

\bref{prop:population-size} implies that it is impossible to determine who benefits more from anti-discrimination regulation without restrictions on the size of consumer groups. Indeed, if one group has vanishing size, the seller will (almost) perfectly tailor the prices to the other group and leave members of that group with no surplus.

\begin{figure}[h!]
    \centering
    \includegraphics{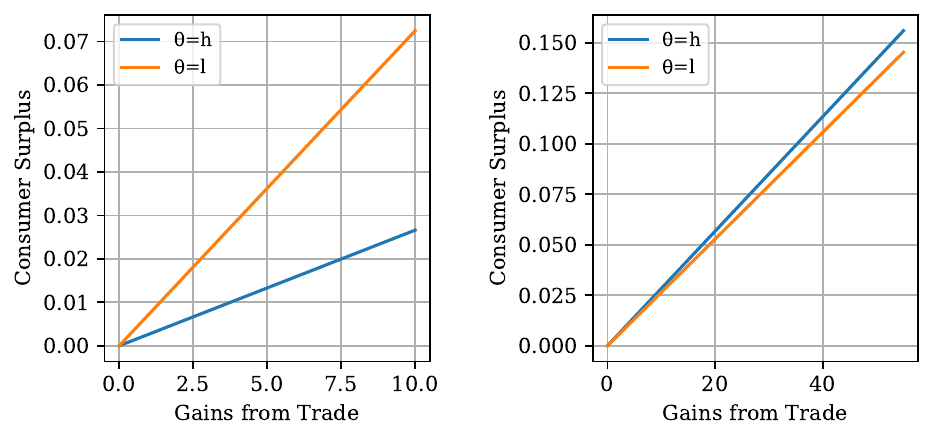}
    \caption{Consumer Surplus for $\alpha_c=\nicefrac{1}{2}$, $\lambda_l=1$. The left panel has $\lambda_h=3$ and the right panel has $\lambda_h=12$.}
    \label{fig4}
\end{figure}

In fact, even with fixed population sizes, with different value distributions $F_{c,l}$ and $F_{c,h}$, it could be that either $l$-consumers benefit more or $h$-consumers benefit more, as illustrated by \bref{fig4}, in the context of the insurance example with exponential distributions introduced in \bref{sec:optimal}.  

As a result, while non-discrimination regulations could strictly benefit consumers of both types, it is noteworthy that consumers' gains might be disproportional across groups with different protected characteristics, and either $h$-consumers or $l$-consumer could benefit more, depending on the relative population and the underlying value distributions. Meanwhile, although non-discriminatory requirements could benefit consumers, such requirements would also create deadweight losses, according to \bref{prop:surplus}. In other words, the benefit consumers get under the non-discrimination regulations is partly due to the fact that some low-value consumers are excluded from the market. 

Together, perhaps contrary to goal of non-discrimination regulations, which typically seeks to protect the socially disadvantaged group of consumers, our results suggest that it is not immediately clear whether such regulations always benefit the disadvantaged group that regulators wish to protect the most. Depending on the underlying value distributions and population sizes, it is possible that the other group of consumers could benefit even more, at the expense of some consumers from the disadvantaged group being excluded.

\subsection{Non-Discriminatory Outcomes}
Our notion of non-discriminatory pricing applies to the distribution of prices \emph{offered} to consumers. In a sense, this requires that consumers in different protected groups are given \emph{equal opportunities}, so that they face, on average, the same prices \emph{before} they decide whether to purchase. Alternatively, one could consider a stronger notion of non-discriminatory pricing, which requires consumers in different protected groups to face \emph{equal outcomes}, so that \emph{after} they make purchase decisions, the resulting outcomes (i.e., transaction outcomes and transaction prices) must be the same. 

Specifically, given a pricing rule $p$, denote by $y(\omega) = \mathbf{1}\{v \geq p\}$ the random variable that indicates whether or not the product is sold to a given consumer.

\begin{definition}
A pricing rule $p$ induces \emph{non-discriminatory outcomes} if for all $c \geq 0$ and $M \subseteq \RR \times \{0,1\}$,
\[
    \PP[ (p,y) \in M \mid c,\theta=0] = \PP[ (p,y) \in M \mid c,\theta=1] \,.
\]
\end{definition}

In other words, a pricing rule $p$ induces non-discriminatory outcomes if the event of transaction and the transaction price, $(p,y)$, are independent of protected characteristic $\theta$ conditional on cost $c$. Clearly, any pricing rule that induces non-discriminatory outcomes must be non-discriminatory. 
Recall that in \eqref{eq:assortative-matching} we defined $p^{ass}$ to be the pricing rule which matches $h$ and $l$ consumers associatively and charges each pair the lower of their values.

\begin{proposition}[Optimal Pricing Rule with Non-Discriminatory Outcomes]\label{prop:compare}
\mbox{}
\begin{compactenum}[(i)]
\item $p^{ass}$ induces non-discriminatory outcomes and maximizes the seller's profit among all pricing rules that induce non-discriminatory outcomes. 
\item $p^{ass}$ yields a lower profit than the optimal non-discriminatory pricing rule: $\Pi(p^\star) \geq \Pi(p^{ass})$, and the inequality is strict if and only if $\alpha_c (\overline{v}_c-c)>\underline{v}_c-c$ for a positive measure of $c \in C$.  
\item The surplus of $h$-consumers is higher under $p^{ass}$ than under $p^{\star}$ and the surplus of $l$-consumers is lower. That is, for all $c$,
\begin{align*}
    CS(c,h; p^\star ) \leq CS(c,l; p^{ass}) \,\,\,\,\text{ and }\,\,\,\, CS(c,l; p^\star ) \geq CS(c,h; p^{ass}) \,,
\end{align*}
where the inequalities are strict if and only if $\alpha_c (\overline{v}_c-c)>\underline{v}_c-c$.
\end{compactenum}
\end{proposition}

\bref{prop:compare} thus establishes that the pricing policy used by the seller, and thus the welfare implications, depend delicately on the notion of non-discriminatory pricing policies. Strengthening the protection, and require non-discriminatory outcomes, instead of non-discriminatory prices, would hurt the group who has lower values. In some settings (e.g., when the disadvantaged group has lower values because they are poorer), this would mean that stricter non-discrimination regulations may actually make the disadvantaged group worse-off and the seller worse-off, while benefiting the advantaged group.  

\begin{remark}[Equalizing Consumer Surplus]
While more difficult to implement, one might also wonder---as a theoretical benchmark---what would the welfare implications be if the notion of non-discrimination is based on welfare directly, as opposed to of observable outcomes such as prices and transactions. In other words, we could also consider another notion of non-discriminatory pricing that requires the average consumer surplus across groups to be the same conditional on costs: $CS(c,h;p)=CS(c,l;p)$. Clearly, under such notion, perfect price discrimination $p=\max\{v,c\}$ is feasible and both $h$-consumers and $l$-consumers would have zero surplus, which is even worse compared to non-discriminatory outcomes defined above. 
\end{remark}
Overall, the above analyses suggest that the welfare implications of \bref{thm1} may serve as a cautionary tale and underlines the importance of more careful analyses for the welfare implications of non-discrimination regulations.

\section{Extensions}\label{sec:extensions}

\subsection{Imperfect Price Discrimination}\label{sec:4.1}
Thus far, we assumed that every pricing rule that satisfy the non-discrimination constraint is feasible. This requires the seller knowing each consumer's type. In practice, sometimes the seller may not have access to enough of data to perfectly estimate consumers' types, and can only obtain a noisy signal.

Our method can still be applied to characterize the profit-maximizing non-discriminatory pricing rule in this environment. Specifically, suppose now that consumers' true values are denoted by $w \geq 0$, whose distribution depends on an observable type $v$. The observable types $v$ are distributed according to $F_{c,l}$, $F_{c,h}$ among consumers with protected characteristics $l$ and $h$ conditional on cost $c$, respectively. A pricing rule $p:V\times C \times \Theta \times[0,1] \to \RR_+$ is defined as before, except that $v$ does not stand for a consumer's true value but only provides noisy information about a consumer's value.  Given a pricing rule $p$, the seller's profit is given by 
\[
\Pi(p):=\EE\left[\max_{p \geq 0} (p-c) \cdot \mathbf{1}\{w \geq p\}\right]\,.
\]
By the same arguments as the proof of \bref{prop:opt-transport},\footnote{Alternatively, this can be derived from Lemma 3 of \citet{SY24}.} we may still recast the problem into an optimal transport:
\begin{equation}\label{eq:transport2}
\tilde{\pi}^\star:= \int_C \left( \max_{\rho_c \in \mathcal{R}_c} \int_{V^2} \widetilde{\pi}_c(v_l,v_h)\diff \rho_c\right) G(\diff c)\,,
\end{equation}
where 
\[
\widetilde{\pi}_c(v_l,v_h):=\max_{p \geq 0} \left[(p-c)\cdot (\alpha_c\PP[w \geq p\mid v_h]+(1-\alpha_c)\PP[w\geq p\mid v_l])\right]\,.
\]
As a result, the profit-maximizing pricing rules can still be found by solving the optimal transport problem \eqref{eq:transport2}. 

To illustrate the solution, suppose that there are no cost to serve consumers (i.e., $c=0$ almost surely), $\alpha_c=\nicefrac{1}{2}$, and that consumers' values $w$ are distributed uniformly on $[0,2v]$ conditional on $v$. It then follows that 
\[
\widetilde{\pi}_c(v_l,v_h)=\max\left\{\frac{v_l}{4},\frac{v_h}{4}, \frac{v_lv_h}{v_l+v_h}\right\}\,.
\]
The solution to the optimal transport problem \eqref{eq:transport2} in this case is qualitatively similar to the solution in the baseline model. To describe the solution, let $\tilde{\kappa}_c \in [0,\ub{v}_c]^5$ be the unique increasing vector with $\tilde{\kappa}_c^4<v^\star_c<\tilde{\kappa}_c^5$ that solves following system of equations 
\begin{align}\label{eq:system2}
&\tilde{\kappa}_c^2=F_{c,l}^{-1}(\Delta_c(\tilde{\kappa}_c^3)+F_{c,h}(\tilde{\kappa}_c^1))=F_{c,l}^{-1}(\Delta_c(\tilde{\kappa}_c^4))=F_{c,l}^{-1}(\Delta_c(\tilde{\kappa}_c^5)) \notag\\
&\frac{\tilde{\kappa}_c^1\tilde{\kappa}_c^2}{\tilde{\kappa}_c^1+\tilde{\kappa}_c^2}=\frac{\tilde{\kappa}_c^3}{4}-\int_{\tilde{\kappa}_c^2}^{\tilde{\kappa}_c^3} \left(\frac{\lb{\beta}_c(z)}{z+\lb{\beta}_c(z)}\right)^2\diff z=\int_{\tilde{\kappa}_c^4}^{\tilde{\kappa}_c^5} \left(\frac{\ub{\beta}_c(z)}{z+\ub{\beta}_c(z)}\right)^2\diff z\,,
\end{align}
where 
\[
\lb{\beta}_c(z):=F_{c,h}^{-1}(F_{c,l}(z)-\Delta_c(\tilde{\kappa}_c^3))
\]
for all $z \in [\tilde{\kappa}_c^2,\tilde{\kappa}_c^3]$, and 
\[
\ub{\beta}_c(z):=F_{c,h}^{-1}(F_{c,l}(z)-\Delta_c(\tilde{\kappa}_c^4))\,.
\]
for all $z \in [\tilde{\kappa}_c^4,\tilde{\kappa}_c^5]$. Then, let 
\begin{align*}
    \tilde{p}^\star(v,c,l,r) & :=\begin{cases}
     \overline{\Delta}_c^{-1}(\Delta_c(\tilde{\kappa}^5_c) - F_{c,l}(v)),&\mbox{if } v < \tilde{\kappa}_c^2\\
    F_{c,h}^{-1}(F_{c,l}(v)-F_{c,l}(\tilde{\kappa}_c^2)+F_{c,h}(\tilde{\kappa}_c^1)),&\mbox{if } v \in [\tilde{\kappa}_c^2,\tilde{\kappa}_c^3)\\
    v,&\mbox{if } v \geq \tilde{\kappa}_c^3
    \end{cases}\,;\\
 \tilde{p}^\star(v,c,h,r) & :=\begin{cases}
    \lb{\Delta}_{c}^{-1}\left(F_{c,h}(v)+\Delta_c(\tilde{\kappa}_c^3)\right),\hspace{1.75cm}&\mbox{if } v < \tilde{\kappa}_c^1\\
    F_{c,l}^{-1}(F_{c,h}(v)+\Delta_c(\tilde{\kappa}_c^4)),&\mbox{if } v \in [\tilde{\kappa}_c^4,\tilde{\kappa}_c^5)\\
    v,&\mbox{if } v \in [\tilde{\kappa}_c^5,\infty) \cup (\tilde{\kappa}_c^1,\tilde{\kappa}_c^4)
    \end{cases}\,,
\end{align*}
for all $v \in V$ and $r \in  [0,1]$. Then, we have:

\begin{proposition}\label{prop:extension1}
$\tilde{p}^\star$ is an optimal non-discriminatory pricing rule. 
\end{proposition}
The optimal pricing implied by \bref{prop:extension1} is qualitatively identical to the optimal pricing rule given by \bref{thm1}, with the only difference being how the thresholds $\tilde{\kappa}_c$ are defined. 

\subsection{Implementable Welfare Outcomes}\label{sec:4.2}
While we have so far focused on how the seller can maximize their profits using a non-discriminatory pricing rule, another natural question is what consumer welfare can be achieved. To explore this question, recall that from \bref{rem:segmentation}, we may view pricing rules as price discriminating consumers based on a given market segmentation. In what follows, we explore the welfare outcomes (i.e., consumer surplus and seller profit) that can be induced by a non-discriminatory pricing rule that charges an optimal price in each market segment. That is, we calculate the average consumer surplus and the seller's profit that can be induced by some non-discriminatory pricing rule $p$ that is
\begin{compactenum}[(i)]
    \item measurable with respect to some segmentation $s:\Omega \to S$, and 
    \item is optimal given each segment for the seller.\footnote{That is, the price $p(s)$ in each segment $s$ satisfies $p(s) \in \argmax_{x \geq 0} \EE[(x-c)\mathbf{1}\{v \geq x\}\mid s]$.}
\end{compactenum}

The set of feasible pairs of seller profit and average consumer surplus without the non-discrimination constraint is characterized by \cite{BBM15} for $c=0$, as the triangle spanned by the points $$(\E{v},0),\,\, (r^\star,0),\,\, (r^\star,\E{v} - r^\star)\,,$$ where $r^\star = \max_p p\cdot \Pr{v \geq p}$ is the optimal uniform pricing revenue. That is, \citet{BBM15} show that any surplus division where the seller's revenue is between $r^\star$ and $\E{v}$, and the consumers' average surplus is below the total surplus $\E{v}$ net of the seller's revenue, is implementable by some segmentation. With the non-discrimination constraint, however, not every outcome in this triangle is feasible. 
\begin{figure}[t]
    \centering
    \includegraphics{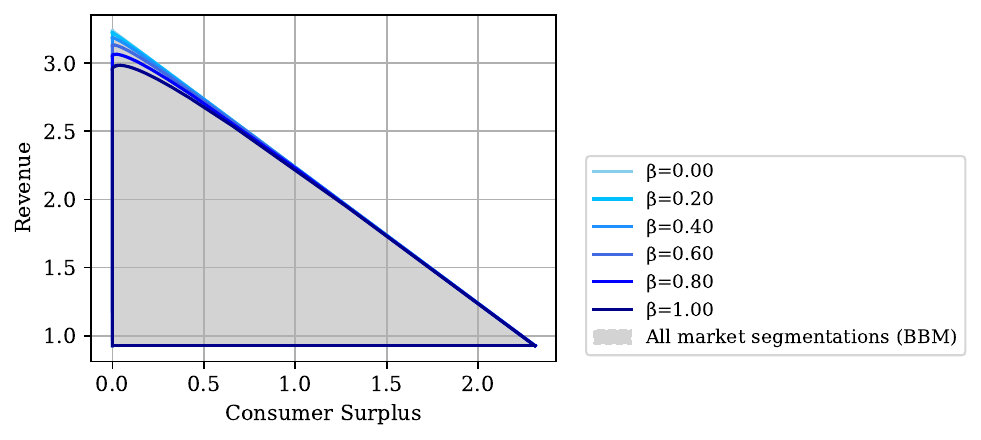}
    \caption{Feasible Welfare Outcomes for $c=0$, $\alpha_c=\nicefrac{1}{4}$, $F_{c,h}=(\beta+\alpha(1-\beta))\ub{F}+(1-(\beta+(1-\beta)\alpha))\lb{F}$ and $F_{c,l}=(1-\beta)\alpha \ub{F}+(1-(1-\beta)\alpha)\lb{F}$. }
    \label{fig:welfare-outcomes}
\end{figure}

As an example, suppose that $c=0$ and $\alpha_c=\nicefrac{1}{4}=:\alpha$ of consumers are of protected characteristic $h$, and let $\ub{F},\lb{F}$ be exponential distributions with means $10$ and $1$. Consider a parameterization of $F_{c,h}$ and $F_{c,l}$ where for some $\beta \in [0,1]$,
\begin{align*}
F_{c,h}&:=(\beta+\alpha(1-\beta))\ub{F}+(1-(\beta+(1-\beta)\alpha))\lb{F}\\
F_{c,l}&:=(1-\beta)\alpha \ub{F}+(1-(1-\beta)\alpha)\lb{F}\,.
\end{align*}
By construction, $F_{c,h}$ dominates $F_{c,l}$ in the likelihood ratio order for all $\beta \in [0,1]$, and the total variation distance $\|F_{c,h}-F_{c,l}\|$ increases in $\beta$. Meanwhile, the overall distribution of values $\alpha F_{c,h}+(1-\alpha)F_{c,l}$ in the population remains unchanged in $\beta$. As a consequence, the surplus triangle of \citet{BBM15} remains the same for all $\beta \in [0,1]$.

\bref{fig:welfare-outcomes} plots the set of feasible pairs of seller profit and average consumer surplus with the non-discrimination constraint for this parameterized family of type distributions. The shaded triangle corresponds to the feasible surplus division given by \citet{BBM15}, whereas the colored curves depict the boundary of the set of feasible welfare outcomes for a different values of the parameter $\beta$. 

As noted above, when $c=0$ almost surely, \bref{ass:no-full-surplus-extraction} holds, and hence the highest feasible revenue is less than the total gains from trade $\E{v}$, due to the non-discrimination constraint. This is reflected in \bref{fig:welfare-outcomes} by the fact that the top-left corner of the triangle not included in feasible surplus region.
One notable feature of the example is the relatively minor restriction on implementable welfare outcomes even when the distributions of values become vastly different across consumer groups (recall that in this example, $h$-consumers value the good 10 times more than $l$-consumers at $\beta=1$ ).

In this parametric example, there exists a segmentation that keeps the seller's revenue the same as the uniform pricing revenue, induces a non-discriminatory pricing rule in which all consumers buy, and thus the boundaries all reach the bottom-right corner of the triangle. However, this is not the case in general. Characterizing explicitly the feasible welfare outcomes in general, and in particular, when can the consumer-optimal outcome be attained under the non-discrimination constraint, is an exciting question for future research. 

\subsection{All Profit-Maximizing Pricing Rules}\label{sec:all}

So far, we have focused on revenue maximizing pricing rules that are undominated, in the sense that there does not exist another revenue maximizng pricing rule that generates a higher surplus for all consumer groups.
We now explore the welfare outcomes of all optimal non-discriminatory pricing rules, including the dominated ones. In particular, we characterize the surplus of consumers with each protected characteristics under all optimal non-discriminatory pricing rules.  To state our welfare results, for all $c \in C$, we say that $(\sigma_{c,l},\sigma_{c,h})$ is a surplus outcome induced by an optimal non-discriminatory pricing rule if there exists a non-discriminatory pricing rule $p$ such that $\Pi(p)=\Pi(p^{\star})$ and that $\sigma$ is the induced consumer surplus $CS(c,\theta;p)=\sigma_{c,\theta}$ for all $c \in C, \theta \in \{l,h\}$. 

\begin{proposition}[Welfare Outcomes]\label{prop:welfare}
\mbox{}
$(\sigma_{c,l},\sigma_{c,h})$ is a surplus outcome induced by an optimal non-discriminatory pricing rule if and only if 
\[
0 \leq \sigma_{c,l} \leq CS(c,l;p^{\star}) 
\,\,\,\,\text{ and }\,\,\,\,
\sigma_{c,h}=CS(c,h;p^{\star})
\]
\end{proposition}
The proof of \bref{prop:welfare} relies on the optimality of $p^\star$, as well as the duality theorem of the optimal transports \eqref{eq:transport}. Details of the proof can be found in the Appendix. According to \bref{prop:welfare}, $h$-consumers retain the same amount of surplus under every optimal non-discriminatory pricing rule, whereas the average surplus of $l$-consumers range from zero to $\E{CS(c,\theta;p^\star)}{\theta=l}$ across all optimal non-discriminatory pricing rules. Moreover, since profits are the same across all optimal non-discriminatory pricing rules, \bref{prop:welfare} in turn implies that $h$-consumers' deadweight losses are the same across all optimal non-discrimiantory pricing rules, whereas the average surplus of $l$-consumers range from $WL(c,l,p^\star)$ to $\E{(v-c)^+}-\Pi(p^\star)-\E{CS(c,\theta;p^\star)+WL(c,\theta;p^\star)}{\theta=h}$ across all optimal non-discriminatory pricing rules.

\subsection{General Distributions}\label{sec:general}
We now relax \bref{ass:no-full-surplus-extraction} and characterize the undominated profit-maximizing pricing rules for all distributions of values. According to \bref{prop:opt-transport}, the optimal pricing rule can be found by solving an optimal transport problem for each $c \in C$. To this end, let 
\[
\begin{aligned}
C_1:=&\{c \in C: F_{c,l}(c) < \|F_{c,l}-F_{c,h}\|\}\\
C_2:=&\{c \in C: F_{c,l}(c) \geq \|F_{c,l}-F_{c,h}\|\,, c<v^\star_c\}\\
C_3:=&\{c \in C: F_{c,l}(c) \geq \|F_{c,l}-F_{c,h}\|\,, c\geq v^\star_c\}\,.
\end{aligned}
\]
By definition, $C_1,C_2,C_3$ partitions the set of possible costs $C$ into three regions. Note that \bref{ass:no-full-surplus-extraction} imposes that $c \in C_1$ almost surely. 

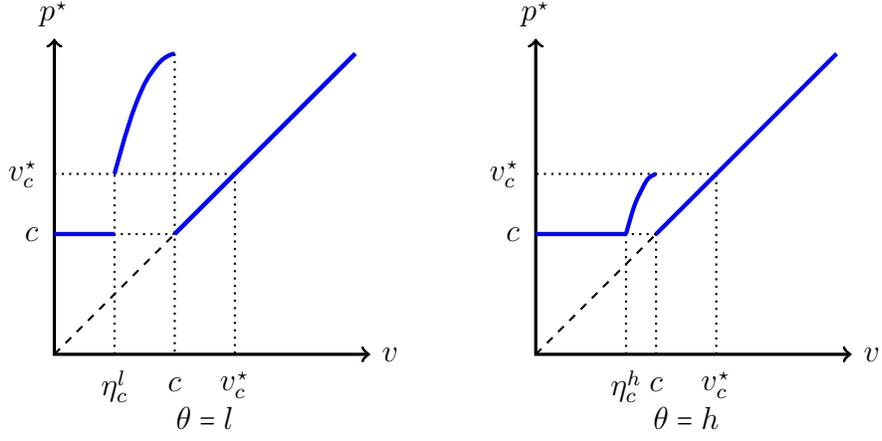
\begin{figure}[t!]
\centering
\tikzset{
solid node/.style={circle,draw,inner sep=1.25,fill=black},
hollow node/.style={circle,draw,inner sep=1.25}
}
\begin{tikzpicture}[scale=4]
\draw [<->, very thick] (-1.1,1.05) node (yaxis) [above] {  $p^\star$}
        |- (-0.05,0) node (xaxis) [right] { $v$};
\draw [dashed, thick] (-1.1,0)--(-0.1,1); 
\draw [dotted, thick] (-0.9,0)--(-0.9,0.6);
\draw [dotted, thick] (-0.7,1)--(-0.7,0);
\draw [dotted, thick] (-0.7,0.4)--(-1.1,0.4); 
\draw [dotted, thick] (-1.1,0.6)--(-0.5,0.6);
\draw [dotted, thick] (-0.5,0.6)--(-0.5,0);
\draw (-0.9,0) node [below=2pt] {  $\eta_c^l$}; 
\draw (-0.7,-0.025) node [below=2pt] {  $c$}; 
\draw (-0.5,0) node [below=2pt] {  $v^\star_c$}; 
\draw (-1.1,0.4) node [left=2pt] {  $c$};
\draw (-1.1,0.6) node [left=2pt] {  $v^\star_c$}; 
\draw [blue, ultra thick] (-1.1,0.4)--(-0.9,0.4);
\draw [blue, ultra thick] (-0.7,0.4)--(-0.1,1); 
\draw [blue, ultra thick] plot [smooth, tension=0.8] coordinates {(-0.9,0.6) (-0.83,0.83) (-0.76,0.96) (-0.7,1)};
\draw [blue, ultra thick] (-0.7,0.4)--(-0.1,1); 
\draw [<->, very thick] (0.5,1.05) node (yaxis) [above] {  $p^{\star}$}
        |- (1.55,0) node (xaxis) [right] { $v$};
\draw [dashed, thick] (0.5,0)--(1.5,1); 
\draw [dotted, thick] (0.9,0)--(0.9,0.4);
\draw [dotted, thick] (0.9,0.4)--(0.5,0.4);
\draw [dotted, thick] (1.1,0)--(1.1,0.6);
\draw [dotted, thick] (1.1,0.6)--(0.5,0.6);
\draw [dotted, thick] (0.8,0)--(0.8,0.4);
\draw (0.8,0) node [below=2pt] {  $\eta_c^h$};
\draw (0.9,-0.025) node [below=2pt] {  $c$};
\draw (1.1,0) node [below=2pt] {  $v^\star_c$}; 
\draw (0.5,0.4) node [left=2pt] {  $c$};
\draw (0.5,0.6) node [left=2pt] {  $v^\star_c$};
\draw [blue, ultra thick] (0.5,0.4)--(0.8,0.4);
\draw [blue, ultra thick] plot [smooth, tension=0.5] coordinates {(0.8,0.4) (0.83,0.5) (0.87,0.58) (0.9,0.6)};
\draw [blue, ultra thick] (0.9,0.4)--(1.5,1);
\draw (1,-0.12) node [below=2pt] {  $\theta=h$};
\draw (-0.6,-0.12) node [below=2pt] {  $\theta=l$}; 
\end{tikzpicture}
\caption{Optimal Pricing Rule when $c \in C_2$.}
\label{figure2}
\end{figure}

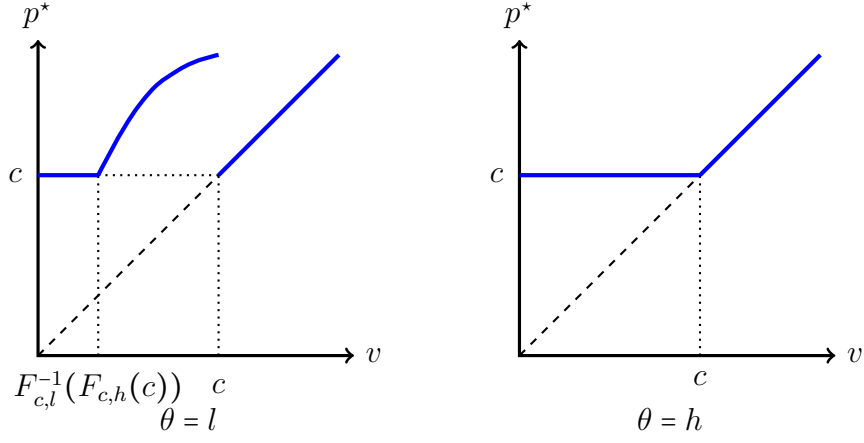
\begin{figure}[t!]
\centering
\tikzset{
solid node/.style={circle,draw,inner sep=1.25,fill=black},
hollow node/.style={circle,draw,inner sep=1.25}
}
\begin{tikzpicture}[scale=4]
\draw [<->, very thick] (-1.1,1.05) node (yaxis) [above] {  $p^\star$}
        |- (-0.05,0) node (xaxis) [right] { $v$};
\draw [dashed, thick] (-1.1,0)--(-0.1,1); 
\draw [dotted, thick] (-0.9,0)--(-0.9,0.6);
\draw [dotted, thick] (-0.9,0.6)--(-1.1,0.6);
\draw [dotted, thick] (-0.5,0)--(-0.5,0.6);
\draw [dotted, thick] (-0.5,0.6)--(-1.1,0.6);
\draw (-0.9,0) node [below=2pt] {  $F_{c,l}^{-1}(F_{c,h}(c))$}; 
\draw (-0.5,-0.025) node [below=2pt] {  $c$};
\draw (-1.1,0.6) node [left=2pt] {  $c$}; 
\draw [blue, ultra thick] (-1.1,0.6)--(-0.9,0.6);
\draw [blue, ultra thick] plot [smooth, tension=0.8] coordinates {(-0.9,0.6) (-0.766, 0.83) (-0.633, 0.95) (-0.5,1)};
\draw [blue, ultra thick] (-0.5,0.6)--(-0.1,1);
\draw [<->, very thick] (0.5,1.05) node (yaxis) [above] {  $p^\star$}
        |- (1.55,0) node (xaxis) [right] { $v$};
\draw [dashed, thick] (0.5,0)--(1.5,1); 
\draw [dotted, thick] (1.1,0)--(1.1,0.6);
\draw (1.1,0) node [below=2pt] {  $c$}; 
\draw (0.5,0.6) node [left=2pt] {  $c$}; 
\draw [blue, ultra thick] (0.5,0.6)--(1.1,0.6);
\draw [blue, ultra thick] (1.1,0.6)--(1.5,1); 
\draw (1,-0.12) node [below=2pt] {  $\theta=h$};
\draw (-0.6,-0.12) node [below=2pt] {  $\theta=l$}; 
\end{tikzpicture}
\caption{Optimal Pricing Rule when $c \in C_3$.}
\label{figure3}
\end{figure}

We now define an optimal pricing rule $p^{\star}$ for general distributions conditioning on different realizations of $c$. When $c \in C_1$, let $p^\star$ be defined in \eqref{eq:p-star}. When $c \in C_2$, since $F_{c,l}(c) \geq \Delta_c(v^\star_c)$, there exists a unique value $\eta_c^l \leq c$ such that $F_{c,l}(c)-F_{c,l}(\eta_c^l)=\Delta_c(v^\star_c)$. Let $\eta_c^h:=F_{c,h}^{-1}(F_{c,l}(\eta_c^l))$. The pricing rule $p^{\star}$, when $c \in C_2$, is defined as follows:
\begin{align*}
p^{\star}(v,c,l,r)&:=\begin{cases}
c,&\mbox{if } v < \eta_c^l\\
\overline{\Delta}_c^{-1}(\Delta_c(v^\star_c)+F_{c,l}(\eta_c^l)-F_{c,l}(v)),&\mbox{if }v \in [\eta_c^l,c)  \\
v,&\mbox{if } v \geq c
\end{cases}\\
p^\star(v,c,h,r)&:=\begin{cases}
c,&\mbox{if } v<\eta_h\\
\underline{\Delta}_c^{-1}(F_{c,h}(v)-F_{c,h}(\eta_c^h)+\Delta_c(c)),&\mbox{if } v \in [\eta_c^h,c)\\
v,&\mbox{if } v \geq c
\end{cases}
\end{align*}
\bref{figure2} depicts the optimal pricing rule $p^{\star}$ when $c \in C_2$.

When $c \in C_3$, the pricing rule $p^\star$ is defined as follows: 
\begin{align*}
p^\star(v,c,l,r)&:=\begin{cases}
c,&\mbox{if } v<F_{c,l}^{-1}(F_{c,h}(c))\\
\ub{\Delta}_c^{-1}(\Delta_c(c)+F_{c,h}(c)-F_{c,l}(v)),&\mbox{if } v \in [F_{c,l}^{-1}(F_{c,h}(c)),c)\\
v,&\mbox{if } v \geq c
\end{cases}\\
p^\star(v,c,h,r)&:= \max\{v,c\}
\end{align*}
\bref{figure3} depicts the optimal pricing rule $p^\star$ when $c \in C_3$.

\bref{thm2} below establishes that $p^\star$ is an optimal non-discriminatory pricing rule. 

\begin{theorem}[Optimal Pricing for General Distributions]\label{thm2}$ $
\begin{compactenum}[(i)]
\item $p^\star$ is a profit-maximizing non-discriminatory pricing rule, that is, $p^\star$ solves \eqref{eq:optimal-pricing}. 
\item Every undominated profit-maximizing non-discriminatory pricing rule induces the same average surplus $CS(\theta,c;p^\star)$ for consumer of each protected characteristic $\theta$ and cost $c$.
\end{compactenum}
\end{theorem}

We note that \bref{thm2} immediately implies \bref{thm1}, since $c \in C_1$ almost surely under \bref{ass:no-full-surplus-extraction}. As another immediate consequence of \bref{thm2}, under the optimal pricing rule $p^\star$, the seller is able to fully extract all gains from trade whenever $c \in C_2\cup C_3$. 

\begin{corollary}\label{cor:full-surplus-extraction}
For all $\tilde{c} \in C_2 \cup C_3$, 
\[
\EE[(p^\star-c)\mathbf{1}\{v \geq p^\star\}\mid c=\tilde{c}]=\EE[(v-c)^+\mid c=\tilde{c}]\,.
\]
As a result, 
\[
CS(c,\theta;p^\star)=WL(c,\theta;p^\star)=0\,,
\]
for all $c \in C_2\cup C_3$ and $\theta \in \{l,h\}$.
\end{corollary}

According to \bref{cor:full-surplus-extraction}, the restrictions on the value distributions imposed by \bref{ass:no-full-surplus-extraction} are in fact equivalent to restricting attention to distributions where the seller cannot fully extract all gains from trade.

\section{A Proof Sketch for Theorem \ref{thm1} and Theorem \ref{thm2}}\label{sec:4.3}

In this section, we outline the main steps of the proof for \bref{thm1} and \bref{thm2}. Details of the proof can be found in the appendix. From \bref{prop:opt-transport}, optimal pricing rules can be identified by solving the optimal transport problem  
\begin{equation}\label{eq:transport-point}
\pi^\star(c):=\max_{\rho \in \mathcal{R}_c} \int_{V^2} \pi_c(v_l,v_h) \diff \rho_c\,,
\end{equation}
for each cost $c \in C$. We solve \eqref{eq:transport-point} by a duality argument. The dual problem corresponding to \eqref{eq:transport-point} is given by 
\begin{align}\label{eq:dual}
\pi_\star(c):=\inf_{\phi_c,\psi_c} &\left[\int_V \phi_c(v_l)  F_{c,l}(\diff v_l)+\int_V \psi_c(v_h) F_{c,h}(\diff v_h)\right] \notag\\
\mbox{s.t. }& \phi_c(v_l)+\psi_c(v_h) \geq \pi_c(v_l,v_h)\,,
\end{align}
where the infimum is taken over all measurable functions $\phi_c,\psi_c:V \to \RR$. Since $\pi_c$ is continuous, the Kantorovich duality theorem holds \citep[see, e.g.,][Theorem 5.10]{V09}.

\begin{lemma}[Kantorovich Duality]\label{lem:dual}
$\pi_\star(c)=\pi^\star(c)$ for all $c \in C$. Moreover, for any $c \in C$, for any measurable $\phi_c,\psi_c$ such that $\phi_c(v_l)+\psi_c(v_h) \geq \pi_c(v_l,v_h)$ for all $(v_l,v_h) \in V \times V$, and for any $\rho_c \in \mathcal{R}_c$, $\rho_c$ is a solution of \eqref{eq:transport-point} and $\phi_c,\psi_c$ is a solution of \eqref{eq:dual} if and only if 
\begin{equation}\label{eq:cs}
\phi_c(v_l)+\psi_c(v_h)=\pi_c(v_l,v_h)
\end{equation}
for all $(v_l,v_h) \in \supp(\rho_c)$. 
\end{lemma}

\begin{figure}[t!]
\centering
\begin{subfigure}[t]{0.48\textwidth}
\centering
\tikzset{
solid node/.style={circle,draw,inner sep=1.25,fill=gray2},
hollow node/.style={circle,draw,inner sep=1.25}
}
\begin{tikzpicture}[scale=6]
    \draw [<->, very thick] (0,0.85) node (yaxis) [above] {  }
        |- (1.05,0) node (xaxis) [right] {  $v$};
    \draw (0,0) node [left=2pt] { $0$};
    \draw [blue, very thick] (0,0.1)--(0.2,0.1);
    \draw [blue, very thick] (0.2,0.1)--(0.35,0.175);
    \draw [blue, very thick] (0.35,0.175)--(0.66,0.475);
    \draw [blue, very thick] (0.66,0.475)--(1,0.645);
    \draw [red, very thick] (0,0)--(0.1,0);
    \draw [red, very thick] (0.1,0)--(0.2,0.1);
    \draw [red, very thick] (0.2,0.1)--(0.35,0.175);
    \draw [red, very thick] (0.35,0.175)--(0.66,0.175);
    \draw [red, very thick] (0.66,0.175)--(1,0.345);
    \draw [mixed, very thick] (0.2,0.1)--(0.35,0.175);
    \draw [dotted, thick] (0.35,0.175)--(0.35,0);
    \draw [dotted, thick] (0.2,0.1)--(0.2,0);
    \draw [dotted, thick] (0.66,0.475)--(0.66,0);
    \draw (0.1,0) node [below=2pt] {  $\kappa_c^1$};
    \draw (0.2,0) node [below=2pt] {  $\kappa_c^3$};
    \draw (0.35,0) node [below=2pt] {  $\kappa_c^4$};
    \draw (0.66,0) node [below=2pt] {  $\kappa_c^5$}; 

    \matrix [draw=black, thin, anchor=north west, row sep=1pt] at (0.05,0.8) {
        \draw [blue, very thick] (0,0) -- (0.2,0); & \node[right] {  $\phi_c^\star$}; \\
        \draw [red, very thick] (0,0) -- (0.2,0); & \node[right] {  $\psi_c^\star$}; \\
    };
\end{tikzpicture}
\caption{Optimal Duals $\phi^\star$, $\psi^\star$}
\label{fig:dual}
\end{subfigure}
\hfill
\begin{subfigure}[t]{0.48\textwidth}
\centering
\begin{tikzpicture}[scale=6]
    \draw [ultra thick] (0,0) rectangle (1,1);
    \draw (1,0) node [right=2pt] {   $v_l$};
    \draw (0,1) node [above=2pt] {   $v_h$};
    \draw [thick] (0.1,-0.01)--(0.1,0.01);
    \draw [thick] (0.2,-0.01)--(0.2,0.01);
    \draw [thick] (0.35,-0.01)--(0.35,0.01);
    \draw [thick] (0.66,-0.01)--(0.66,0.01);
    \draw [thick] (-0.01,0.1)--(0.01,0.1);
    \draw [thick] (-0.01,0.2)--(0.01,0.2);
    \draw [thick] (-0.01,0.35)--(0.01,0.35);
    \draw [thick] (-0.01,0.66)--(0.01,0.66);
    \draw (0.215,0) node [below=2pt] {   $\kappa_c^3$};
    \draw (0.35,0) node [below=2pt] {   $\kappa_c^4$};
    \draw (0.66,0) node [below=2pt] {   $\kappa_c^5$};
    \draw (0,0.1) node [left=2pt] {   $\kappa_c^1$};
    \draw (0,0.2) node [left=2pt] {   $\kappa_c^3$};
    \draw (0,0.35) node [left=2pt] {   $\kappa_c^4$};
    \draw (0,0.66) node [left=2pt] {   $\kappa_c^5$};
    \draw (0.135,0) node [below=2pt] {   $\kappa_c^2$};

    \draw [dashed, thick] (0.1,0)--(0.1,1);
    \draw [dashed, thick] (0.2,0)--(0.2,1);
    \draw [dashed, thick] (0.35,0)--(0.35,1);
    \draw [dashed, thick] (0.66,0)--(0.66,1);
    \draw [dashed, thick] (0,0.1)--(1,0.1);
    \draw [dashed, thick] (0,0.2)--(1,0.2);
    \draw [dashed, thick] (0,0.35)--(1,0.35);
    \draw [dashed, thick] (0,0.66)--(1,0.66);
    \draw [dashed, thick] (0.15,0)--(0.15,1);

    \fill [pattern={north east lines},pattern color=red] (0.2,0) rectangle (0.35,0.1);
    \fill [pattern={north east lines},pattern color=red] (0,0.66) rectangle (0.2,1);
    \fill [pattern={north east lines},pattern color=red] (0.1,0.1)--(0.2,0.1)--(0.2,0.2);
    \fill [pattern={north east lines},pattern color=red] (0.35,0.35)--(0.35,0.66)--(0.66,0.66);
    \draw [red] (0.2,0.2)--(0.35,0.35);
    \draw [red] (0.66,0.66)--(1,1);

    \draw [blue, thick] plot [smooth, tension=0.75] coordinates {(0.15,0.1) (0.175,0.13) (0.2,0.2) };
    \draw [blue, thick] plot [smooth, tension=0.75] coordinates { (0,0.66) (0.05,0.8) (0.15,1)};
    \draw [blue, thick] (0.2,0.2)--(0.35,0.35);
    \draw [blue, thick] (0.66,0.66)--(1,1);
    \draw [blue, thick] plot [smooth, tension=0.75] coordinates { (0.35,0.35) (0.5,0.575) (0.66,0.66)};
    \draw [blue, thick] plot [smooth, tension=0.75] coordinates { (0.2,0) (0.25,0.03) (0.3,0.08) (0.35,0.1)};
\end{tikzpicture}
\caption{Complementary Slackness: Blue = support of $\rho_c^\star$, Red = dual binding regions}
\label{fig:joint}
\end{subfigure}
\caption{Structure of the optimal dual variables and complementary slackness conditions.}
\label{fig:combined}
\end{figure}
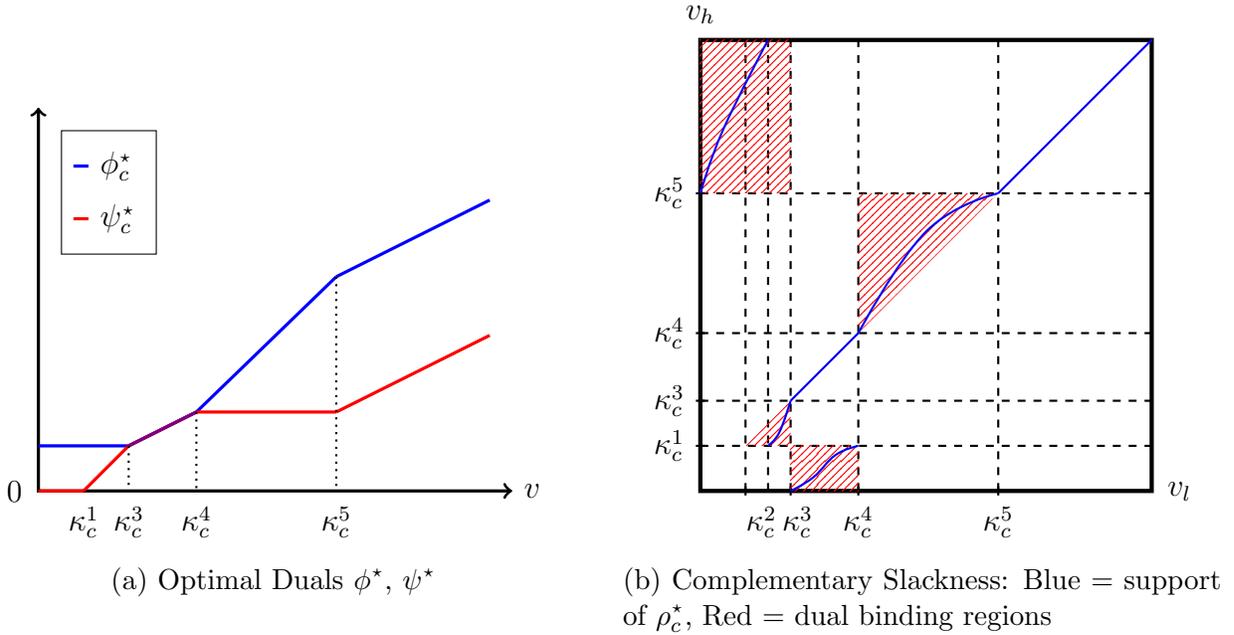

Therefore, to solve \eqref{eq:transport-point}, it suffices to find, for each $c \in C$, a joint distribution $\rho_c^\star \in \mathcal{R}_c$ and a pair of functions $\phi_c^\star$ and $\psi_c^\star$ such that $(\phi_c^\star,\psi_c^\star)$ is feasible in the dual problem \eqref{eq:dual} and that the complementary slackness condition holds: $\psi_c^\star(v_l)+\psi_c^\star(v_h)=\pi_c(v_l,v_h)$ for all $(v_l,v_h) \in \supp(\rho_c^\star)$. In the appendix, we construct explicitly the optimal dual variables $(\phi^\star_c,\psi^\star_c)$. \bref{fig:dual} illustrates the functions $\phi_c^\star$ and $\psi_c^\star$ when $\alpha_c=\nicefrac{1}{2}$ and $c=0$. 

Then, we show that the complementary slackness condition \eqref{eq:cs} holds under the joint distribution $\rho^\star_c$ associated with the pricing rule $p^\star$. \bref{fig:joint} illustrates the support of the joint distribution of $\rho^\star_c$ when $c \in C_1$, where the blue region indicates the support of $\rho_c^\star$ and the dashed red region indicates the set of $(v_l,v_h)$ at which $\phi_c^\star(v_l)+\psi_c^\star(v_h)=\pi_c(v_l,v_h)$.

\section{Conclusion}\label{sec:con}
We characterize the profit-maximizing non-discriminatory pricing rules. We show that the pricing problem can be represented by a family of optimal transport problems and explicitly solve the optimal transports. Under the optimal non-discriminatory pricing rule, consumers could retain a positive surplus given the non-discrimination constraint, even if the seller observes their types can engage in personalized pricing. This is reminiscent of information rents in screening problems, since some information is prohibited from being used even though it is not private. The distribution of information rents, however, differ qualitatively from standard screening problems: surplus is allocated to consumers with intermediate values, while low-value consumers are excluded, and high-value consumers are extracted. Furthermore, welfare gains could be distributed unevenly between protected groups. Depending on the value distribution and the population size, it is possible that the advantaged group benefits more from non-discrimination regulations than the disadvantaged group, at the expense of low-value consumers from the disadvantaged group being excluded. When strengthening the notion of non-discrimination, and requiring both the transaction outcomes and the transaction prices to be the same across protected groups, the protected group with lower values are worse-off whereas the protect group with higher values are better-off. We also consider several extensions to the baseline model, including imperfect price discrimination and implementable outcomes. 

\bibliography{lit.bib}

\section*{Appendix}
\renewcommand{\theequation}{A.\arabic{equation}}
\renewcommand{\thesubsection}{A.\arabic{subsection}}
\setcounter{equation}{0}
\subsection{Proof of Main Results}
\begin{proof}[Proof of \bref{lem:lem1}]
Consider any non-discriminatory pricing rule $p$. By definition, $p$ is independent of $\theta$ conditional on $c$. Therefore, for any $\tilde{c} \in C$ and for any $z \subseteq V$, 
\begin{align*}
\rho_{\tilde{c}}([0,z] \times V)=&\EE[\PP[v \in [0,z] \mid p, c=\tilde{c},\theta=l] \times \PP[v \in V\mid p,\theta=h,c=\tilde{c}]\mid c=\tilde{c}]\\
=& \EE[\PP[v \in [0,z]\mid p,c=\tilde{c},\theta=l]\mid c=\tilde{c}]\\
=& \EE[\PP[v \in [0,z]\mid c=\tilde{c},\theta=l]]\\
=&F_{\tilde{c},l}(z)\,.
\end{align*}
Likewise, 
\begin{align*}
\rho_{\tilde{c}}(V \times [0,z])=&\EE[\PP[v \in V \mid p, c=\tilde{c},\theta=l] \times \PP[v \in [0,z]\mid p,\theta=h,c=\tilde{c}]\mid c=\tilde{c}]\\
=& \EE[\PP[v \in [0,z]\mid p,c=\tilde{c},\theta=h]\mid c=\tilde{c}]\\
=& \EE[\PP[v \in [0,z]\mid c=\tilde{c},\theta=h]]\\
=&F_{\tilde{c},h}(z)\,.
\end{align*}
Therefore, the marginals of $\rho_c$ equals $F_{c,l}$ and $F_{c,h}$ for all $c \in C$, as desired. 
\end{proof}

\begin{proof}[Proof of \bref{prop:opt-transport}]
Consider any non-discriminatory pricing rule $p$, for each $c \in C$, let $\rho_c$ be defined by \eqref{eq:match}. We first claim that 
\[
\Pi(p) \leq \int_C \left(\int_{V^2} \pi_c(v_l,v_h) \rho_c(\diff v_l,\diff v_h)\right) G(\diff c)\,.
\]
Indeed, note that, for all $\hat{c} \in C$, by the definition of $\pi_{\hat{c}}$,
\[
    \int_{V^2} \pi_{\hat{c}} (v_l,v_h) \rho_{\hat{c}}(\diff v_l,v_h) = \int_{V^2} \max_{\tilde{p}} (\tilde{p}-\hat{c})\left[ (1-\alpha_{\hat{c}}) \mathbf{1}\{v_l \geq \tilde{p}\} + \alpha_{\hat{c}} \mathbf{1}\{v_h \geq \tilde{p}\} \right]  \rho_{\hat{c}}(\diff v_l,\diff v_h)\,.
\]
By the definition of $\rho_{\hat{c}}$, for all $\hat{c} \in C$, we have
\begin{equation}\label{eq:upperbound}
\begin{aligned}
    &\int_{V^2} \max_{\tilde{p}} (\tilde{p}-\hat{c})\left[ (1-\alpha_{\hat{c}}) \mathbf{1}\{v_l \geq \tilde{p}\} + \alpha_{\hat{c}} \mathbf{1}\{v_h \geq \tilde{p}\} \right]  \rho_{\hat{c}}(\diff v_l,\diff v_h)\\
    =&\E{ \max_{\tilde{p}} \E{(\tilde{p}-c) (1-\alpha_c) \mathbf{1}\{v \geq \tilde{p}\} }{ p, \theta=l, c = \hat{c} } \times \E{(\tilde{p}-c) \alpha_c \mathbf{1}\{v \geq \tilde{p}\}}{p, \theta= h,  c = \hat{c} } }{c = \hat{c}}\\
    =&\E{ \max_{\tilde{p}} \E{(\tilde{p}-c) \mathbf{1}\{v \geq \tilde{p}\} }{ p, c = \hat{c} } }{c = \hat{c}}\\
    \geq& \E{ \E{({p}-c) \mathbf{1}\{v \geq {p}\} }{ p, c = \tilde{c} } }{c = \hat{c}} \\
    =& \E{ ({p}-c) \mathbf{1}\{v \geq {p}\}  }{c = \hat{c}}\,.
\end{aligned}
\end{equation}
Therefore, 
\[
\Pi(p)=\EE[(p-c)\mathbf{1}\{v \geq p\}] \leq \int_{C}\left(\int_{V^2} \pi_c(v_l,v_h) \rho_c(\diff v_l,\diff v_h)\right)G(\diff c)\,,
\]
as desired. 

Next, we show that 
\[
\sup_{p \in \mathcal{D}} \Pi(p) \geq \int_C\left(\int_{V^2}\max_{\rho_c \in \mathcal{R}_c} \pi_c(v_l,v_h) \rho_c(\diff v_l,\diff v_h)\right)G(\diff c)\,.
\]
To this end, we show that any $\{\rho_c\}_{c\in C}$ such that $\rho_c \in \mathcal{R}_c$ for all $c \in C$, we can construct a non-discriminatory pricing rule $p$ such that 
\[
\Pi(p)=\int_C \left(\int_{V^2} \pi_c(v_l,v_h) \rho_c(\diff v_l,\diff v_h)\right)G(\diff c)\,.
\]
Indeed, consider any $\{\rho_c\}_{c \in C}$ such that $\rho_c \in \mathcal{R}_c$ for all $c \in C$. Since $V \subseteq \RR_+$ is a standard Borel space, by the disintegration theorem \citep[see, e.g.,][Theorem 2.17, pp. 151]{cinlarprobability2010}, for each $c \in C$, there exists transition probabilities $\gamma_{c,l}:V \to \Delta(V)$ and $\gamma_{c,h}:V \to \Delta(V)$ such that for all measurable $V_l,V_h \subseteq V$
\begin{equation}\label{eq:kernel}
\int_{V_h} \gamma_{c,l}(V_l\mid v_h) F_{c,h}(\diff v_h)=\rho_c(V_l \times V_h)=\int_{V_l} \gamma_{c,h}(V_h\mid v_l)F_{c,l}(\diff v_l)\,.
\end{equation}
Let $\Gamma_{c,\theta}(\cdot\mid v)$ be the CDF associated with $\gamma_{c,\theta}(\cdot\mid v)$, for all $v \in V$, $c \in C$ and $\theta \in \{l,h\}$. 
In the meantime, let $\xi_c:V^2 \to \RR_+$ be a measurable selection of 
\[
\argmax_{\tilde{p} \geq 0} (\tilde{p}-c)(\alpha_c \mathbf{1}\{v_h \geq \tilde{p}\}+(1-\alpha_c) \mathbf{1}\{v_l \geq \tilde{p}\})\,.
\]
Then, let 
\begin{equation}\label{eq:price}
p(v,c,\theta,r):=\begin{cases}
\xi_c(v,\Gamma^{-1}_{c,h}(r\mid v)),&\mbox{if } \theta=l\\
\xi_c(\Gamma^{-1}_{c,l}(r\mid v),v),&\mbox{if } \theta=h
\end{cases}\,,
\end{equation}
for all $(v,c,\theta,r) \in \Omega$. By construction, 
\begin{equation}\label{eq:lowerbound}
\begin{aligned}
\Pi(p)=&\EE[(p-c)\mathbf{1}\{v \geq p\}]\\
=&\EE[\EE[\EE[(p-c)\mathbf{1}\{v \geq p\}\mid c,\theta]\mid c]]\\
=&\int_C \bigg(\alpha_c \int_{V \times [0,1]} (\xi_c(\Gamma_{c,l}^{-1}(r\mid v_h),v_h)-c)\mathbf{1}\{ v_h\geq \xi_c(\Gamma_{c,l}^{-1}(r\mid v_h),v_h)\}\diff rF_{c,h}(\diff v_h)\\
&+(1-\alpha_c) \int_{V \times [0,1]} (\xi_c(v_l,\Gamma_{c,h}^{-1}(r\mid v_l))-c)\mathbf{1}\{v_l \geq \xi_c(v,\Gamma_{c,h}^{-1}(r\mid v_l))\}\diff rF_{c,l}(\diff v_l) \bigg)G(\diff c)\\
=& \int_C \bigg(\alpha_c \int_{V^2} (\xi_c(v_l,v_h)-c)\mathbf{1}\{v_h \geq \xi_c(v_l,v_h)\}\Gamma_{c,l}(\diff v_l\mid v_h)F_{c,h}(\diff v_h)\\
&+(1-\alpha_c)\int_{V^2}(\xi_c(v_l,v_h)-c)\mathbf{1}\{v_l \geq \xi_c(v_l,v_h)\}\Gamma_{c,h}(\diff v_h\mid v_l)F_{c,l}(\diff v_l)\bigg)G(\diff c)\\
=&\int_C \left(\int_{V^2} (\xi_c(v_l,v_h)-c)(\alpha_c\mathbf{1}\{v_h \geq \xi_c(v_l,v_h)+(1-\alpha_c)\mathbf{1}\{v_l \geq \xi_c(v_l,v_h)\})\rho_c(\diff v_l,\diff v_h)\right)G(\diff c)\\
=&\int_C\left(\int_{V^2}\pi_c(v_l,v_h)\rho_c(\diff v_l,\diff v_h)\right)G(\diff c)\,,
\end{aligned}
\end{equation}
where the second equality follows from the law of iterated expectation, the fourth equality follows from changing variables of the integration, the fifth equality follows from \eqref{eq:kernel}, and the last equality follows from the definition of $\xi_c$. 

Moreover, for any $c \in C$ and for any measurable $M \subseteq V$, 
\begin{align*}
\PP[p \in M\mid c,\theta=l]=&\PP[\xi_c(v,\Gamma_{c,h}^{-1}(r\mid v)) \in M\mid c, \theta=l]\\
=&\int_{V \times [0,1]} \mathbf{1}\{\xi_c(v_l,\Gamma_{c,h}^{-1}(r\mid v_l))\in M\}\diff r \diff F_{c,l}(\diff v_l)\\
=&\int_{V^2} \mathbf{1}\{\xi_c(v_l,v_h) \in M\} \Gamma_{c,h}(\diff v_h\mid v_l)F_{c,l}(\diff v_l)\\
=&\int_{V^2} \mathbf{1}\{\xi_c(v_l,v_h) \in M\} \rho_c(\diff v_l,\diff v_h)\,,
\end{align*}
where the third equality again follows from changing variables of the integration, and the last equality follows from \eqref{eq:kernel}. Likewise, 
\begin{align*}
\PP[p \in M\mid c,\theta=h]=&\PP[\xi_c(\Gamma_{c,l}^{-1}(r\mid v),v) \in M\mid c, \theta=h]\\
=&\int_{V \times [0,1]} \mathbf{1}\{\xi_c(\Gamma_{c,l}^{-1}(r\mid v_h),v_h)\in M\}\diff r \diff F_{c,h}(\diff v_h)\\
=&\int_{V^2} \mathbf{1}\{\xi_c(v_l,v_h) \in M\} \Gamma_{c,l}(\diff v_l\mid v_h)F_{c,h}(\diff v_h)\\
=&\int_{V^2} \mathbf{1}\{\xi_c(v_l,v_h) \in M\} \rho_c(\diff v_l,\diff v_h)\,.
\end{align*}
As a result, for all $c \in C$, and for all measurable $M \subseteq [0,1]$,
\[
\PP[p \in M\mid c,\theta=l]=\int_{V^2} \mathbf{1}\{\xi_c(v_l,v_h) \in M\}\rho_c(\diff v_l, \diff v_h) =\PP[p \in M\mid c,\theta=h]\,,
\]
and thus $p$ is indeed non-discriminatory. 

Together, we have 
\begin{equation}\label{eq:prop1}
\sup_{p \in \mathcal{D}}\Pi(p)=\int_C\left(\max_{\rho_c \in \mathcal{R}_c}\int_{V^2} \pi_c(v_l,v_h)\rho_c(\diff v_l,\diff v_h)\right)\,.
\end{equation}
Furthermore, for any profit-maximizing non-discriminatory pricing rule $p$, let $\rho_c$ be defined by \eqref{eq:match}, \eqref{eq:upperbound} and \eqref{eq:prop1} then implies that $\{\rho_c\}_{c \in C}$ solves \eqref{eq:transport}. Conversely, for any $\{\rho_c\}_{c \in C}$ that solves \eqref{eq:transport}, let $p$ be defined by \eqref{eq:price}. Then $p$ solves \eqref{eq:optimal-pricing} by \eqref{eq:lowerbound} and \eqref{eq:prop1}. This completes the proof. 
\end{proof}

\begin{proof}[Proof of \bref{lem:system}]
Since $\Delta_c$ is continuous and quasi-concave, it has a unique maximizer. Let $v^\star_c$ be the unique maximizer of $\Delta_c$ for all $c \in C$. Since $F_{c,l}$ and $F_{c,h}$ are CDFs on $\RR_+$ that are absolutely continuous, $\Delta_c(0)=0$ and $\lim_{v \to\infty } \Delta_c(v)=0$, and 
\[
\|F_{c,l}-F_{c,h}\|=\max_{A \subseteq \RR_+} \bigg|\int_A [f_{c,l}(v)-f_{c,h}(v)] \diff v\bigg|=\max_v \Delta_c(v)=\Delta_c(v^\star_c)\,.
\]
Moreover, since $\Delta_c$ is continuous and quasi-concave, for any $v \geq v_c^\star$, there exists a unique $g_c(v) \in [\lb{v}_c, v_c^\star]$ such that $\Delta_c(v)=\Delta_c(g_c(v))$. Moreover, the function $g_c:[v_c^\star,\infty) \to [\lb{v}_c,v_c^\star]$ is continuous and decreasing in $v$, with $g_c(v_c^\star)=v_c^\star$ and $\lim_{v \to \infty } g_c(v)=\lb{v}_c$. For any $v \geq v_c^\star$, let 
\[
h_c(v):=v-g_c(v)\,.
\]
Note that $h_c$ is increasing on $[v_c^\star,\ub{v}_c]$ and $h_c(v_c^\star)=0$, $\lim_{v \to \infty} h_c(v)=\infty$. In particular, since $F_{c,l}(c)<\|F_{c,l}-F_{c,h}\|=\Delta_c(v^\star_c)$, and thus $c<v^\star_c$, there exists a unique $\tilde{v}_c>v^\star_c$ such that  $\alpha_c h_c(\tilde{v}_c) =(1-\alpha_c)(v_c^\star-c)$. Meanwhile let $\hat{v}_c:=\inf\{v \geq v^\star_c: \alpha_ch_c(v) \geq (1-\alpha_c)(\lb{v}_c-c)\}$. Since $h_c$ is nondecreasing, it must be that $\hat{v}_c \in[v^\star_c, \tilde{v}_c]$. 

Note that, if $\hat{v}_c=v^\star_c$, then it must be that 
\[
\Delta_c\left(\frac{\alpha_c}{1-\alpha_c}h_c(\hat{v}_c)+c\right)+F_{c,h}(\alpha_c h_c(\hat{v}_c)+c)=F_{c,l}(c)<\|F_{c,l}-F_{c,h}\|=\Delta_c(v^\star_c)=\Delta_c(\hat{v}_c)\,;
\]
If $\hat{v}_c \in (v^\star_c,\ub{v}_c)$, then it must be that $0\leq \alpha_ch_c(\hat{v}_c)=(1-\alpha_c)(\lb{v}_c-c)$, and thus $\alpha_ch_c(\hat{v}_c)+c=(1-\alpha_c)\lb{v}_c+\alpha_c c \leq \lb{v}_c$. Therefore, 
\[
\Delta_c\left(\frac{\alpha_c}{1-\alpha_c}h_c(\hat{v}_c)+c\right)+F_{c,h}(\alpha_c h_c(\hat{v}_c)+c)=0<\Delta_c(\hat{v}_c)\,.
\]
If $\hat{v}_c \geq \ub{v}_c$, then 
\[
\Delta_c\left(\frac{\alpha_c}{1-\alpha_c}h_c(\hat{v}_c)+c\right)+F_{c,h}(\alpha_c h_c(\hat{v}_c)+c)=0=\Delta_c(\hat{v}_c)\,.
\]
Since $\Delta_c$ is quasi-concave, and hence is decreasing on $[v^\star_c,\ub{v}_c]$ while $\Delta_c(v)=0$ for all $v \geq \ub{v}_c$. In the meantime, since $\alpha_ch_c(v)/(1-\alpha_c)+c \leq v^\star_c$ for all $v \in [v^\star_c,\tilde{v}_c]$ and since $h_c$ is increasing in $v$, the function 
\[
v \mapsto \Delta_c\left(\frac{\alpha_c}{1-\alpha_c}h_c(v)+c\right)+F_{c,h}(\alpha_c h_c(v)+c)
\]
is increasing on $[v_c^\star,\tilde{v}_c]$. Together, there exists a unique $\kappa_c^5 \in [\hat{v}_c,\tilde{v}_c]$ such that 
\begin{equation}\label{eq:fixedpoint}
\Delta_c(\kappa^5_c)=\Delta_c\left(\frac{\alpha_c}{1-\alpha_c} h_c(\kappa^5_c)+c\right)+F_{c,h}(\alpha_c h_c(\kappa^5_c)+c)\,,
\end{equation}
and that 
\[
\frac{\alpha_c}{1-\alpha_c} h_c(\kappa^5_c)\leq \frac{\alpha_c}{1-\alpha_c} h_c(\tilde{v}_c)=v_c^\star-c\,.
\]
Let $\kappa^4_c:=g_c(\kappa^5_c)$ and $\kappa^1_c:=\alpha_c h_c(\kappa^5_c)+c$, $\kappa^3_c:=\alpha_c h_c(\kappa_c^5)/(1-\alpha_c)+c$, and $\kappa_c^2:=F_{c,l}^{-1}(\Delta_c(\kappa^5_c))$. By construction, $\kappa_c^5 \geq \hat{v}_c \geq v^\star_c>\kappa_c^4$, with at least one of the first two inequalities being strict, and $\kappa_c^3 \geq \kappa_c^1$. Moreover, since $\kappa_c^3=\alpha_ch_c(\kappa_c^5)/(1-\alpha_c)+c \leq v^\star_c$ and since $\Delta_c(\kappa_c^3) \leq \Delta_c(\kappa_c^4)<\Delta_c(v^\star_c)$, $\kappa_c^3 \leq \kappa_c^4$. In the meantime, since $F_{c,l}(\kappa_c^2)=\Delta_c(\kappa_c^3)+F_{c,h}(\kappa_c^1)$, 
\[
F_{c,l}(\kappa_c^3)-F_{c,l}(\kappa_c^2)=F_{c,h}(\kappa_c^3)-F_{c,h}(\kappa_c^1) \geq 0\,,
\]
and hence $\kappa_c^2 \leq \kappa_c^3$. Lastly, since $\kappa_c^1 \leq \kappa_c^3 \leq v^\star_c$, it must be that $\Delta_c(\kappa_c^1)\leq \Delta_c(\kappa_c^3)$. Therefore, 
\[
F_{c,l}(\kappa_c^2)-F_{c,h}(\kappa_c^1)=\Delta_c(\kappa_c^3) \geq \Delta_c(\kappa_c^1)=F_{c,l}(\kappa_c^1)-F_{c,h}(\kappa_c^1)\,,
\]
 and hence $F_{c,l}(\kappa_c^2) \geq F_{c,l}(\kappa_c^1)$, which in turn implies $\kappa_c^2 \geq \kappa_c^1$. 
 
Together, it then follows that $\kappa^1_c\leq \kappa_c^2 \leq \kappa^3_c\leq \kappa^4_c<v^\star_c < \kappa^5_c$. Moreover, for any $\tilde{\kappa}_c \in \RR^5$ that solves \eqref{eq:system} such that $\tilde{\kappa}_c^4<v^\star_c<\tilde{\kappa}_c^5$, it must be that $\tilde{\kappa}_c^4 \geq \tilde{\kappa}_c^2 \geq \lb{v}_c$. Since $\kappa_c^5$ is the unique solution of \eqref{eq:fixedpoint} among $v \in [\hat{v}_c,\tilde{v}_c]$, for which $g_c(v) \geq \lb{v}_c$, it must be that $\tilde{\kappa}_c^5=\kappa_c^5$. Meanwhile, since $\tilde{\kappa}_c$ solves \eqref{eq:system}, it must be that $\tilde{\kappa}_c=\kappa_c$. Thus, $\kappa_c$ is the unique increasing vector in $\RR^5$ with $\kappa_c^4<v^\star_c<\kappa_c^5$ that solves \eqref{eq:system}. 
\end{proof}

\begin{proof}[Proof of \bref{thm2}]
Note that \bref{thm2} immediately implies \bref{thm1}, and therefore we prove \bref{thm2} directly. For any $c \in C$, if $F_{c,l}(c) \geq \|F_{c,l}-F_{c,h}\|$, let 
\[
\phi^\star_c(v_l):=(1-\alpha_c)\cdot (v_l-c)^+ \quad \mbox{ and } \quad \psi^\star_c(v_h):=\alpha_c \cdot (v_h-c)^+\,,
\]
for all $(v_l,v_h) \in V \times V$. Meanwhile, if $F_{c,l}(c)<\|F_{c,l}-F_{c,h}\|$, let 
\[
\phi_c^\star(v_l):=\begin{cases}
\kappa_c^1-c,& \mbox{if } v_l \leq \kappa_c^3\\
(1-\alpha_c)\cdot(v_l-c),&\mbox{if } v_l \in \left(\kappa_c^3,\kappa_c^4\right]\\
v_l-c-\alpha_c\cdot(\kappa_c^4-c),&\mbox{if } v_l \in (\kappa_c^4,\kappa_c^5]\\
(1-\alpha_c)(v_l-c)+\kappa_c^1-c, &\mbox{if } v_l>\kappa_c^5
\end{cases}\,,
\]
and let 
\[
\psi_c^\star(v_h):=\begin{cases}
0,& \mbox{if } v_h \leq \kappa_c^1\\
v_h-\kappa_c^1,&\mbox{if } v_h \in \left(\kappa_c^1,\kappa_c^3\right]\\
\alpha_c\cdot (v_h-c),&\mbox{if } v_h \in \left(\kappa_c^3,\kappa_c^4\right]\\
\alpha_c\cdot (\kappa_c^4-c),&\mbox{if } v_h \in (\kappa_c^4,\kappa_c^5]\\
\alpha_c(v_h-c)-(\kappa_c^1-c), &\mbox{if } v_h>\kappa_c^5
\end{cases}\,.
\]

\begin{taggedlemma}{A.1}\label{lem:feasible}
For any $(v_l,v_h) \in V \times V$ and for any $c \in C$, 
\[
\phi_c^\star(v_l)+\psi_c^\star(v_h) \geq \pi_c(v_l,v_h)\,.
\]
\end{taggedlemma}
The proof of \bref{lem:feasible} is by inspection, using the system of equation \eqref{eq:system} that defines $\kappa_c$. Details of the proof can be found in \bref{sec:a.3}. Next, we define a joint distribution $\rho_c^\star \in \Delta(V^2)$. When $F_{c,l}<\|F_{c,l}-F_{c,h}\|$, define a transition probability $\gamma_c^\star:V \to \Delta(V)$ as follows: 
\begin{align*}
&\gamma^\star_c(v_l \leq x \mid v_h):=\\
&\begin{cases}
\mathbf{1}\{\lb{\Delta}_c^{-1}(F_{c,h}(v_h+\Delta_c(\kappa_c^3)) \leq x\},& \mbox{if } v_h \leq \kappa_c^1\\
\mathbf{1}\{F_{c,l}^{-1}(F_{c,h}(v_h)+\Delta_c(\kappa_c^3)) \leq x\},&\mbox{if } v_h \in (\kappa_c^1,\kappa^3_c]\\
\mathbf{1}\{v_h \leq x\},& \mbox{if } v_h \in (\kappa_c^3,\kappa_c^4]\\
\mathbf{1}\{F_{c,l}^{-1}(F_{c,h}(v_h)+\Delta_c(\kappa_c^4)) \leq x\}, &\mbox{if } v_h \in (\kappa_c^4,\kappa_c^5]\\
\frac{f_{c,l}(v_h)}{f_{c,h}(v_h)}\cdot \mathbf{1}\{v_h \leq x\}+\frac{f_{c,h}(v_h)-f_{c,l}(v_h)}{f_{c,h}(v_h)} \cdot \mathbf{1}\{F_{c,l}^{-1}(\Delta_c(\kappa_c^5)-\Delta_c(v_h)) \leq x\},&\mbox{if } v_h > \kappa_c^5,
\end{cases}\,,
\end{align*}
for all $x \in V$ and for all $v_h \in V$. Meanwhile, when $F_{c,l} \geq \|F_{c,l}-F_{c,h}\|$ and $c<v^\star_c$, define a transition probability $\gamma^\star_c:V \to \Delta(V)$ as: 
\begin{align*}
&\gamma_c^\star(v_l \leq x \mid v_h):=\\
&\begin{cases}
\mathbf{1}\left\{F_{c,l}^{-1}(F_{c,h}(v_h)) \leq x\right\},&\mbox{if } v_h \leq \eta_c^h\\
\mathbf{1}\left\{\lb{\Delta}_c^{-1}(F_{c,h}(v_h)-F_{c,h}(\eta_c^h)+\Delta_c(c)) \leq x\right\},&\mbox{if }  v_h \in (\eta_c^h,c]\\
\mathbf{1}\{v_h \leq x\},&\mbox{if } v_h \in (c,v^\star_c]\\
\frac{f_{c,l}(v_h)}{f_{c,h}(v_h)}\mathbf{1}\{v_h \leq x\}+\frac{f_{c,h}(v_h)-f_{c,l}(v_h)}{f_{c,h}(v_h)} \mathbf{1}\{F_{c,l}^{-1}(\Delta_c(v^\star_c)-\Delta_c(v_h)+F_{c,l}(\eta_c^l)) \leq x\}, &\mbox{if } v_h >v_c^\star
\end{cases}\,,
\end{align*}
for all $x \in V$ and for all $v_h \in V$. When $F_{c,l} \geq \|F_{c,l}-F_{c,h}\|$ and $c \geq v^\star_c$, define $\gamma^\star_c$ as: 
\begin{align*}
&\gamma^\star_c(v_l \leq x\mid v_h):=\\
&\begin{cases}
\mathbf{1}\left\{F_{c,l}^{-1}(F_{c,h}(v_h)) \leq x\right\},&\mbox{if } v_h \leq c\\
\frac{f_{c,l}(v_h)}{f_{c,h}(v_h)}\mathbf{1}\{v_h \leq x\}+\frac{f_{c,h}(v_h)-f_{c,l}(v_h)}{f_{c,h}(v_h)} \mathbf{1}\{F_{c,l}^{-1}(F_{c,l}(c)-\Delta_c(v_h)) \leq x\}, &\mbox{if } v_h >c
\end{cases}\,,
\end{align*}
for all $x \in V$ and for all $v_h \in V$.

Then, for all $c \in C$, let $\rho_c^\star \in \Delta(V \times V)$ be defined as 
\begin{equation}\label{eq:sol2}
\rho_c^\star(v_l \in A, v_h \in B):=\int_B \gamma_c^\star(A\mid v_h) F_{c,h}(\diff v_h)\,,
\end{equation}
for all measurable sets $A,B \subseteq V$. By construction, the marginals of $\rho_c^\star$ are exactly $F_{c,l}$ and $F_{c,h}$. That is,

\begin{taggedlemma}{A.2}\label{lem:rfeas}
$\rho_c^\star \in \mathcal{R}_c$ for all $c \in C$. 
\end{taggedlemma}

Combining \bref{lem:dual}, \bref{lem:feasible} and \bref{lem:rfeas} with   \bref{lem:complementary_slackness} below, it then follows that $\rho_c^\star$ is a solution of \eqref{eq:transport-point}. 

\begin{taggedlemma}{A.3}\label{lem:complementary_slackness}
For any $c \in C$, $\phi_c^\star(v_l)+\psi_c^\star(v_h)=\pi_c(v_l,v_h)$ for all $(v_l,v_h) \in \mathrm{supp}(\rho_c^\star)$.
\end{taggedlemma}

Since $\rho_c^\star$ is a solution of \eqref{eq:transport-point} for all $c$, \bref{prop:opt-transport} implies that one can construct an optimal non-discriminatory pricing rule from $\{\rho^\star_c\}_{c \in C}$. To this end, for any $c \in C$, let $\beta_c^\star: V \to \Delta(V)$ be the conditional distribution of $v_h$ given $v_l$ implied by $\rho_c^\star$. That is, $\beta_c^\star$ is a version of the regular conditional probability defined by 
\begin{equation}\label{eq:beta}
\rho_c^\star(v_l \in A,v_h \in B)=\int_B \beta_c^\star(v_l \in A\mid v_h ) F_{c,h}(\diff v_h)\,,
\end{equation}
for all measurable $A,B \subseteq V$. Next, let $\gamma_c^{-1}$ and $\beta_c^{-1}$ be the quantile function defined by $\gamma_c^\star$ and $\beta_c^\star$, respectively. That is, 
\[
\gamma_c^{-1}(r\mid v_h):=\inf\{v_l \in V: \gamma_c^\star([0,v_l]\mid v_h ) \geq r\} \mbox{ and } \beta_c^{-1}(r \mid v_l):=\inf\{v_h \in V: \beta_c^\star([0,v_h]\mid v_l) \geq r\}
\]
for all $r \in [0,1]$ and for all $(v_l,v_h) \in V^2$. Meanwhile, for any $(v_l,v_h) \in V^2$, let $\ub{p}_c(v_l,v_h)$ be the minimum element of 
\[
\argmax_{\tilde{p} \geq 0} (\tilde{p}-c)(\alpha_c \mathbf{1}\{v_h \geq \tilde{p}\}+(1-\alpha_c)\mathbf{1}\{v_l \geq \tilde{p}\})\,.
\]
It then follows that $p^\star$ can be written as 
\[
p^\star(v,l,c,r)=\ub{p}_c(v,\beta_c^{-1}(r\mid v)) \mbox{ and } p^\star (v,h,c,r)=\ub{p}_c(\gamma_c^{-1}(r\mid v),v)
\]
for all $v \in V$, $c \in C$ and $r \in [0,1]$. By construction, 
\begin{align*}
\PP[p^\star \in M\mid c,\theta=l]=&\PP[\ub{p}_c(v,\beta_{c}^{-1}(r\mid v)) \in M\mid \theta=l,c]\\
=&\int_{V \times [0,1]} \mathbf{1}\{\ub{p}_c(v_l,\beta_{c}^{-1}(r\mid v_l))\in M\}\diff r \diff F_{c,l}(\diff v_l)\\
=&\int_{V^2} \mathbf{1}\{\ub{p}_c(v_l,v_h) \in M\} \beta^\star_{c}(\diff v_h\mid v_l)F_{c,l}(\diff v_l)\\
=&\int_{V^2} \mathbf{1}\{\ub{p}_c(v_l,v_h) \in M\} \rho^\star_c(\diff v_l,\diff v_h)\,,
\end{align*}
where the third equality again follows from changing variables of the integration, and the last equality follows from \eqref{eq:sol2} and \eqref{eq:beta}. Likewise, 
\begin{align*}
\PP[p^\star \in M\mid c,\theta=h]=&\PP[\ub{p}_c(\gamma_{c}^{-1}(r\mid v),v) \in M\mid \theta=h,c]\\
=&\int_{V \times [0,1]} \mathbf{1}\{\ub{p}_c(\gamma_{c}^{-1}(r\mid v_h),v_h)\in M\}\diff r \diff F_{c,h}(\diff v_h)\\
=&\int_{V^2} \mathbf{1}\{\ub{p}_c(v_l,v_h) \in M\} \gamma^\star_{c}(\diff v_l\mid v_h)F_{c,h}(\diff v_h)\\
=&\int_{V^2} \mathbf{1}\{\ub{p}_c(v_l,v_h) \in M\} \rho^\star_c(\diff v_l,\diff v_h)\,.
\end{align*}
As a result, for all $c \in C$, and for all measurable $M \subseteq [0,1]$,
\[
\PP[p^\star \in M\mid c,\theta=l]=\int_{V^2} \mathbf{1}\{\ub{p}_c(v_l,v_h) \in M\}\rho^\star_c(\diff v_l \diff v_h) =\PP[p^\star \in M\mid c,\theta=h]\,,
\]
and thus $p^\star$ is indeed non-discriminatory. Moreover, 
\begin{align*}
\Pi(p^\star)=&\EE[(p^\star-c)\mathbf{1}\{v \geq p^\star\}]\\
=&\EE[\EE[\EE[(p^\star-c)\mathbf{1}\{v \geq p^\star\}\mid c,\theta]\mid c]]\\
=&\int_C \bigg(\alpha_c \int_{V \times [0,1]} (\ub{p}_c(\gamma_{c}^{-1}(r\mid v_h),v_h)-c)\mathbf{1}\{ v_h\geq \ub{p}_c(\gamma_{c}^{-1}(r\mid v_h),v_h)\}\diff rF_{c,h}(\diff v_h)\\
&+(1-\alpha_c) \int_{V \times [0,1]} (\ub{p}_c(v_l,\beta_{c}^{-1}(r\mid v_l))-c)\mathbf{1}\{v_l \geq \ub{p}_c(v,\beta_{c}^{-1}(r\mid v_l))\}\diff rF_{c,l}(\diff v_l) \bigg)G(\diff c)\\
=& \int_C \bigg(\alpha_c \int_{V^2} (\ub{p}_c(v_l,v_h)-c)\mathbf{1}\{v_h \geq \ub{p}_c(v_l,v_h)\}\gamma^\star_{c}(\diff v_l\mid v_h)F_{c,h}(\diff v_h)\\
&+(1-\alpha_c)\int_{V^2}(\ub{p}_c(v_l,v_h)-c)\mathbf{1}\{v_l \geq \ub{p}_c(v_l,v_h)\}\beta_{c}^\star(\diff v_h\mid v_l)F_{c,l}(\diff v_l)\bigg)G(\diff c)\\
=&\int_C \left(\int_{V^2} (\ub{p}_c(v_l,v_h)-c)(\alpha_c\mathbf{1}\{v_h \geq \ub{p}_c(v_l,v_h)\}+(1-\alpha_c)\mathbf{1}\{v \geq \ub{p}_c(v_l,v_h)\})\rho^\star_c(\diff v_l,\diff v_h)\right)G(\diff c)\\
=&\int_C\left(\int_{V^2}\pi_c(v_l,v_h)\rho^\star_c(\diff v_l,\diff v_h)\right)G(\diff c)\,,
\end{align*}
where the second equality follows from the law of iterated expectation, the fourth equality follows from changing variables of the integration, the fifth equality follows from \eqref{eq:sol2} and \eqref{eq:beta}, and the last equality follows from the definition of $\ub{p}_c$. Thus, by \bref{prop:opt-transport}, $p^\star$ is indeed an optimal non-discriminatory pricing rule. This completes the proof of $(i)$. Part $(ii)$ then immediately follows from \bref{prop:welfare}. 
\end{proof}

\begin{proof}[Proof of \bref{prop:surplus}]
For $(i)$, by \bref{thm1}, to show that $CS(c,h;p^\star)>0$, it suffices to show $F_{c,h}(\kappa_c^5)>F_{c,h}(\kappa_c^4)$ and that $F_{c,h}(\kappa_c^1)>0$. To see this, since $\kappa_c^4<v^\star_c<\kappa_c^5$ and since $v_c^\star \in (\lb{v}_c, \ub{v}_c)$, it must be that $F_{c,h}(\kappa_c^5)>F_{c,h}(\kappa_c^4)$, as desired. In the meantime, to show that $WL(c,h;p^\star)$, it suffices to show $F_{c,h}(\kappa_c^1)>0$ by \bref{thm1}. According to \eqref{eq:system}, $\kappa_c^1>c$ for all $c$. Therefore, whenever $\lb{v}_c \leq c$, $F_{c,h}(\kappa_c^1)>F_{c,h}(c) \geq F_{c,h}(\lb{v}_c)=0$, as desired. 

For $(ii)$, suppose that $\alpha_c \ub{v}_c>\lb{v}_c-(1-\alpha_c)c$. By \bref{thm1}, it suffices to show that $F_{c,l}(\kappa_c^3)>F_{c,l}(\kappa_c^2)>0$. We first claim that $\Delta_c(\kappa_c^4)>0$. To see this, suppose the contrary. $\Delta_c(\kappa_c^4)=0$. Then it must be that $\kappa_c^4 \leq \lb{v}_c$ and $\kappa_c^5 \geq \ub{v}_c$. Moreover, $\kappa_c^2=F_{c,l}^{-1}(\Delta_c(\kappa_c^4))=F_{c,l}^{-1}(0)=\lb{v}_c$, and $\Delta_c(\kappa_c^3)=F_{c,h}(\kappa_c^1)=0$. Since $\kappa_c$ is increasing, it must be that $\kappa_c^3 \leq \kappa_c^4 \leq \lb{v}_c$. Together, 
\[
\lb{v}_c=\kappa_c^3 \leq \kappa_c^3 \leq \kappa_c^4 \leq \lb{v}_c\,,
\]
Therefore, $\kappa_c^3=\kappa_c^4=\lb{v}_c$. Thus, 
\[
(1-\alpha_c)(\kappa_c^3-c)=(1-\alpha_c)(\lb{v}_c-c)=\alpha_c(\kappa_c^5-\kappa_c^4) \geq \alpha_c(\ub{v}_c-\lb{v}_c)\,,
\]
and hence 
\[
\alpha_c \ub{v}_c \leq \lb{v}_c-(1-\alpha_c)c\,,
\]
a contradiction. As a result, $\Delta_c(\kappa_c^4)>0$. This implies that $F_{c,l}(\kappa_c^2)>0$. Together, we have that $CS(c,l;p^\star)>0$ and $WL(c,h;p^\star)>0$.

Conversely, suppose that $\alpha_c(\overline{v}_c-c) \leq \underline{v}_c-c$. Then, it must be that $\kappa_c^1 \leq \underline{v}_c$, since otherwise, as $\kappa_c^3 \leq \kappa_c^4<v^\star_c<\ub{v}_c$, 
\[
\alpha_c(\kappa_c^3-c) \leq \alpha_c(\ub{v}_c) \leq \lb{v}_c-c < \kappa_c^1-c\,,
\]
contradicting to \eqref{eq:system}. Since $\kappa_c^1 \leq \lb{v}_c$, \eqref{eq:system} then implies that  $0=\Delta_c(\kappa_c^3)=\Delta_c(\kappa_c^4)$, and hence $\kappa_c^3=\kappa_c^4=\lb{v}_c$, which in turn implies that $\kappa_c^2=\lb{v}_c$ and $\kappa_c^5=\ub{v}_c$. Thus, by \bref{thm1}, $CS(c,l;p^\star)=WL(c,l;p^\star)=0$, as desired. 

\end{proof}

\begin{proof}[Proof of \bref{prop:profit-bound}]
Recall that the assortative pricing rule $p^{ass}$, defined by \eqref{eq:assortative-matching} gives a profit of $\E{(v-c)^+}{c=\tilde{c},\theta=l}$ for all $\tilde{c} \in C$. Since $p^{ass}$ is non-discriminatory, and since 
\[
\E{(v-c)^+}{c=\tilde{c}}=\alpha_{\tilde{c}}\E{(v-c)^+}{c=\tilde{c},\theta=h}+(1-\alpha_{\tilde{c}})\E{(v-c)^+}{c=\tilde{c},\theta=l}
\]
we have
    \begin{equation}\label{eq:lb1}
        \frac{\Pi^\star_{\tilde{c}}}{\E{(v-c)^+}{c=\tilde{c}}} \geq \frac{\E{(v-c)^+}{c=\tilde{c},\theta=l}}{\E{(v-c)^+}{c=\tilde{c}}} \geq \frac{1}{\alpha_{\tilde{c}}r_{\tilde{c}} + 1} \,,
    \end{equation}
    for all $\tilde{c} \in C$. In the meantime, since the partly anti-assortative pricing rule $p^{anti}$ defined by \eqref{eq:partly}, with $q_c=1$ for all $c$, gives a profit of $\E{\alpha_c(v-c)^+}{c=\tilde{c},\theta=h}$ conditional on every $\tilde{c} \in C$, and is also non-discriminatory, we have  
    \begin{equation}\label{eq:lb2}
          \frac{\Pi^\star_{\tilde{c}}}{\E{(v-c)^+}{c=\tilde{c}}} \geq \frac{\alpha_{\tilde{c}}\E{(v-c)^+}{c=\tilde{c},\theta=h}}{\E{(v-c)^+}{c=\tilde{c}}} = \frac{\alpha_{\tilde{c}}(r_{\tilde{c}}+1)}{\alpha_{\tilde{c}}r_{\tilde{c}} + 1}\,,
    \end{equation}
 for all $\tilde{c} \in C$. Since the right-hand side of \eqref{eq:lb1} is decreasing in $\alpha_{\tilde{c}}$ and the left-hand side of \eqref{eq:lb2} is increasing in $\alpha_{\tilde{c}}$, the maximum of the two is minimizes at $\alpha_{\tilde{c}}=\nicefrac{1}{(r_{\tilde{c}}+1)}$, we have 
\[
\frac{\Pi^\star_{\tilde{c}}}{\E{(v-c)^+}{c=\tilde{c}}} \geq \max\left\{\frac{1}{\alpha_{\tilde{c}}r_{\tilde{c}} + 1},\frac{\alpha_{\tilde{c}}(r_{\tilde{c}}+1)}{\alpha_{\tilde{c}}r_{\tilde{c}} + 1}\right\} \geq \frac{r_{\tilde{c}}+1}{2\,r_{\tilde{c}} + 1}\,.
\]
as desired. 
\end{proof}

\begin{proof}[Proof of \bref{prop:population-size}]
For $(i)$, consider each $c \in C$, note that since $\tilde{v}_c$ defined in the proof of \bref{lem:system} is decreasing in $\alpha_c$, and since the function 
\[
\alpha_c \mapsto \Delta_c\left(\frac{\alpha_c}{1-\alpha_c} h_c(v)+c\right)+F_{c,h}(\alpha_c h_c(v)+c)
\]
is increasing in $\alpha_c$ for all $v \in [v^\star_c,\tilde{v}_c]$, $\kappa_c^5$ defined in \eqref{eq:fixedpoint} is decreasing in $\alpha_c$, which in turn implies that $\kappa_c^4$ is increasing in $\alpha_c$. Therefore, since 
\[
CS(c,h;p^\star)=\int_{F_{c,h}(\kappa_c^4)}^{F_{c,h}(\kappa_c^5)}(F_{c,h}^{-1}(q)-F_{c,l}^{-1}(q-\Delta_c(\kappa_c^4)))\diff q, 
\]
$CS(c,h;p^\star)$ is decreasing in $\alpha_c$. 

For $(ii)$, note that since $\Delta_c(\kappa_c^5)=\Delta_c(\kappa_c^4)$ is increasing in $\alpha_c$ as established above, and since the function 
\[
v \mapsto \Delta_c\left(\frac{\alpha_c}{1-\alpha_c} h_c(v)+c\right)+F_{c,h}(\alpha_c h_c(v)+c)
\]
is increasing on $[v^\star_c,\tilde{v}_c]$, $\kappa_c^3=\alpha_c h_c(v)/(1-\alpha_c)+c$ is also increasing in $\alpha_c$. Together with $\kappa_c^3 \leq v^\star_c$ for all $\alpha_c \in [0,1]$, it follows that $\Delta_c(\kappa_c^3)$ is increasing in $\alpha_c$ as well. In the meantime, since $F_{c,l}(\kappa_c^4)-F_{c,l}(\kappa_c^2)=F_{c,h}(\kappa_c^4)$, it is also increasing in $\alpha_c$. 

Moreover, note that since $\Delta_c$ is increasing on $[0,v^\star_c]$ and since $\kappa_c^4 \leq v^\star_c$, $F_{c,h}(v) \leq F_{c,l}(v)-\Delta_c(\kappa_c^3)$ for all $v \in [\kappa_c^3,\kappa_c^4]$,
\begin{align*}
CS(c,l;p^\star)=\int_{\kappa_c^2}^{\kappa_c^3} (v-F_{c,h}^{-1}(F_{c,l}(v)-\Delta_c(\kappa_c^3)))F_{c,l}(\diff v)=\int_{\kappa_c^2}^{\kappa_c^4} (v-\tau(v))F_{c,l}(\diff v)\,,
\end{align*}
where 
\[
\tau(v):=\min\{v, F_{c,h}^{-1}(F_{c,l}(v)-\Delta_c(\kappa_c^3))\}
\]
for all $v \in [\kappa_c^2,\kappa_c^4]$. Note that $\tau(v)$ is decreasing in $\alpha_c$ for all $v \in [\kappa_c^2,\kappa_c^4]$ since $\Delta_c(\kappa_c^3)$ is increasing in $\alpha_c$. Together, $CS(c,l;p^\star)$ is increasing in $\alpha_c$. 

For $(iii)$, suppose first that $\alpha_c \to 1$. Since $\kappa_c^5$ is decreasing in $\alpha_c$ and is bounded from below by $v^\star$. Thus, the limit of $\kappa_c^5$ exists. As a result, the limits of $\kappa_c^1,\kappa_c^2, \kappa_c^3$ and $\kappa_c^4$ exist as well. 
Moreover, since $\kappa_c^3 \leq v^\star_c$ for all $\alpha_c \in (0,1)$, $\kappa_c^3$ must converge to a finite value as $\alpha_c \to 1$. Thus, since $\kappa_c^3-c=\nicefrac{\kappa_c^1-c}{(1-\alpha_c)}$, for all $\alpha_c \in (0,1)$, $\kappa_c^1$ must converge to $c$ as $\alpha_c \to 1$. This in turn implies that the limits of $\kappa_c^5$ and $\kappa_c^4$ coincide and equal $v_c^\star$. This implies $CS(c,h,p^\star)=0$. 

Now suppose that $\alpha_c \to 0$. Since $\kappa_c^4$ is decreasing as $\alpha_c \to 0$ and is bounded from below by $0$, the limit of $\kappa_c^4$ exists, and hence the limits of $\kappa_c^1,\kappa_c^2,\kappa_c^3$ and $\kappa_c^5$ exist as well. Moreover, since $\kappa_c^1-c=\alpha_c(\kappa_c^5-\kappa_c^4)$, the limit of $\kappa_c^1$ as $\alpha_c \to 0$ must be $c$, which in turn implies that the limit of $\kappa_c^3$ as $\alpha_c \to 0$ equals $c$ as well. Since $\kappa_c^2 \in [\kappa_c^1,\kappa_c^3]$, it then follows that the limit of $\kappa_c^2$ equals $c$ as well. As a result, $CS(c,l;p^\star)=0$. This completes the proof.
\end{proof}

\begin{proof}[Proof of \bref{prop:compare}]
To prove $(i)$, consider any non-discriminatory pricing rule $p$ that induces non-discriminatory outcomes. For each $c \in C$, define a matching scheme $\rho_c \in \Delta(V^2)$ as 
\[
\rho_c(V_l \times V_h):= \EE\left[\PP[v \in V_l\mid p,c,\theta=l,y] \times \PP[v \in V_h \mid p,c,\theta=h,y]\mid c\right]\,,
\]
for all measurable $V_l,V_h \subseteq V$. Since $p$ induces non-discriminatory outcomes, 
\[
\rho_c([0,z] \times V)=\EE[\PP[v \in [0,z] \mid p,\theta=l,c,y]\mid c]=F_{c,l}(z)\,,
\]
and 
\[
\rho_c(V\times [0,z])=\EE[\PP[v \in [0,z]\mid p,\theta=h,c,y]\mid c]=F_{c,h}(z)\,,
\]
for all $z\geq 0$. Therefore, $\rho_c \in \mathcal{R}_c$. 

For each $c$, given such matching scheme $\rho_c$, since $(p,y)$ is independent of $\theta$ conditional on $c$, each matched pair $(v_l,v_h) \in \mathrm{supp}(\rho_c)$ must face the same price under $p$; and either both purchase or both do not purchase. Therefore, the seller's profit under $p$ must be weakly lower than selling to each matched pair $(v_l,v_h) \in \mathrm{supp}(\rho_c)$ at a price $\min\{v_l,v_h\}$ whenever $\min\{v_l,v_h\} \geq c$, and not selling to the pair otherwise. That is, 
\[
\Pi(p) \leq \EE\left[\int_{V^2} (\min\{v_l,v_h\}-c)^+\diff \rho_c\right]\,.
\]
As a result, for any pricing $p$ that induces non-discriminatory outcomes, 
\[
\Pi(p) \leq \EE\left[\max_{\rho_c \in \mathcal{R}_c} \int_{V^2}(\min\{v_l,v_h\}-c)^+\diff \rho_c \right]=:\ub{\pi}\,.
\]
Note that the objective $(\min\{v_l,v_h\}-c)^+$ of the optimal transport problem 
\[
\max_{\rho_c \in \mathcal{R}_c} \int_{V^2} (\min\{v_l,v_h\}-c)^+\diff \rho_c
\]
if supermodular for all $c \in C$, the assortative matching must be a solution. Therefore, 
\[
\max_{\rho_c \in \mathcal{R}_c} \int_{V^2} (\min\{v_l,v_h\}-c)^+\diff \rho_c=\int_V (v-c)^+ F_{c,l}(\diff v)\,.
\]
Thus, by construction, under the pricing rule $p^{ass}$, 
\[
\Pi(p^{ass})=\EE\left[\int_V (v-c)^+ F_{c,l}(\diff v) \right]=\ub{\pi}\,.
\]
Since $p^{ass}$ also induces non-discriminatory outcomes, $p^{ass}$ is optimal. Furthermore, as $p^{ass}$ is also non-discriminatory, $\Pi(p^\star) \geq \Pi(p^{ass})$. Lastly, since the solution of 
\[
\max_{\rho_c \in \mathcal{R}_c} \int_{V \times V} (\min\{v_l,v_h\}-c)^+\rho_c(\diff v_l, \diff v_h)
\]
must correspond to a pricing rule that is outcome-equivalent to the assortative matching for all $c \in C$, any profit-maximizing pricing rule that induces non-discriminatory outcomes must yield the same surplus to consumers. 

For $(iii)$, since $F_{c,h}$ dominates $F_{c,l}$ in the likelihood ratio order, under $p^{ass}$, each matched pair of consumers who are purchasing consumers must buy at the value of the consumer with $\theta=l$, and hence $0=CS(c,l;p^{ass}) \leq CS(c,h;p^\star)$; while the price distribution faced by purchasing consumers with $\theta=h$ equals $(F_{c,l}(\cdot)-F_{c,l}(c))^+/(1-F_{c,l}(c))$, which in turn is dominated by the price distribution faced by purchasing consumers with $\theta=h$ in the sense of first-order-stochastic dominance under $p^\star$. Therefore, $CS(c,h;p^{ass}) \geq CS(c,h;p^\star)$. Moreover, note that as established in the proof of $(ii)$ of \bref{prop:surplus}, $p^\star \neq p^{ass}$ if and only if $\alpha_c(\overline{v}_c-c)>\underline{v}_c-c$ for a positive measure of $c$. Thus, $CS(c,h;p^{ass}) > CS(c,h;p^\star)$ if and only if $\alpha_c(\overline{v}_c-c)>\underline{v}_c-c$ for a positive measure of $c$.

For $(ii)$, since $p^{ass}$ is non-discriminatory, \bref{thm1} implies that $\Pi(p^\star) \geq \Pi(p^{ass})$. Moreover, by $(iii)$, since $CS(c,h;p^{ass})>CS(c,h;p^\star)$ if and only if $\alpha_c(\ub{v}_c-c)\lb{v}_c-c$, and since every undominated profit-maximizing rule $p$ must yield $CS(c,\theta;p)=CS(c,\theta;p^\star)$ for all $c \in C$ and for all $\theta \in \{l,h\}$ by \bref{thm1}, it must be that $p^{ass}$ is not optimal if and only if $\alpha_c(\ub{v}_c-c)>\lb{v}_c-c$ for a positive measure of $c \in C$. Therefore, $\Pi(p^\star)>\Pi(p^{ass})$ if and only if $\alpha_c(\ub{v}_c-c)>\lb{v}_c-c$. This completes the proof.    
\end{proof}

\subsection{Proofs of Omitted Auxiliary Lemmas}\label{sec:a.3}
\begin{proof}[Proof of \bref{lem:feasible}] 
We show that $\phi_c^\star(v_l)+\psi_c^\star(v_h) \geq \pi_c(v_l,v_h)$ for all $(v_l, v_h)$ by discussing all cases separately. Suppose first that $F_{c,l}(c)<\Delta_c(v^\star_c)=\|F_{c,l}-F_{c,h}\|$.  \\
\mbox{}

\noindent \textbf{Case 1: }$v_l \leq \kappa_c^3$.

\noindent When $v_h \leq \kappa_c^1$, $\pi_c(v_l,v_h)$ either equals $\min\{v_l,v_h\}-c \leq v_h-c \leq \kappa_c^1-c$, or $\alpha_c (v_h-c)^+ \leq \kappa_c^1-c$, or $(1-\alpha_c) (v_l-c)^+ \leq (1-\alpha_c) (\kappa_c^3-c)=\kappa_c^1-c$. Therefore, 
\[
\phi_c^\star(v_l)+\psi_c^\star(v_h)=\kappa_c^1-c \geq \pi_c(v_l,v_h)\,.
\]
When $v_h \in (\kappa_c^1,\kappa_c^3]$, $v_h-c>\kappa_c^1-c=(1-\alpha_c) (\kappa_c^3-c) \geq (1-\alpha_c)(v_l-c)^+$ and hence $\min\{v_h,v_l\}-c \geq (1-\alpha_c)(v_l-c)^+$. Therefore, $\pi_c(v_l,v_h)$ either equals $\min\{v_l,v_h\}-c \leq v_h-c$ or $\alpha_c (v_h-c) \leq v_h-c$. As a result, 
\[
\phi_c^\star(v_l)+\psi_c^\star(v_h)=v_h-c \geq \pi_c(v_l,v_h)\,.
\]
When $v_h \in (\kappa_c^3,\kappa_c^4]$, $v_h>v_l$, and therefore $\pi_c(v_l,v_h)=\max\{v_l-c,\alpha_c(v_h-c)\}$. Moreover, since $(1-\alpha_c) (\kappa_c^3-c)=\kappa_c^1-c$, we have $v_l-c\leq \kappa_c^3-c = \alpha_c(\kappa_c^3-c)+\kappa_c^1-c \leq \alpha_c (v_h-c)+\kappa_c^1-c$. Thus, 
\[
\phi_c^\star(v_l)+\psi_c^\star(v_h)=\alpha_c(v_h-c)+\kappa_c^1-c \geq \max\{v_l-c,\alpha_c(v_h-c)\}=\pi_c(v_l,v_h)\,.
\]
When $v_h \in (\kappa_c^4,\kappa_c^5]$, $v_h>v_l$, and therefore $\pi_c(v_l,v_h)=\max\{v_l-c,\alpha_c(v_h-c)\}$. As argued above, we have $v_l -c\leq \kappa_c^3-c = \alpha_c (\kappa_c^3-c)+\kappa_c^1-c \leq \alpha_c (\kappa_c^4-c)+(\kappa_c^1-c)$. Furthermore, by \eqref{eq:system},
\[
\alpha_c(\kappa_c^4-c)+\kappa_c^1-c=\alpha_c (\kappa_c^5-c) \geq \alpha_c (v_h-c)\,.
\]
Together, 
\[
\phi_c^\star(v_l)+\psi_c^\star(v_h)=\alpha_c(\kappa_c^4-c)+\kappa_c^1-c \geq \max\{v_l-c,\alpha_c(v_h-c)\}=\pi_c(v_l,v_h)\,.
\]
When $v_h>\kappa_c^5$, note that since $\kappa_c^1-c=(1-\alpha_c)(\kappa_c^3-c) \leq (1-\alpha_c)(\kappa_c^4-c)$, and since $\alpha_c(\kappa_c^5-\kappa_c^4)=\kappa_c^1-c$, we have that $\alpha_c (v_h-c) \geq \alpha_c(\kappa_c^5-c)=\alpha_c(\kappa_c^4-c)+(1-\alpha_c)(\kappa_c^3-c)\geq \kappa_c^3-c \geq v_l-c$. Therefore, $\pi_c(v_l,v_h)=\alpha_c(v_h-c)$, and hence 
\[
\phi_c^\star(v_l)+\psi_c^\star(v_h)=\alpha_c (v_h-c) =\pi_c(v_l,v_h)\,.
\]
\mbox{}

\noindent \textbf{Case 2:} $v_l \in (\kappa_c^3,\kappa_c^4]$. 

\noindent When $v_h \leq \kappa_c^1$, $(1-\alpha_c)(v_l-c)>(1-\alpha_c)(\kappa_c^3-c)=\kappa_c^1-c \geq v_h-c$, and hence $\pi_c(v_l,v_h)=(1-\alpha_c) (v_l-c)$. Therefore, 
\[
\phi_c^\star(v_l)+\psi_c^\star(v_h)=(1-\alpha_c)(v_l-c) = \pi_c(v_l,v_h)\,.
\]
When $v_h \in (\kappa_c^1,\kappa_c^3]$, $v_h \leq \kappa_c^3<v_l$ and hence $\min\{v_l,v_h\}-c=v_h-c$. Therefore, $\pi_c(v_l,v_h)=\max\{v_h-c,(1-\alpha_c)(v_l-c)\}$. Since $v_h-\kappa_c^1+(1-\alpha_c)(v_l-c)>v_h-\kappa_c^1+(1-\alpha_c)(\kappa_c^3-c)=v_h-c$, we have 
\[
\phi_c^\star(v_l)+\psi_c^\star(v_h)=v_h-\kappa_c^1+(1-\alpha_c)(v_l-c) \geq \max\{v_h-c,(1-\alpha_c)(v_l-c)\}\,.
\]
When $v_h \in (\kappa_c^3,\kappa_c^4]$, 
\begin{align*}
\phi_c^\star(v_l)+\psi_c^\star(v_h)=&\alpha_c(v_h-c)+(1-\alpha_c)(v_l-c) \\
\geq& \max\{\min\{v_l,v_h\}-c,(1-\alpha_c)(v_l-c),\alpha_c(v_h-c)\}\,.
\end{align*}
When $v_h \in (\kappa_c^4,\kappa_c^5]$, since $\kappa_c^1-c=\alpha_c(\kappa_c^5-\kappa_c^4)$ and since $(1-\alpha_c)(\kappa_c^3-c)=\kappa_c^1-c$, we have 
\begin{align*}
\alpha_c(v_h-c) \leq \alpha_c(\kappa_c^5-c) =&\alpha_c(\kappa_c^4-c)+\kappa_c^1-c\\
=&\alpha_c(\kappa_c^4-c)+(1-\alpha_c)(\kappa_c^3-c)\\
\leq& \alpha_c(\kappa_c^4-c)+(1-\alpha_c)(v_l-c) \,.
\end{align*}
Together, 
\[
\phi_c^\star(v_l)+\psi_c^\star(v_h)=\alpha_c(\kappa_c^4-c) + (1-\alpha_c) (v_l-c) \geq \max\{v_l-c,\alpha_c(v_h-c)\}=\pi_c(v_l,v_h)\,.
\]
When $v_h>\kappa_c^5$, since $\kappa_c^1-c=\alpha_c(\kappa_c^5-\kappa_c^4)$, we have 
\[
\alpha_c(v_h-c)-(\kappa_c^1-c) \geq \alpha_c(\kappa_c^5-c)-(\kappa_c^1-c)=\alpha_c(\kappa_c^4-c) \geq \alpha_c(v_l-c)\,.
\]
Therefore, 
\[
\alpha_c(v_h-c)-(\kappa_c^1-c)+(1-\alpha_c) v_l \geq v_l-c\,.
\]
Meanwhile, since $(1-\alpha_c)(\kappa_c^3-c)=\kappa_c^1-c$, 
\[
(1-\alpha_c)(v_l-c) \geq (1-\alpha_c) (\kappa_c^3-c)=\kappa_c^1-c\,,
\]
and thus
\[
\alpha_c (v_h-c)-(\kappa_c^1-c)+(1-\alpha_c)(v_l-c) \geq \alpha_c (v_h-c)\,.
\]
Together, 
\begin{align*}
\phi_c^\star(v_l)+\psi_c^\star(v_h)=&\alpha_c(v_h-c)-(\kappa_c^1-c)+(1-\alpha_c)(v_l-c)\\ 
\geq& \max\{v_l-c,\alpha_c (v_h-c)\}\\
=&\pi_c(v_l,v_h)\,.
\end{align*}
\mbox{}

\noindent \textbf{Case 3:} $v_l \in (\kappa_c^4,\kappa_c^5]$.

\noindent When $v_h \leq \kappa_c^1$, we have
\[
(1-\alpha_c)(v_l-c) > (1-\alpha_c) (\kappa_c^4-c) \geq  (1-\alpha_c)(\kappa_c^3-c) =\kappa_c^1-c \geq (v_h-c)^+\,.
\]
Therefore, $\pi_c(v_l,v_h)=(1-\alpha_c) (v_l-c)$ and hence 
\begin{align*}
\phi_c^\star(v_l)+\psi_c^\star(v_h)=v_l-c-\alpha_c(\kappa_c^4-c)=&\alpha_c(v_l-\kappa_c^4)+(1-\alpha_c)(v_l-c)\\
>&(1-\alpha_c)(v_l-c)\\
=&\pi_c(v_l,v_h)\,.
\end{align*}
When $v_h \in (\kappa_c^1,\kappa_c^3]$, 
\[
v_l-c-\alpha_c(\kappa_c^4-c)+v_h-\kappa_c^1>v_l-c-\alpha_c(\kappa_c^4-c)>(1-\alpha_c)(v_l-c)\,.
\]
Moreover, since $(1-\alpha_c)(\kappa_c^3-c)=\kappa_c^1-c$ and since $\kappa_c^4 \geq \kappa_c^3$, 
\begin{align*}
v_l-c-\alpha_c(\kappa_c^4-c)+v_h-\kappa_c^1=&v_l-c-\alpha_c(\kappa_c^4-c)+v_h-c-(\kappa_c^1-c)\\
> & (1-\alpha_c)(\kappa_c^4-c)+v_h-c-(\kappa_c^1-c)\\
\geq & (1-\alpha_c)(\kappa_c^3-c)+v_h-c-(\kappa_c^1-c)\\
=& v_h-c\,.
\end{align*}
Together, 
\[
\phi_c^\star(v_l)+\psi_c^\star(v_h)=v_l-c-\alpha_c(\kappa_c^4-c)+v_h-\kappa_c^1 \geq \max\{v_h-c,(1-\alpha_c)(v_l-c)\}=\pi_c(v_l,v_h)\,.
\]
When $v_h \in (\kappa_c^3,\kappa_c^4]$, since $v_l>\kappa_c^4 \geq v_h$,
\[
v_l-c-\alpha_c(\kappa_c^4-c)+\alpha_c(v_h-c) > (1-\alpha_c)(\kappa_c^4-c)+\alpha_c(v_h-c) \geq v_h-c
\]
In the meantime, since $v_h>\kappa_c^3 \geq \kappa_c^1 \geq c$, 
\[
v_l-c-\alpha_c(\kappa_c^4-c)+\alpha_c(v_h-c) \geq v_l-c-\alpha_c(\kappa_c^4-c) > (1-\alpha_c)(v_l-c)\,.
\]
Therefore, 
\[
\phi_c^\star(v_l)+\psi_c^\star(v_h)=v_l-c-\alpha_c(\kappa_c^4-c)+\alpha_c(v_h-c) \geq \max\{v_h-c, (1-\alpha_c)(v_l-c)\}=\pi_c(v_l,v_h)\,.
\]
When $v_h \in (\kappa_c^4,\kappa_c^5]$, since $\kappa_c^1-c=\alpha_c(\kappa_c^5-\kappa_c^4)$  and since $\kappa_c^4-c \geq \kappa_c^3-c=\nicefrac{(\kappa_c^1-c)}{(1-\alpha_c)}$, we have 
\[
\alpha_c(\kappa_c^5-c)=\alpha_c(\kappa_c^4-c)+\kappa_c^1-c= \alpha_c (\kappa_c^4-c) + (1-\alpha_c)(\kappa_c^3-c) \leq \kappa_c^4-c\,.
\]
Therefore, 
\[
\alpha_c(v_h-c) \leq \alpha_c(\kappa_c^5-c) \leq \kappa_c^4-c \leq v_l-c\,,
\]
and hence $\pi_c(v_l,v_h)=\max\{\min\{v_l,v_h\}-c,(1-\alpha_c)(v_l-c)\}$. Together,
\[
\phi_c^\star(v_l)+\psi_c^\star(v_h)=v_l-c \geq \max\{\min\{v_l,v_h\}-c, (1-\alpha_c)(v_l-c)\}=\pi_c(v_l,v_h)\,.
\]
When $v_h>\kappa_c^5$, since $\kappa_c^4-c \geq \kappa_c^3-c=\nicefrac{(\kappa_c^1-c)}{(1-\alpha_c)}$, 
\begin{align*}
&\alpha_c (v_h-c)+v_l-c-\alpha_c(\kappa_c^4-c)-(\kappa_c^1-c)\\
\geq &\alpha_c(v_h-c)+(1-\alpha_c)(\kappa_c^4-c)-(\kappa_c^1-c) \\
\geq& \alpha_c(v_h-c) +(1-\alpha_c)(\kappa_c^3-c)-(\kappa_c^1-c)\\
=&\alpha_c (v_h-c)\,. 
\end{align*}
Moreover, since $\kappa_c^1-c=\alpha_c(\kappa_c^5-\kappa_c^4)$, 
\begin{align*}
&\alpha_c(v_h-c)+v_l-c-(\kappa_c^1-c)-\alpha_c(\kappa_c^4-c) \\
> &\alpha_c(\kappa_c^5-c)+v_l-c-(\kappa_c^1-c)-\alpha_c (\kappa_c^4-c) \\
=&\alpha_c(\kappa_c^5-\kappa_c^4)-(\kappa_c^1-c)+v_l-c\\
=&v_l-c\,.
\end{align*}
Together, 
\begin{align*}
\phi_c^\star(v_l)+\psi_c^\star(v_h)=&\alpha_c( v_h-c)-(\kappa_c^1-c)+v_l-c-\alpha_c(\kappa_c^4-c) \\
\geq& \max\{v_l-c,\alpha_c(v_h-c)\}\\
=&\pi_c(v_l,v_h)\,.
\end{align*}
\mbox{}

\noindent \textbf{Case 4: }$v_l>\kappa_c^5$.

\noindent When $v_h \leq \kappa_c^1$, $(v_h-c)^+ \leq \kappa_c^1-c=(1-\alpha_c)(\kappa_c^3-c) \leq (1-\alpha_c) (v_l-c)$. Thus, $\pi_c(v_l,v_h)=(1-\alpha_c)(v_l-c)$, and hence, 
\[
\phi_c^\star(v_l)+\psi_c^\star(v_h)=(1-\alpha_c)(v_l-c)+\kappa_c^1-c\geq (1-\alpha_c)(v_l-c)=\pi_c(v_l,v_h)\,.
\]
When $v_h \in (\kappa_c^1,\kappa_c^3]$, $v_h \leq \kappa_c^3 \leq \kappa_c^4<v_l$. Thus,
\[
\phi_c^\star(v_l)+\psi_c^\star(v_h)=v_h-c+(1-\alpha_c)(v_l-c) \geq \max\{v_h-c,(1-\alpha_c)(v_l-c)\}=\pi_c(v_l,v_h)\,.
\]
When $v_h \in (\kappa_c^3,\kappa_c^4]$, since $\kappa_c^1 \geq c$, 
\begin{align*}
\phi_c^\star(v_l)+\psi_c^\star(v_h)=&\alpha_c(v_h-c)+(1-\alpha_c)(v_l-c)+\kappa_c^1-c \\
\geq&\alpha_c(v_h-c)+(1-\alpha_c)(v_l-c)\\
\geq &\max\{\min\{v_l,v_h\}-c,\alpha_c (v_h-c), (1-\alpha_c)(v_l-c)\}\\
=&\pi_c(v_l,v_h)\,.
\end{align*}
When $v_h \in (\kappa_c^4,\kappa_c^5]$, since $\kappa_c^1-c=\alpha_c(\kappa_c^5-\kappa_c^4)$. 
\begin{align*}
\phi_c^\star(v_l)+\psi_c^\star(v_h)=&\alpha_c(\kappa_c^4-c)+(1-\alpha_c)(v_l-c)+\kappa_c^1-c\\
=&\alpha_c(\kappa_c^5-c)-(\kappa_c^1-c)+(1-\alpha_c)(v_l-c)+\kappa_c^1-c\\
=& \alpha_c(\kappa_c^5-c)+(1-\alpha_c)(v_l-c)\\
\geq& \alpha_c(v_h-c) +(1-\alpha_c)(v_l-c)\\
\geq& \max\{\min\{v_l,v_h\}-c,\alpha_c(v_h-c),(1-\alpha_c)(v_l-c)\}\\
=&\pi_c(v_l,v_h)\,.
\end{align*}
When $v_h>\kappa_c^5$, 
\begin{align*}
\phi_c^\star(v_l)+\psi_c^\star(v_h)
=&\alpha_c(v_h-c)+(1-\alpha_c)(v_l-c) \\
\geq& \max\{\min\{v_l,v_h\}-c,\alpha_c(v_h-c),(1-\alpha_c)(v_l-c)\}\\
=&\pi_c(v_l,v_h)\,.
\end{align*}

Together, it follows that 
\[
\phi_c^\star(v_l)+\psi_c^\star(v_h) \geq \pi_c(v_l,v_h)\,,
\]
for all $v_l,v_h \geq 0$, as desired.

Now suppose that $F_{c,l}(c) \geq \Delta_c(v^\star_c)$. Then clearly. 
\begin{align*}
\phi^\star_c(v_l)+\psi_c^\star(v_h)=&(1-\alpha_c) (v_l-c)^++\alpha_c (v_h-c)^+\\ \geq & \max\{(1-\alpha_c) (v_l-c)^+, \alpha_c (v_h-c)^+, \min\{v_l,v_h\}-c\}\\
=&\pi_c(v_l,v_h)\,,
\end{align*}
for all $v_l,v_h \in V$. This completes the proof. 
\end{proof}

\begin{proof}[Proof of \bref{lem:rfeas}]
We first note that for all $c \in C$,  $\gamma_c^\star(\cdot\mid v_h)$ is a probability measure for all $v_h \in V$. Indeed, for all $v_h \in V$ $\lim_{x \to \infty} \gamma^\star_c(v_l \leq x\mid v_h)=1$ and $\gamma^\star_c(v_l \leq 0 \mid v_h)=0$; $x \mapsto \gamma^\star(v_l \leq x\mid v_h)$ is right-continuous. Moreover, for any $c \in C$ and for any measurable set $A \subseteq V$, $\gamma_c^\star(A\mid \cdot)$ is a measurable function. Therefore, $\gamma_c^\star$ is a transition probability for all $c \in C$. 

Next, we show that the marginals of $\rho_c^\star$ equal $F_{c,l}$ and $F_{c,h}$, respectively. By construction, 
\[
\rho^\star_c(v_l \in V, v_h \leq x)=\int_0^x 1 F_{c,h}(\diff v_h)=F_{c,h}(x)\,.
\]
To show that $\rho^\star_c(v_l \leq x, v_h \in V)=F_{c,l}(x)$ for all $c \in C$ and for all $x \in V$, consider first the case when $F_{c,l}(c)<\Delta_c(v^\star_c)$. For all $x \leq \kappa_c^2$, 
\[
\gamma_c^\star(v_l \leq x \mid v_h)=
\begin{cases}
0,&\mbox{if } v_h \leq \kappa_c^5\\
\frac{f_{c,h}(v_h)-f_{c,l}(v_h)}{f_{c,h}(v_h)} \mathbf{1}\{F_{c,l}^{-1}(\Delta_c(\kappa_c^5)-\Delta_c(v_h)) \leq x\},&\mbox{if } v_h>\kappa_c^5
\end{cases}\,.
\]
Note that the derivative of $v \mapsto \Delta_c(\kappa_c^5)-\Delta_c(v)$ equals $f_{c,h}-f_{c,l}$. Therefore,
\begin{align*}
\int_0^\infty \gamma_c^\star(v_l \leq x\mid v_h)F_{c,h}(\diff v_h)=& \int_{\kappa_c^5}^\infty \frac{f_{c,h}(v_h)-f_0(v_h)}{f_{c,h}(v_h)} \mathbf{1}\{F_{c,l}^{-1}(\Delta_c(\kappa_c^5)-\Delta_c(v_h)) \leq x\}F_{c,h}(\diff v_h)\\
=& \int_{\kappa_c^5}^\infty (f_{c,h}(v_h)-f_{c,l}(v_h)) \mathbf{1}\{\Delta_c(\kappa_c^5)-\Delta_c(v_h) \leq F_{c,l}(x)\}\diff v_h\\
=& \int_{0}^{\Delta_c(\kappa_c^5)} \mathbf{1}\{z \leq F_{c,l}(x)\}\diff z\\
=& \min\{\Delta_c(\kappa_c^5),F_{c,l}(x)\}\\
=& F_{c,l}(x)\,,
\end{align*}
where the third equality follows from changing variables of integration, and the last inequality follows from $F_{c,l}(x) \leq F_{c,l}(\kappa_c^2)=\Delta_c(\kappa_c^5)$, which in turn follows from \eqref{eq:system}. 

For all $x \in (\kappa_c^2,\kappa_c^3]$, 
\[
\gamma_c^\star(v_l \leq x\mid v_h)=
\begin{cases}
0,&\mbox{if } v_h \in [0,\kappa_c^1] \cup (\kappa_c^3,\kappa_c^5]\\
\mathbf{1}\{F_{c,l}^{-1}(F_{c,h}(v_h)+\Delta_c(\kappa_c^3)) \leq x\},&\mbox{if } v_h \in (\kappa_c^1,\kappa_c^3]\\
\frac{f_{c,h}(v_h)-f_{c,l}(v_h)}{f_{c,h}(v_h)},&\mbox{if } v_h >\kappa_c^5
\end{cases}\,.
\]
Therefore, 
\begin{align*}
&\int_0^\infty \gamma_c^\star(v_l \leq x\mid v_h)F_{c,h}(\diff v_h)\\=& \int_{\kappa_c^1}^{\kappa_c^3} \mathbf{1}\{F_{c,l}^{-1}(F_{c,h}(v_h)+\Delta_c(\kappa_c^3)) \leq x\} F_{c,h}(\diff v_h)+\int_{\kappa_c^5}^\infty \frac{f_{c,h}(v_h)-f_{c,l}(v_h)}{f_{c,h}(v_h)} F_{c,h}(\diff v_h)\\
=&\int_{\kappa_c^1}^{\kappa_c^3} \mathbf{1}\{F_{c,h}(v_h)\leq F_{c,l}(x)-\Delta_c(\kappa_c^3)\}F_{c,h}(\diff v_h)+F_{c,l}(\kappa_c^5)-F_{c,h}(\kappa_c^5)\\
=& \int_{F_{c,h}(\kappa_c^1)}^{F_{c,h}(\kappa_c^3)} \mathbf{1}\{z \leq F_{c,l}(x)-\Delta_c(\kappa_c^3)\} \diff z+\Delta_c(\kappa_c^5)\\
=& \min \{F_{c,l}(x)-\Delta_c(\kappa_c^3),F_{c,h}(\kappa_c^3)\}-F_{c,h}(\kappa_c^1)+\Delta_c(\kappa_c^5)\\
=&F_{c,l}(x)-\Delta_c(\kappa_c^3)-F_{c,h}(\kappa_c^1)+\Delta_c(\kappa_c^5)\\
=& F_{c,l}(x)\,,
\end{align*}
where the third equality follows from changing variables for integration, the fifth equality follows from $F_{c,l}(x)-\Delta_c(\kappa_c^3) \leq F_{c,l}(\kappa_c^3)-\Delta_c(\kappa_c^3)=F_{c,h}(\kappa_c^3)$, and the last equality also follows from \eqref{eq:system}.

For all $x \in (\kappa_c^3,\kappa_c^4]$,
\[
\gamma_c^\star(v_l \leq x\mid v_h)=
\begin{cases}
\mathbf{1}\{\lb{\Delta}_c^{-1}(F_{c,h}(v_h)+\Delta_c(\kappa_c^3)) \leq x\},& \mbox{if } v_h \leq \kappa_c^1\\
1,&\mbox{if } v_h \in (\kappa_c^1,\kappa_c^3]\\
\mathbf{1}\{v_h \leq x\},&\mbox{if } v_h \in (\kappa_c^3,\kappa_c^4]\\
0,&\mbox{if } v_h \in (\kappa_c^4,\kappa_c^5]\\
\frac{f_{c,h}(v_h)-f_{c,l}(v_h)}{f_{c,h}(v_h)},&\mbox{if } v_h >\kappa_c^5
\end{cases}\,.
\]
Thus, 
\begin{align*}
&\int_0^\infty \gamma_c^\star(v_l \leq x\mid v_h)F_{c,h}(\diff v_h)\\
=&\int_0^{\kappa_c^1}\mathbf{1}\{\lb{\Delta}^{-1}_c(F_{c,h}(v_h)+\Delta_c(\kappa_c^3)) \leq x\}F_{c,h}(\diff v_h)+F_{c,h}(\kappa_c^3)-F_{c,h}(\kappa_c^1)\\
&+\int_{\kappa_c^3}^{\kappa_c^4} \mathbf{1}\{v_h \leq x\}F_{c,h}(\diff v_h)+\int_{\kappa_c^5}^\infty \frac{f_{c,h}(v_h)-f_{c,l}(v_h)}{f_{c,h}(v_h)}F_{c,h}(\diff v_h)\\
=& \int_0^{\kappa_c^1}\mathbf{1}\{F_{c,h}(v_h)+\Delta_c(\kappa_c^3) \leq \Delta_c(x)\}F_{c,h}(\diff v_h)\\
&+F_{c,h}(\kappa_c^3)-F_{c,h}(\kappa_c^1)+F_{c,h}(x)-F_{c,h}(\kappa_c^3)+F_{c,l}(\kappa_c^5)-F_{c,h}(\kappa_c^5)\\
=&\int_{0}^{F_{c,h}(\kappa_c^1)}\mathbf{1}\{z \leq \Delta_c(x)-\Delta_c(\kappa_c^3)\}\diff z+F_{c,h}(x)-F_{c,h}(\kappa_c^1)+\Delta_c(\kappa_c^5)\\
=& \min\{\Delta_c(x)-\Delta_c(\kappa_c^3),F_{c,h}(\kappa_c^1)\}+F_{c,h}(x)-F_{c,h}(\kappa_c^1)+\Delta_c(\kappa_c^5)\\
=& \Delta_c(x)-\Delta_c(\kappa_c^3)+F_{c,h}(x)-F_{c,h}(\kappa_c^1)+\Delta_c(\kappa_c^5)\\
=&F_{c,l}(x)-(\Delta_c(\kappa_c^3)+F_{c,h}(\kappa_c^1))+\Delta_c(\kappa_c^5)\,,
\end{align*}
where the third equality follows from changing variables of the integration, the fifth equality follows from $\Delta_c(x)-\Delta_c(\kappa_c^3) \leq \Delta_c(\kappa_c^4)-\Delta_c(\kappa_c^3) = F_{c,h}(\kappa_c^1)$, which in turn follows from \eqref{eq:system}; whereas the last equality also follows from \eqref{eq:system}. 

For all $x \in (\kappa_c^4,\kappa_c^5]$, 
\[
\gamma_c^\star(v_l \leq x \mid v_h)=
\begin{cases}
1,&\mbox{if } v_h \leq \kappa_c^4\\
\mathbf{1}\{F_{c,l}^{-1}(F_{c,h}(v_h)+\Delta_c(\kappa_c^4)) \leq x\},&\mbox{if } v_h \in (\kappa_c^4,\kappa_c^5]\\
\frac{f_{c,h}(v_h)-f_{c,l}(v_h)}{f_{c,h}(v_h)},&\mbox{if } v_h>\kappa_c^5
\end{cases}\,.
\]
Thus, 
\begin{align*}
&\int_0^\infty \gamma_c^\star(v_l \leq x\mid v_h) F_{c,h}(\diff v_h)\\
=& \int_0^{\kappa_c^4} 1 F_{c,h}(\diff v)+\int_{\kappa_c^4}^{\kappa_c^5} \mathbf{1}\{F_{c,l}^{-1}(F_{c,h}(v_h)+\Delta_c(\kappa_c^4)) \leq x\}F_{c,h}(\diff v_h)
+\int_{\kappa_c^5}^{\infty} \frac{f_{c,h}(v_h)-f_{c,l}(v_h)}{f_{c,h}(v_h)} F_{c,h}(\diff v_h)\\
=&F_{c,h}(\kappa_c^4)+\int_{\kappa_c^4}^{\kappa_c^5}\mathbf{1}\{F_{c,h}(v_h) \leq F_{c,l}(x)-\Delta_c(\kappa_c^4)\}F_{c,h}(\diff v_h)
+\int_{\kappa_c^5}^{\infty} [f_{c,h}(v_h)-f_{c,l}(v_h)] \diff v_h\\
=& F_{c,h}(\kappa_c^4)+\min\{F_{c,l}(x)-\Delta_c(\kappa_c^4),F_{c,h}(\kappa_c^5)\}-F_{c,h}(\kappa_c^4)+\Delta_c(\kappa_c^5)\\
=&F_{c,h}(\kappa_c^4)+F_{c,l}(x)-\Delta_c(\kappa_c^4)-F_{c,h}(\kappa_c^4)+\Delta_c(\kappa_c^5)\\
=&F_{c,l}(x)\,,
\end{align*}
where the third equality follows from changing variables of  the integration, the fourth equality follows from $F_{c,l}(x)-\Delta_c(\kappa_c^4) \leq F_{c,l}(\kappa_c^5)-\Delta_c(\kappa_c^4)=F_{c,l}(\kappa_c^5)-\Delta_c(\kappa_c^5)=F_{c,h}(\kappa_c^5)$, which in turn follows from \eqref{eq:system}; and the last equality follows from \eqref{eq:system}.

For all $x> \kappa_c^5$, 
\[
\begin{cases}
1,&\mbox{if } v_h \leq \kappa_c^5\\
\frac{f_{c,l}(v_h)}{f_{c,h}(v_h)}\mathbf{1}\{v_h \leq x\}+\frac{f_{c,h}(v_h)-f_{c,l}(v_h)}{f_{c,h}(v_h)},&\mbox{if } v_h>\kappa_c^5
\end{cases}\,.
\]
Therefore, 
\begin{align*}
&\int_0^\infty \gamma_c^\star (v_l \leq x\mid v_h) F_{c,h}(\diff v_h)\\
=& \int_0^{\kappa_c^5} 1 F_{c,h}(\diff v_h)+\int_{\kappa_c^5}^\infty \left(\frac{f_{c,l}(v_h)}{f_{c,h}(v_h)}\mathbf{1}\{v_h \leq x\}+\frac{f_{c,h}(v_h)-f_{c,l}(v_h)}{f_{c,h}(v_h)}\right) F_{c,h}(\diff v_h)\\
=& F_{c,h}(\kappa_c^5)+\int_{\kappa_c^5}^x f_{c,l}(v_h)\diff v_h+\int_{\kappa_c^5}^\infty [f_{c,h}(v_h)-f_{c,l}(v_h)]\diff v_h\\
=& F_{c,h}(\kappa_c^5)+F_{c,l}(x)-F_{c,l}(\kappa_c^5)+F_{c,l}(\kappa_c^5)-F_{c,h}(\kappa_c^5)\\
=& F_{c,l}(x)\,.
\end{align*}
Together, we have that 
\[
\rho_c^\star(v_l \leq x,v_h \in V)=\int_0^\infty \gamma^\star(v_l \leq x\mid v_h)F_1(\diff v_h)=F_{c,l}(x)\,,
\]
for all $x \in V$ and and hence $\rho_c^\star \in \mathcal{R}_c$ for all $c \in C$ such that $F_{l,c}(c)<\|F_{c,l}-F_{c,h}\|$.

Now consider the case when $F_{c,l}(c) \geq \|F_{c,l}-F_{c,h}\|$ and $c \leq v^\star_c$. If $x \leq \eta_c^l$, then 
\[
\gamma^\star_c(v_l\leq x \mid v_h)=\begin{cases}
0,&\mbox{if } v_h>\eta_c^h\\
\mathbf{1}\{F_{c,l}^{-1}(F_{c,h}(v_h))\leq x\},&\mbox{if } v_h \leq \eta_c^h
\end{cases}\,.
\]
Thus, 
\begin{align*}
\int_0^\infty \gamma_c^\star (v_l \leq x \mid v_h)=&\int_0^{\eta_c^h} \mathbf{1}\{F_{c,l}^{-1}(F_{c,h}(v_h))\leq x\}F_{c,h}(\diff v_h)\\
=& \int_{0}^{\eta_c^1} \mathbf{1}\{F_{c,h}(v_h) \leq F_{c,l}(x)\}F_{c,h}(\diff v_h)\\
=& \int_0^{F_{c,h}(\eta_c^h)} \mathbf{1}\{z \leq F_{c,l}(x)\}\diff z\\
=& \min\{F_{c,h}(\eta_c^h), F_{c,l}(x)\}\\
=& F_{c,l}(x)\,,
\end{align*}
where the third equality follows from changing variables of the integration, and the last equality follows from $F_{c,l}(x) \leq F_{c,l}(\eta_c^l)=F_{c,h}(\eta_c^h)$. 

If $x \in (\eta_c^l,c]$, then 
\[
\gamma_c^\star(v_l\leq x\mid v_h)=\begin{cases}
1,&\mbox{if } v_h \leq \hat{v}_c^1\\
\frac{f_{c,h}(v_h)-f_{c,l}(v_h)}{f_{c,h}(v_h)} \mathbf{1}\{F_{c,l}^{-1}(\Delta_c(v^\star_c)-\Delta_c(v_h)+F_{c,l}(\eta_c^l))\leq x\},&\mbox{if } v_h>v^\star_c\\
0,&\mbox{otherwise}
\end{cases}\,.
\]
Thus, 
\begin{align*}
&\int_0^\infty \gamma_c^\star(v_l\leq x\mid v_h)F_{c,h}(\diff v_h)\\
=& F_{c,h}(\eta_c^h) +\int_{v^\star_c}^\infty \frac{f_{c,h}(v_h)-f_{c,l}(v_h)}{f_{c,h}(v_h)} \mathbf{1}\{F_{c,l}^{-1}(\Delta_c(v^\star_c)-\Delta_c(v_h)+F_{c,l}(\eta_c^l)\leq x\} F_{c,h}(\diff v_h)\\
=& F_{c,h}(\eta_c^h)+\int_{v^\star_c}^\infty (f_{c,h}(v_h)-f_{c,l}(v_h)) \mathbf{1}\{\Delta_c(v^\star_c)-\Delta_c(v_h) \leq F_{c,l}(x)-F_{c,l}(\eta_c^l)\}\diff v_h\\
=& F_{c,h}(\eta_c^h)+\int_{0}^{\Delta_c(v^\star_c)} \mathbf{1}\{z \leq F_{c,l}(x)-F_{c,l}(\eta_c^l)\}\\
=& F_{c,h}(\eta_c^h)+\min\{\Delta_c(v^\star_c),F_{c,l}(x)-F_{c,l}(\eta_c^l)\}\\
=&F_{c,h}(\eta_c^h)+F_{c,l}(x)-F_{c,l}(\eta_c^l)\\
=& F_{c,l}(x)\,,
\end{align*}
where the third equality follows from changing variables of the integration, the fifth equality follows from $F_{c,l}(x)-F_{c,l}(\eta_c^l) \leq F_{c,l}(c)-F_{c,l}(\eta_c^l)=\Delta_c(v^\star_c)$, which in turn follows from the definition of $\eta_c^l$; and the last equality follows from $F_{c,h}(\eta_c^h)=F_{c,l}(\eta_c^l)$. 

If $x \in (c,v^\star_c]$, then 
\[
\gamma^\star_c(v_l \leq x\mid v_h)=\begin{cases}
1,&\mbox{if } v_h \leq \hat{v}_c^1\\
\mathbf{1}\{\lb{\Delta}_c^{-1}(F_{c,h}(v_h)-F_{c,h}(\eta_c^h)+\Delta_c(c))\leq x\},&\mbox{if } v_h \in (\eta_c^h,c]\\
\mathbf{1}\{v_h \leq x\},&\mbox{if } v_h \in (c,v^\star_c]\\
\frac{f_{c,h}(v_h)-f_{c,l}(v_h)}{f_{c,h}(v_h)},&\mbox{if } v_h>v^\star_c
\end{cases}\,.
\]
Thus, 
\begin{align*}
&\int_0^\infty \gamma^\star_c(v_l \leq x\mid v_h)F_{c,h}(\diff v_h)\\
=&\int_0^{\eta_c^h}1F_{c,h}(\diff v_h) + \int_{\eta_c^h} \mathbf{1}\{\lb{\Delta}_c^{-1}(F_{c,h}(v_h)-F_{c,h}(\eta_c^h)+\Delta_c(c))\leq x\}F_{c,h}(\diff v_h)\\
&+ \int_c^{v^\star_c} \mathbf{1}\{v_h \leq x\}F_{c,h}(\diff v_h)+\int_{v^\star_c}^\infty \frac{f_{c,h}(v_h)-f_{c,l}(v_h)}{f_{c,h}(v_h)} F_{c,h}(\diff v_h)\\
=& F_{c,h}(\eta_c^h) + \int_{\eta_c^h}^c \mathbf{1}\{F_{c,h}(v_h)-F_{c,h}(\eta_c^h) \leq \Delta_c(x)-\Delta_c(c))\}F_{c,h}(\diff v_h)+F_{c,h}(x)-F_{c,h}(c)+ \Delta_c(v^\star_c)\\
=& F_{c,h}(\eta_c^h)+\int_{0}^{F_{c,h}(c)-F_{c,h}(\eta_c^h)} \mathbf{1}\{z \leq \Delta_c(x)-\Delta_c(c)\}\diff z+F_{c,h}(x)-F_{c,h}(c)+\Delta_c(v_c^\star)\\
=& F_{c,h}(\eta_c^h)+\min\{\Delta_c(x)-\Delta_c(c), F_{c,h}(c)-F_{c,h}(\eta_c^h)\}+F_{c,h}(x)-F_{c,h}(c)+\Delta_c(v^\star_c)\\
=& F_{c,h}(\eta_c^h)+ \Delta_c(x)-\Delta_c(c)+F_{c,h}(x)-F_{c,h}(c)+\Delta_c(v^\star_c)\\
=& F_{c,h}(\eta_c^h)+F_{c,l}(x)+\Delta_c(v^\star_c)-F_{c,l}(c)\\
=& F_{c,h}(\eta_c^h)+F_{c,l}(x)-F_{c,l}(\eta_c^l)\\
=&F_{c,l}(x)\,,
\end{align*}
where the third equality follows from changing variables of the integration, the fifth equality follow from $\Delta_c(x)-\Delta_c(c) \leq \Delta_c(v^\star_c)-\Delta_c(c)=F_{c,h}(c)-F_{c,h}(\eta_c^h)$, which in turn follows from the definition of $\eta_c^h$; and the last equality also follows from the definition of $\eta_c^h$ and $\eta_c^l$. 

If $x>v^\star_c$, then 
\[
\gamma_c^\star(v_l \leq x\mid v_h)=\begin{cases}
1,&\mbox{if } v_h \leq v^\star_c\\
\frac{f_{c,l}(v_h)}{f_{c,h}(v_h)}\mathbf{1}\{v_h \leq x\}+\frac{f_{c,h}(v_h)-f_{c,l}(v_h)}{f_{c,h}(v_h)},&\mbox{if } v_h>v_c^\star. 
\end{cases}\,.
\]
Thus, 
\begin{align*}
&\int_0^\infty \gamma_c^\star (v_l \leq x\mid v_h) F_{c,h}(\diff v_h)\\
=& \int_0^{v^\star_c} 1 F_{c,h}(\diff v_h)+\int_{v^\star_c}^\infty \left(\frac{f_{c,l}(v_h)}{f_{c,h}(v_h)}\mathbf{1}\{v_h \leq x\}+\frac{f_{c,h}(v_h)-f_{c,l}(v_h)}{f_{c,h}(v_h)}\right) F_{c,h}(\diff v_h)\\
=& F_{c,h}(v^\star_c)+\int_{v^\star_c}^x f_{c,l}(v_h)\diff v_h+\int_{v^\star_c}^\infty [f_{c,h}(v_h)-f_{c,l}(v_h)]\diff v_h\\
=& F_{c,h}(v^\star_c)+F_{c,l}(x)-F_{c,l}(v^\star_c)+F_{c,l}(v^\star_c)-F_{c,h}(v^\star_c)\\
=& F_{c,l}(x)\,.
\end{align*}
Together, whenever $F_{c,l}(c) \geq \|F_{c,l}-F_{c,h}\|$ and $c \leq v_c^\star$, 
\[
\rho^\star_c(v_l \leq x, v_h \in V)=\int_0^\infty \gamma_c^\star(v_l \leq x\mid v_h)F_{c,h}(\diff v_h)=F_{c,l}(x)\,,
\]
for all $x \in V$ and hence $\rho^\star_c \in \mathcal{R}_c$. 

Lastly, consider the case when $F_{c,l}(c) \geq \|F_{c,l}-F_{c,h}\|$ and $c>v^\star_c$. If $x \leq F_{c,l}^{-1}(F_{c,h}(c))$, then 
\[
\gamma^\star_c(v_l \leq x \mid v_h)=\begin{cases}
\mathbf{1}\{F_{c,l}^{-1}(F_{c,h}(v_h)) \leq x\},&\mbox{if } v_h \leq c\\
0,&\mbox{if } v_h>c
\end{cases}\,.
\]
Thus, 
\begin{align*}
\int_0^\infty \gamma_c^\star(v_l \leq x\mid v_h)F_{c,h}(\diff v_h)=& \int_0^c \mathbf{1}\{F_{c,l}^{-1}(F_{c,h}(v_h))\leq x\}F_{c,h}(\diff v_h)\\
=& \int_0^{c} \mathbf{1}\{F_{c,h}(v_h) \leq F_{c,l}(x)\}\diff z\\
=& \int_0^{F_{c,h}(c)} \mathbf{1}\{z \leq F_{c,l}(x)\}\diff z\\
=&\min\{F_{c,h}(c),F_{c,l}(x)\}\\
=& F_{c,l}(x)\,,
\end{align*}
where the third equality follows from changing variables of the integration. 

If $x>F_{c,l}^{-1}(F_{c,h}(c))$, then 
\[
\gamma_c^\star(v_l \leq x \mid v_h)=\begin{cases}
1,&\mbox{if } v_h \leq c\\
\frac{f_{c,l}(v_h)}{f_{c,h}(v_h)}\mathbf{1}\{v_h \leq x\}+\frac{f_{c,h}(v_h)-f_{c,l}(v_h)}{f_{c,h}(v_h)},&\mbox{if } v_h > c\\
\end{cases}\,.
\]
Thus, 
\begin{align*}
&\int_0^\infty \gamma_c^\star(v_l \leq x\mid v_h) F_{c,h}(\diff v_h)\\
=&\int_0^c F_{c,h}(\diff v_h)+\int_c^\infty \frac{f_{c,l}(v_h)}{f_{c,h}(v_h)}\mathbf{1}\{v_h \leq x\}F_{c,h}(\diff v_h)+\int_c^\infty \frac{f_{c,l}(v_h)-f_{c,h}(v_h)}{f_{c,h}(v_h)}F_{c,h}(\diff v_h)\\
=&F_{c,h}(c)+\int_{c}^\infty \mathbf{1}\{v_h \leq x\} f_{c,l}(v_h) \diff v_h+\int_c^\infty (f_{c,h}(v_h)-f_{c,l}(v_l))\diff v_h\\
=&F_{c,h}(c)+F_{c,l}(x)-F_{c,l}(c)+F_{c,l}(c)-F_{c,h}(c)\\
=&F_{c,l}(x)\,.
\end{align*}

Together, whenever $F_{c,l}(c) \geq \|F_{c,l}-F_{c,h}\|$ and $c > v^\star_c$, 
\[
\rho^\star_c(v_l \leq x, v_h \in V)=\int_0^\infty \gamma_c^\star(v_l \leq x\mid v_h)F_{c,h}(\diff v_h)=F_{c,l}(x)\,,
\]
for all $x \in V$, and hence $\rho^\star_c \in \mathcal{R}_c$. This completes the proof. 
\end{proof}

\begin{proof}[Proof of \bref{lem:complementary_slackness}]
Consider first the case when $F_{c,l}(c)<\|F_{c,l}-F_{c,h}\|$. By construction,
\begin{align*}
\supp(\rho_c^\star)\subseteq &[0,\kappa_c^2] \times [\kappa_c^5,\infty) \cup [\kappa_c^3,\kappa_c^4] \times [0,\kappa_c^1] \\
\cup&\{(v_l,v_h): v_l=F_{c,l}^{-1}(F_{c,h}(v_h)+\Delta_c(\kappa_c^3))\}\\
\cup& \{(v_l,v_h): v_l=v_h\,, v_l,v_h \in [\kappa_c^3,\kappa_c^4] \cup [\kappa_c^5,\infty)\}\\
\cup& \{(v_l,v_h): v_l=F_{c,l}^{-1}(F_{c,h}(v_h)+\Delta_c(\kappa_c^4))\,, v_l,v_h \in [\kappa_c^4,\kappa_c^5]\}\,.
\end{align*}
For all $(v_l,v_h) \in \supp(\rho_c^\star) \cap [0,\kappa_c^2] \times [\kappa_c^5,\infty)$, since 
\begin{align*}
\alpha_c(v_h-c) \geq \alpha_c(\kappa_c^5-c) =& \alpha_c (\kappa_c^4-c)+\kappa_c^1-c\\
=& \alpha_c(\kappa_c^4-c)+(1-\alpha_c)(\kappa_c^3-c)\\
\geq& \alpha_c(\kappa_c^3-c)+(1-\alpha_c)(\kappa_c^3-c)\\
=&\kappa_c^3-c \\
\geq& \kappa_c^2-c\\
\geq & v_l-c\,,
\end{align*}
$\Pi_c(v_l,v_h)=\alpha_c(v_h-c)$. Thus, 
\[
\phi_c^\star(v_l)+\psi_c^\star(v_h)=\alpha_c (v_h-c)=\Pi_c(v_l,v_h)\,.
\]

For all $(v_l,v_h) \in \supp(\rho^\star_c) \cap [\kappa_c^2,\kappa_c^3] \times [\kappa_c^1,\kappa_c^3]$, it must be that $v_l=F_{c,l}^{-1}(F_{c,h}(v_h)+\Delta_c(\kappa_c^3))$. Moreover, since $v \mapsto \Delta_c(v)$ is increasing on $[0,v_c^\star]$, $\Delta_c(v) \leq \Delta_c(\kappa_c^3)$. Therefore, 
\[
F_{c,l}(v) \leq F_{c,h}(v)+\Delta_c(\kappa_c^3)\,,
\]
for all $v \in [\kappa_c^1,\kappa_c^3]$, and hence 
\[
v \leq F_{c,l}^{-1}(F_{c,h}(v)+\Delta_c(\kappa_c^3))
\]
for all $v \in [\kappa_c^1,\kappa_c^3]$

Therefore, for all $(v_l,v_h) \in \supp(\rho^\star_c) \cap [\kappa_c^2,\kappa_c^3] \times [\kappa_c^1,\kappa_c^3]$, it must be that 
\[
v_l=F_{c,l}^{-1}(F_{c,h}(v_h)+\Delta_c(\kappa_c^3)) \geq v_h\,,
\]
Together with the fact that 
\[
(1-\alpha_c) (v_l-c) \leq (1-\alpha_c)(\kappa_c^3-c) = \kappa_c^1 -c  \leq v_h-c\,,
\]
which follows from \eqref{eq:system}, it must be that 
\[
\phi_c^\star(v_l)+\psi_c^\star(v_h)=v_h-c=\Pi_c(v_l,v_h)\,.
\]

For all $(v_l,v_h) \in \supp(\rho^\star) \cap [\kappa_c^3,\kappa_c^4] \times [0,\kappa_c^1]$, 
\[
v_h-c \leq \kappa_c^1-c=(1-\alpha_c)(\kappa_c^3-c) \leq (1-\alpha_c)(v_l-c)\,.
\]
Therefore, $\Pi_c(v_l,v_h)=(1-\alpha_c)(v_l-c)$. As a result, 
\[
\phi_c^\star(v_l)+\psi_c^\star(v_h)=(1-\alpha_c) (v_l-c)=\Pi_c(v_l,v_h)\,.
\]

For all $(v_l,v_h) \in \supp(\rho_c^\star) \cap \{(v_l,v_h): v_l=v_h\,, v_l,v_h \in [\kappa_c^3,\kappa_c^4] \cup [\kappa_c^5,\infty)\}$, we have $\Pi_c(v_l,v_h)=v_l=v_h$. Therefore, 
\[
\phi^\star(v_l)+\psi^\star(v_h)=(1-\alpha_c)(v_l-c)+\alpha_c(v_h-c)=v_h-c=v_l-c=\Pi_c(v_l,v_h)\,. 
\]

For all $(v_l,v_h) \in \supp(\rho_c^\star) \cap \{(v_l,v_h): v_l=F_{c,l}^{-1}(F_{c,h}(v_h)+\Delta_c(\kappa_c^4))\,, v_l,v_h \in [\kappa_c^4,\kappa_c^5]\}$, it must be that $v_l \leq v_h$. Indeed, since $\Delta_c$ is quasi-concave, $\Delta_c(v) \geq \Delta_c(\kappa_c^4)=\Delta_c(\kappa_c^5)$ for all $v \in [\kappa_c^4,\kappa_c^5]$. As a result, 
\[
F_{c,h}(v)+\Delta_c(\kappa_c^4) \leq F_{c,l}(v)\,,
\]
and hence 
\[
v_l =F_{c,l}^{-1}(F_{c,h}(v_h)+\Delta_c(\kappa_c^4))\leq v_h\,.
\]
Therefore, $\Pi_c(v_l,v_h)=v_l-c$, and hence
\[
\phi^\star(v_l)+\psi^\star(v_h)=v_l-c=\Pi_c(v_l,v_h)\,.
\]

Together, it follows that 
\[
\phi^\star(v_l)+\psi^\star(v_h)=\Pi(v_l,v_h)
\]
for all $(v_l,v_h) \in \supp(\rho^\star)$, as desired. 

Now consider the case when $F_{c,l}(c) \geq \|F_{c,l}-F_{c,h}\|$. Note that for all $(v_l,v_h) \in \supp(\rho_c^\star)$, either $v_l=v_h$ or $\min\{v_l,v_h\}\leq c$. In both cases, we have 
\[
\Pi_c(v_l,v_h)=\alpha_c(v_h-c)^++(1-\alpha_c)(v_l-c)^+=\phi_c^\star(v_l)+\psi_v^\star(v_h)\,,
\]
a desired. 
\end{proof}

\subsection{Proofs for the Extensions}
\begin{proof}[Proof of \bref{prop:extension1}]
Since $c=0$, we slightly abuse the notation  and suppress the subscript $c$. Consider the following pair of functions $\tilde{\phi}$ and $\tilde{\psi}$: 
\[
\tilde{\phi}(v_l):=
\begin{cases}
\frac{\tilde{\kappa}^2\tilde{\kappa}^1}{\tilde{\kappa}^2+\tilde{\kappa}^1},&\mbox{if } v_l \leq \tilde{\kappa}^2\\
\frac{\tilde{\kappa}^1\tilde{\kappa}^2}{\tilde{\kappa}^1+\tilde{\kappa}^2} + \int_{\tilde{\kappa}^2}^{\tilde{\kappa}^3}\left(\frac{\lb{\beta}(z)}{z+\lb{\beta}(z)}\right)^2 \diff z,&\mbox{if } v_l \in (\tilde{\kappa}^2,\tilde{\kappa}^3]\\
\frac{v_l}{4},&\mbox{if } v_l \in (\tilde{\kappa}^3,\tilde{\kappa}^4]\\
\frac{\tilde{\kappa}^4}{4}+\int_{\tilde{\kappa}^4}^{\tilde{\kappa}^5}\left(\frac{\ub{\beta}(z)}{z+\ub{\beta}(z)}\right)^2 \diff z,&\mbox{if } v_l \in (\tilde{\kappa}^4,\tilde{\kappa}^5]\\
\frac{v_l}{4}
+\frac{\tilde{\kappa}^2\tilde{\kappa}^1}{\tilde{\kappa}^2+\tilde{\kappa}^1},&\mbox{if } v_l>\tilde{\kappa}^5
\end{cases}
\]
and 
\[
\tilde{\psi}(v_h):=
\begin{cases}
0,&\mbox{if } v_h \leq \tilde{\kappa}^1\\
\frac{\lb{\beta}^{-1}(v_h)v_h}{v_h+\lb{\beta}^{-1}(v_h)}-\frac{\tilde{\kappa}^2\tilde{\kappa}^3}{\tilde{\kappa}^2+\tilde{\kappa}^1}-\int_{\tilde{\kappa}^1}^{\lb{\beta}^{-1}(v_h)} \left(\frac{\lb{\beta}(z)}{z+\lb{\beta}(z)}\right)^2 \diff z,&\mbox{if } v_h \in (\tilde{\kappa}^1,\tilde{\kappa}^3]\\
\frac{v_h}{4},&\mbox{if } v_h \in (\tilde{\kappa}^3,\tilde{\kappa}^4]\\
\frac{v_h\ub{\beta}^{-1}(v_h)}{v_h+\ub{\beta}^{-1}(v_h)}-\frac{\tilde{\kappa}^4}{4}-\int_{\tilde{\kappa}^4}^{\ub{\beta}^{-1}(v_h)}\left(\frac{\ub{\beta}^{-1}(z)}{z+\ub{\beta}^{-1}(z)}\right)^2 \diff z,&\mbox{if } v_h \in [\tilde{\kappa}^4,\tilde{\kappa}^5)\\
\frac{v_h}{4}-\frac{\tilde{\kappa}^2\tilde{\kappa}^1}{\tilde{\kappa}^2+\tilde{\kappa}^1},&\mbox{if } v_h>\tilde{\kappa}^5
\end{cases}\,,
\]
where 
\[
\lb{\beta}(z):=F_h^{-1}(F_l(z)-F_l(\tilde{\kappa}^3)+F_h(\tilde{\kappa}^3)) 
\]
for all $z \in [\tilde{\kappa}^1,\tilde{\kappa}^3]$, and 
\[
\ub{\beta}(z):=F_h^{-1}(F_l(z)-F_l(\tilde{\kappa}^4)+F_h(\tilde{\kappa}^4))
\]
for all $z \in [\tilde{\kappa}^4,\tilde{\kappa}^5]$. We now show that 
\begin{equation}\label{eq:feasible_dual}
\tilde{\phi}_0(v_l)+\tilde{\psi}(v_h) \geq \widetilde{\pi}(v_l,v_h)
\end{equation}
for all $v_l,v_h$. To this end, we first establish three inequalities that follow from \eqref{eq:system2}. First, note that the function 
\[
v \mapsto \frac{\tilde{\kappa}^1v}{\tilde{\kappa}^1+v}+\int_{v}^{\tilde{\kappa}^3}\left(\frac{\lb{\beta}(z)}{z+\lb{\beta}(z)}\right)^2 \diff z
\]
is decreasing in $v$. This implies that 
\begin{equation*}
\frac{\tilde{\kappa}^1\tilde{\kappa}^3}{\tilde{\kappa}^1+\tilde{\kappa}^3} \leq \frac{\tilde{\kappa}^1\tilde{\kappa}^2}{\tilde{\kappa}^1+\tilde{\kappa}^2}+\int_{\tilde{\kappa}^2}^{\tilde{\kappa}^3}\left(\frac{\lb{\beta}(z)}{z+\lb{\beta}(z)}\right)^2 \diff z=\frac{\tilde{\kappa}^3}{4}\,.
\end{equation*}
Next, note that since the function 
\[
v \mapsto \frac{v}{4}-\int_{\tilde{\kappa}^2}^v \left(\frac{\lb{\beta}(z)}{z+\lb{\beta}(z)}\right)^2 \diff z
\]
is increasing in $v$, it follows that 
\[
\frac{\tilde{\kappa}^1\tilde{\kappa}^2}{\tilde{\kappa}^1+\tilde{\kappa}^2}=\frac{\tilde{\kappa}^3}{4}-\int_{\tilde{\kappa}^2}^{\tilde{\kappa}^3} \left(\frac{\lb{\beta}(z)}{z+\lb{\beta}(z)}\right)^2 \diff z \geq \frac{\tilde{\kappa}^2}{4}\,. 
\]
Together, the above two inequalities imply 
\begin{equation}\label{eq:fact1}
\tilde{\kappa}^2 \leq 3\tilde{\kappa}^1 \leq \tilde{\kappa}^3\,.
\end{equation}
Lastly, note that since 
\[
\frac{\tilde{\kappa}^5}{4}=\int_{\tilde{\kappa}^4}^{\tilde{\kappa}^5} \left(\frac{\ub{\beta}(z)}{z+\ub{\beta}(z)}\right)^2+\frac{1}{4}(\tilde{\kappa}^4-\tilde{\kappa}^3)+\int_{\tilde{\kappa}^2}^{\tilde{\kappa}^3}\left(\frac{\lb{\beta}(z)}{z+\lb{\beta}(z)}\right)^2\diff z \leq \int_{\tilde{\kappa}^2}^{\tilde{\kappa}^5}\left(\frac{\tilde{\kappa}^5}{z+\tilde{\kappa}^5}\right)^2 \diff z=\int_{\tilde{\kappa}^4}^{\tilde{\kappa}^5} \frac{\diff }{\diff z} \left(\frac{\tilde{\kappa}^5z}{\tilde{\kappa}^5+z}\right) \diff z \,,
\]
it follows that 
\begin{equation}\label{eq:fact2}
\frac{\tilde{\kappa}^5}{4}=\frac{\tilde{\kappa}^5}{2}-\frac{\tilde{\kappa}^5}{4} \geq \frac{\tilde{\kappa}^5}{2}-\int_{\tilde{\kappa}^4}^{\tilde{\kappa}^5} \frac{\diff }{\diff z} \left(\frac{\tilde{\kappa}^5z}{\tilde{\kappa}^5+z}\right) \diff z=\frac{\tilde{\kappa}^5\tilde{\kappa}^2}{\tilde{\kappa}^5+\tilde{\kappa}^2} \iff \tilde{\kappa}^5 \geq 3\tilde{\kappa}^2\,.
\end{equation}

Then, we discuss all the cases.

\mbox{}

\noindent \textbf{Case 1: }$v_l \leq \tilde{\kappa}^2$.

\noindent When $v_h \leq \tilde{\kappa}^1$, 
\[
\tilde{\phi}(v_l)+\tilde{\psi}(v_h)=\frac{\tilde{\kappa}^2\tilde{\kappa}^1}{\tilde{\kappa}^2+\tilde{\kappa}^1} \geq \frac{v_lv_h}{v_l+v_h}\,.
\]
Meanwhile, $3\tilde{\kappa}^2 \geq \tilde{\kappa}^2 \geq \tilde{\kappa}^1$ implies that 
\[
\tilde{\phi}(v_l)+\tilde{\psi}(v_h)=\frac{\tilde{\kappa}^2\tilde{\kappa}^1}{\tilde{\kappa}^2+\tilde{\kappa}^1} \geq \frac{\tilde{\kappa}^1}{4} \geq \frac{v_h}{4}\,.
\]
Moreover, \eqref{eq:fact1} implies that 
\[
\tilde{\phi}(v_l)+\tilde{\psi}(v_h)=\frac{\tilde{\kappa}^2\tilde{\kappa}^1}{\tilde{\kappa}^2+\tilde{\kappa}^1} \geq \frac{\tilde{\kappa}^2}{4} \geq \frac{v_l}{4}\,.
\]
Together, 
\[
\tilde{\phi}(v_l)+\tilde{\psi}(v_h) \geq \max\left\{\frac{v_lv_h}{v_l+v_h},\frac{v_h}{4},\frac{v_l}{4}\right\}=\widetilde{\pi}(v_l,v_h)\,,
\]
as desired. 

When $v_h \in (\tilde{\kappa}^1,\tilde{\kappa}^3]$, from \eqref{eq:fact1}, it follows that 
\[
3v_h \geq 3\tilde{\kappa}^1 \geq \tilde{\kappa}^2 \geq v_l\,,
\]
and hence 
\[
\widetilde{\pi}(v_l,v_h)=\max\left\{\frac{v_lv_h}{v_l+v_h},\frac{v_h}{4}\right\}\,.
\]
Moreover, since 
\[
v_h \mapsto \tilde{\phi}(v_l)+\tilde{\psi}(v_h)-\frac{v_h}{4}
\]
is increasing in $v_h$ for all $v_l \leq \tilde{\kappa}^2$, 
\[
\tilde{\phi}(v_l)+\tilde{\psi}(v_h)-\frac{v_h}{4} \geq \tilde{\phi}(v_l)+\tilde{\psi}(\tilde{\kappa}^1)-\frac{\tilde{\kappa}^1}{4}=\frac{\tilde{\kappa}^2\tilde{\kappa}^1}{\tilde{\kappa}^2+\tilde{\kappa}^1}-\frac{\tilde{\kappa}^1}{4} \geq 0\,,
\]
where the last inequality follows from \eqref{eq:fact1}.  Meanwhile, since the function 
\[
(v_l,v_h) \mapsto \tilde{\phi}(v_l)+\tilde{\psi}(v_h)-\frac{v_lv_h}{v_l+v_h}
\]
is decreasing in $v_l$ and increasing in $v_h$, it follows that 
\[
\tilde{\phi}(v_l)+\tilde{\psi}(v_h)-\frac{v_lv_h}{v_l+v_h} \geq \tilde{\phi}(\tilde{\kappa}^2)+\tilde{\psi}(\tilde{\kappa}^1)-\frac{\tilde{\kappa}^2\tilde{\kappa}^1}{\tilde{\kappa}^2+\tilde{\kappa}^1}=0\,.
\]
Together, 
\[
\tilde{\phi}(v_l)+\tilde{\psi}(v_h) \geq \widetilde{\pi}(v_l,v_h)=\max\left\{\frac{v_lv_h}{v_l+v_h},\frac{v_h}{4}\right\}\,,
\]
as desired. 

When $v_h \in (\tilde{\kappa}^3,\tilde{\kappa}^4]$, 
\begin{align*}
\tilde{\phi}(v_l)+\tilde{\psi}(v_h)=&\frac{\tilde{\kappa}^3}{4}-\int_{\tilde{\kappa}^2}^{\tilde{\kappa}^3} \left(\frac{\lb{\beta}(z)}{z+\lb{\beta}(z)}\right)^2 \diff z+\frac{v_h}{4}\\
=&\int_{\tilde{\kappa}^2}^{\tilde{\kappa}^3} \left[\frac{1}{4}-\left(\frac{\lb{\beta}(z)}{z+\lb{\beta}(z)}\right)^2\right]\diff z+\frac{v_h}{4}+\frac{\tilde{\kappa}^2}{4}\\
\geq & \frac{v_h}{4}+\frac{\tilde{\kappa}^2}{4}\\
\geq & \max \left\{\frac{v_lv_h}{v_l+v_h},\frac{v_l}{4},\frac{v_h}{4}\right\}\,,
\end{align*}
as desired. 

When $v_h \in (\tilde{\kappa}^4,\tilde{\kappa}^5]$, since $v_l \leq v_h$, we have 
\[
\widetilde{\pi}(v_l,v_h)=\max\left\{\frac{v_lv_h}{v_l+v_h},\frac{v_h}{4}\right\}\,.
\]
Note that the function 
\[
v_h \mapsto \tilde{\phi}(v_l)+\tilde{\psi}(v_h)-\frac{v_h}{4}
\]
is decreasing in $v_h$ for all $v_l \leq \tilde{\kappa}^2$. Therefore, 
\[
\tilde{\phi}(v_l)+\tilde{\psi}(v_h)-\frac{v_h}{4} \geq \tilde{\phi}(v_l)+\tilde{\psi}(\tilde{\kappa}^5)-\frac{\tilde{\kappa}^5}{4}=0\,.
\]
Meanwhile, note that the function 
\[
(v_l,v_h) \mapsto \tilde{\phi}(v_l)+\tilde{\psi}(v_h)-\frac{v_lv_h}{v_l+v_h}
\]
is decreasing in $v_l$ and increasing in $v_h$. Thus,
\begin{align*}
\tilde{\phi}(v_l)+\tilde{\psi}(v_h)-\frac{v_lv_h}{v_l+v_h} \geq \tilde{\phi}(\tilde{\kappa}^2)+\tilde{\psi}(\tilde{\kappa}^4)-\frac{\tilde{\kappa}^2\tilde{\kappa}^4}{\tilde{\kappa}^2+\tilde{\kappa}^4}=&\frac{\tilde{\kappa}^4}{4}+\frac{\tilde{\kappa}^3}{4}-\frac{\tilde{\kappa}^2\tilde{\kappa}^4}{\tilde{\kappa}^2+\tilde{\kappa}^4}-\int_{\tilde{\kappa}^2}^{\tilde{\kappa}^3}\left(\frac{\lb{\beta}(z)}{z+\lb{\beta}(z)}\right)^2 \diff z\\
\geq & \frac{\tilde{\kappa}^4}{4}+\frac{\tilde{\kappa}^3}{4}-\frac{\tilde{\kappa}^2\tilde{\kappa}^4}{\tilde{\kappa}^2+\tilde{\kappa}^4}-\int_{\tilde{\kappa}^2}^{\tilde{\kappa}^3}\left(\frac{\tilde{\kappa}^4}{z+\tilde{\kappa}^4}\right)^2 \diff z\\
=& \frac{\tilde{\kappa}^4}{4}+\frac{\tilde{\kappa}^3}{4}-\frac{\tilde{\kappa}^2\tilde{\kappa}^4}{\tilde{\kappa}^2+\tilde{\kappa}^4}-\frac{\tilde{\kappa}^3\tilde{\kappa}^4}{\tilde{\kappa}^3+\tilde{\kappa}^4}+\frac{\tilde{\kappa}^2\tilde{\kappa}^4}{\tilde{\kappa}^2+\tilde{\kappa}^4}\\
=&\frac{\tilde{\kappa}^4}{4}+\frac{\tilde{\kappa}^3}{4}-\frac{\tilde{\kappa}^3\tilde{\kappa}^4}{\tilde{\kappa}^3+\tilde{\kappa}^4}\\
\geq & 0\,.
\end{align*}
Together, 
\[
\tilde{\phi}(v_l)+\tilde{\psi}(v_h) \geq \widetilde{\pi}(v_l,v_h)\,,
\]
as desired. 

Lastly, when $v_h>\tilde{\kappa}^5$, by \eqref{eq:fact2}, $v_h>\tilde{\kappa}^5 \geq \tilde{\kappa}^2 \geq 3v_l$. Thus, $\widetilde{\pi}(v_l,v_h)=\nicefrac{v_h}{4}$, and hence 
\[
\tilde{\phi}(v_l)+\tilde{\psi}(v_h)=\frac{v_h}{4} = \widetilde{\pi}(v_l,v_h)\,,
\]
as desired. 

\mbox{}

\noindent \textbf{Case 2: }$v_l \in (\tilde{\kappa}^2,\tilde{\kappa}^3]$.

\mbox{}

\noindent When $v_h \leq \tilde{\kappa}^1$, 
\[
\widetilde{\pi}(v_l,v_h)=\max\left\{\frac{v_lv_h}{v_l+v_h},\frac{v_l}{4}\right\}\,.
\]
Note that 
\[
v_l \mapsto \tilde{\phi}(v_l)+\tilde{\psi}(v_h)-\frac{v_lv_h}{v_l+v_h}
\]
is increasing in $v_l$ and 
\[
v_l \mapsto \tilde{\phi}(v_l)+\tilde{\psi}(v_h)-\frac{v_l}{4}
\]
is decreasing in $v_l$. Therefore, 
\[
\tilde{\phi}(v_l)+\tilde{\psi}(v_h)-\frac{v_lv_h}{v_l+v_h} \geq \tilde{\phi}(\tilde{\kappa}^2)+\tilde{\psi}(v_h)-\frac{\tilde{\kappa}^2v_h}{\tilde{\kappa}^2+v_h}=\frac{\tilde{\kappa}^2\tilde{\kappa}^1}{\tilde{\kappa}^2+\tilde{\kappa}^1}-\frac{\tilde{\kappa}^2v_h}{\tilde{\kappa}^2+v_h} \geq 0\,,
\]
and 
\[
\tilde{\phi}(v_l)+\tilde{\psi}(v_h)-\frac{v_l}{4} \geq \tilde{\phi}(\tilde{\kappa}^3)+\tilde{\psi}(v_h)-\frac{\tilde{\kappa}^3}{4}=0\,,
\]
as desired. 

When $v_h \in (\tilde{\kappa}^1,\tilde{\kappa}^3]$, by \eqref{eq:fact1}, it follows that 
\[
\widetilde{\pi}(v_l,v_h)=\frac{v_lv_h}{v_l+v_h}\,.
\]
In particular, $\widetilde{\pi}$ is supermodular on $(\tilde{\kappa}^2,\tilde{\kappa}^3] \times (\tilde{\kappa}^1,\tilde{\kappa}^3]$. Therefore, since $\lb{\beta}$ is increasing, 
\[
\tilde{\phi}(v_l)=\max_{v_h' \in (\tilde{\kappa}^1,\tilde{\kappa}^3]} \left[\widetilde{\pi}(v_l,v_h')-\tilde{\psi}(v_h')\right] \geq \widetilde{\pi}(v_l,v_h)-\tilde{\psi}(v_h)\,,
\]
as desired. 

When $v_h \in (\tilde{\kappa}^3,\tilde{\kappa}^4]$, 
\[
\widetilde{\pi}(v_l,v_h)=\max\left\{\frac{v_lv_h}{v_l+v_h},\frac{v_h}{4}\right\}\,.
\]
Note that 
\[
\tilde{\phi}(v_l)+\tilde{\psi}(v_h)=\frac{\tilde{\kappa}^1\tilde{\kappa}^2}{\tilde{\kappa}^1+\tilde{\kappa}^2}+\int_{\tilde{\kappa}^2}^{v_l} \left(\frac{\lb{\beta}(z)}{z+\lb{\beta}(z)}\right)^2 \diff z+\frac{v_h}{4} \geq \frac{v_h}{4}\,.
\]
Moreover, since 
\[
(v_l,v_h) \mapsto \tilde{\phi}(v_l)+\tilde{\psi}(v_h)-\frac{v_lv_h}{v_l+v_h}
\]
is increasing in $v_h$ and decreasing in $v_l$, 
\[
\tilde{\phi}(v_l)+\tilde{\psi}(v_h)-\frac{v_lv_h}{v_l+v_h} \geq \tilde{\phi}(\tilde{\kappa}^3)+\tilde{\psi}(\tilde{\kappa}^3)-\frac{\tilde{\kappa}^3}{2}=0\,.
\]
Together, $\tilde{\phi}(v_l)+\tilde{\psi}(v_h)\geq \widetilde{\pi}(v_l,v_h)$, as desired. 

When $v_h \in (\tilde{\kappa}^4,\tilde{\kappa}^5]$, note that 
\[
(v_l,v_h) \mapsto \tilde{\phi}(v_l)+\tilde{\psi}(v_h)-\frac{v_h}{4}
\]
is increasing in both $v_l$ and $v_h$. Also, 
\[
(v_l,v_h) \mapsto \tilde{\phi}(v_l)+\tilde{\psi}(v_h)-\frac{v_lv_h}{v_l+v_h}
\]
is decreasing in $v_l$ and increasing in $v_h$. Therefore,
\[
\tilde{\phi}(v_l)+\tilde{\psi}(v_h)-\frac{v_h}{4} \geq \tilde{\phi}(\tilde{\kappa}^2)+\tilde{\psi}(\tilde{\kappa}^4)-\frac{\tilde{\kappa}^4}{4}=\frac{\tilde{\kappa}^2\tilde{\kappa}^1}{\tilde{\kappa}^2+\tilde{\kappa}^1} \geq 0\,,
\]
and 
\[
\tilde{\phi}(v_l)+\tilde{\psi}(v_h)-\frac{v_lv_h}{v_l+v_h} \geq \tilde{\phi}(\tilde{\kappa}^3)+\tilde{\psi}(\tilde{\kappa}^4)-\frac{\tilde{\kappa}^3\tilde{\kappa}^4}{\tilde{\kappa}^3+\tilde{\kappa}^4}=\frac{\tilde{\kappa}^3}{4}+\frac{\tilde{\kappa}^4}{4}-\frac{\tilde{\kappa}^3\tilde{\kappa}^4}{\tilde{\kappa}^3+\tilde{\kappa}^4} \geq 0\,,
\]
as desired. 

When $v_h>\tilde{\kappa}^5$, 
\[
\widetilde{\pi}(v_l,v_h)=\max\left\{\frac{v_lv_h}{v_l+v_h},\frac{v_h}{4}\right\}\,.
\]
Moreover, 
\[
\tilde{\phi}(v_l)+\tilde{\psi}(v_h)=\int_{\tilde{\kappa}^2}^{v_l}\left(\frac{\lb{\beta}(z)}{z+\lb{\beta}(z)}\right)^2 \diff z+\frac{v_h}{4} \geq \frac{v_h}{4}\,.
\]
Meanwhile, since 
\[
(v_l,v_h) \mapsto \tilde{\phi}(v_l)+\tilde{\psi}(v_h)-\frac{v_lv_h}{v_l+v_h}
\]
is decreasing in $v_l$ and increasing in $v_h$, 
\begin{align*}
\tilde{\phi}(v_l)+\tilde{\psi}(v_h)-\frac{v_lv_h}{v_l+v_h} \geq& \tilde{\phi}(\tilde{\kappa}^3)+\tilde{\psi}(\tilde{\kappa}^5)-\frac{\tilde{\kappa}^3\tilde{\kappa}^5}{\tilde{\kappa}^3+\tilde{\kappa}^5}\\
=& \int_{\tilde{\kappa}^2}^{\tilde{\kappa}^3}\left(\frac{\lb{\beta}(z)}{z+\lb{\beta}(z)}\right)^2 \diff z+\frac{\tilde{\kappa}^5}{4}-\frac{\tilde{\kappa}^3\tilde{\kappa}^5}{\tilde{\kappa}^3+\tilde{\kappa}^5}\\
=& \int_{\tilde{\kappa}^4}^{\tilde{\kappa}^5}\left(\frac{\ub{\beta}(z)}{z+\ub{\beta}(z)}\right)^2 \diff z +\frac{\tilde{\kappa}^4}{4}+\frac{\tilde{\kappa}^3}{4}-\frac{\tilde{\kappa}^3\tilde{\kappa}^5}{\tilde{\kappa}^3+\tilde{\kappa}^5}\\
\geq & \frac{1}{4}(\tilde{\kappa}^5-\tilde{\kappa}^4)+\frac{\tilde{\kappa}^4}{4}+\frac{\tilde{\kappa}^3}{4}-\frac{\tilde{\kappa}^3\tilde{\kappa}^5}{\tilde{\kappa}^3+\tilde{\kappa}^5}\\
=& \frac{\tilde{\kappa}^5}{4}+\frac{\tilde{\kappa}^3}{4}-\frac{\tilde{\kappa}^3\tilde{\kappa}^5}{\tilde{\kappa}^3+\tilde{\kappa}^5}\\
\geq & 0\,,
\end{align*}
where the first inequality follows from $\ub{\beta}(z) \geq z$ for all $z \in [\tilde{\kappa}^4,\tilde{\kappa}^5]$, which in turn is because $\tilde{\kappa}^4 \leq v^\star \leq \tilde{\kappa}^5$. 

\mbox{}

\noindent \textbf{Case 3: }$v_l \in (\tilde{\kappa}^3,\tilde{\kappa}^4]$.

\mbox{}

\noindent When $v_h \leq \tilde{\kappa}^1$, by \eqref{eq:fact1}, $v_l \geq \tilde{\kappa}^4 \geq \tilde{\kappa}^3 \geq 3\tilde{\kappa}^1 \geq 3v_h$. Thus, $\widetilde{\pi}(v_l,v_h)=\nicefrac{v_l}{4}$. Moreover, 
\[
\tilde{\phi}(v_l)+\tilde{\psi}(v_l)=\frac{\tilde{\kappa}^4}{4}+\int_{\tilde{\kappa}^4}^{v_l} \left(\frac{\ub{\beta}(z)}{z+\ub{\beta}(z)}\right)^2 \diff z \geq \frac{\tilde{\kappa}^4}{4}+\frac{1}{4}(v_l-\tilde{\kappa}^4) \geq \frac{v_l}{4}\,,
\]
where the inequality follows from $\ub{\beta}(z) \geq z$ for all $z \in [\tilde{\kappa}^4,\tilde{\kappa}^5]$, which in turn is because $\tilde{\kappa}^4 \leq v^\star \leq \tilde{\kappa}^5$. 

When $v_h \in (\tilde{\kappa}^1,\tilde{\kappa}^3]$, 
\[
\widetilde{\pi}(v_l,v_h)=\max\left\{\frac{v_lv_h}{v_l+v_h},\frac{v_l}{4}\right\}\,.
\]
Note that 
\[
\tilde{\phi}(v_l)+\tilde{\psi}(v_h)=\frac{v_l}{4}+\tilde{\psi}(v_h) \geq \frac{v_l}{4}\,.
\]
Moreover, since 
\[
(v_l,v_h) \mapsto \tilde{\phi}(v_l)+\tilde{\psi}(v_h)-\frac{v_lv_h}{v_l+v_h}
\]
is increasing in $v_l$ and decreasing in $v_h$, 
\[
\tilde{\phi}(v_l)+\tilde{\psi}(v_h)-\frac{v_lv_h}{v_l+v_h} \geq \tilde{\phi}(\tilde{\kappa}^3)+\tilde{\psi}(\tilde{\kappa}^3)-\frac{\tilde{\kappa}^3}{2}=0\,.
\]
Thus, $\tilde{\phi}(v_l)+\tilde{\psi}(v_h) \geq \widetilde{\pi}(v_l,v_h)$. 

When $v_h \in (\tilde{\kappa}^3,\tilde{\kappa}^4]$, 
\[
\tilde{\phi}(v_l)+\tilde{\psi}(v_h)=\frac{v_l}{4}+\frac{v_h}{4} \geq \max\left\{\frac{v_lv_h}{v_l+v_h},\frac{v_l}{4},\frac{v_h}{4}\right\}=\widetilde{\pi}(v_l,v_h)\,.
\]

When $v_h \in (\tilde{\kappa}^4,\tilde{\kappa}^5]$, 
\[
\widetilde{\pi}(v_l,v_h)=\max\left\{\frac{v_lv_h}{v_l+v_h},\frac{v_h}{4}\right\}\,.
\]
First note that the function 
\[
(v_l,v_h) \mapsto \tilde{\phi}(v_l)+\tilde{\psi}(v_h)-\frac{v_h}{4}
\]
is increasing in $v_l$ and decreasing in $v_h$. Thus, 
\begin{align*}
\tilde{\phi}(v_l)+\tilde{\psi}(v_h)-\frac{v_h}{4} \geq &\tilde{\phi}(\tilde{\kappa}^3)+\tilde{\psi}(\tilde{\kappa}^5)-\frac{\tilde{\kappa}^5}{4}\\
=& \frac{\tilde{\kappa}^3}{4}+\frac{\tilde{\kappa}^5}{4}-\frac{\tilde{\kappa}^2\tilde{\kappa}^1}{\tilde{\kappa}^2+\tilde{\kappa}^1}-\frac{\tilde{\kappa}^5}{4}\\
=&\frac{\tilde{\kappa}^3}{4}-\frac{\tilde{\kappa}^3}{4}+\int_{\tilde{\kappa}^2}^{\tilde{\kappa}^3}\left(\frac{\lb{\beta}(z)}{z+\lb{\beta}(z)}\right)^2 \diff z\\\
\geq & 0\,.
\end{align*}
Moreover, 
\[
(v_l,v_h) \mapsto \tilde{\phi}(v_l)+\tilde{\psi}(v_h)-\frac{v_lv_h}{v_l+v_h}
\]
is decreasing in $v_l$ and increasing in $v_h$. Therefore, 
\[
\tilde{\phi}(v_l)+\tilde{\psi}(v_h)-\frac{v_lv_h}{v_l+v_h} \geq \tilde{\phi}(\tilde{\kappa}^4)+\tilde{\psi}(\tilde{\kappa}^4)-\frac{\tilde{\kappa}^4}{2}=0\,,
\]
as desired. 

When $v_h>\tilde{\kappa}^5$, 
\[
\widetilde{\pi}(v_l,v_h)=\max\left\{\frac{v_lv_h}{v_l+v_h},\frac{v_h}{4}\right\}\,.
\]
First note that 
\[
\tilde{\phi}(v_l)+\tilde{\psi}(v_h)=\frac{1}{4}(v_l+v_h)-\frac{\tilde{\kappa}^2\tilde{\kappa}^1}{\tilde{\kappa}^2+\tilde{\kappa}^1} \geq \frac{\tilde{\kappa}^3}{4}-\frac{\tilde{\kappa}^2\tilde{\kappa}^1}{\tilde{\kappa}^2+\tilde{\kappa}^1} +\frac{v_h}{4} =\int_{\tilde{\kappa}^2}^{\tilde{\kappa}^3} \left(\frac{\lb{\beta}(z)}{z+\lb{\beta}(z)}\right)^2 \diff z+\frac{v_h}{4} \geq \frac{v_h}{4}\,.
\]
Moreover, since 
\[
(v_l,v_h) \mapsto \tilde{\phi}(v_l)+\tilde{\psi}(v_h)-\frac{v_lv_h}{v_l+v_h}
\]
is decreasing in $v_l$ and increasing in $v_h$, 
\begin{align*}
\tilde{\phi}(v_l)+\tilde{\psi}(v_h)-\frac{v_lv_h}{v_l+v_h} \geq &\tilde{\phi}(\tilde{\kappa}^4)+\tilde{\psi}(\tilde{\kappa}^5)-\frac{\tilde{\kappa}^4\tilde{\kappa}^5}{\tilde{\kappa}^4+\tilde{\kappa}^5}\\
=&\frac{1}{4}(\tilde{\kappa}^4+\tilde{\kappa}^5)-\frac{\tilde{\kappa}^2\tilde{\kappa}^1}{\tilde{\kappa}^2+\tilde{\kappa}^1}-\frac{\tilde{\kappa}^4\tilde{\kappa}^5}{\tilde{\kappa}^4+\tilde{\kappa}^5}\\
=& \frac{1}{4}(\tilde{\kappa}^4+\tilde{\kappa}^5)+\frac{1}{4}(\tilde{\kappa}^5-\tilde{\kappa}^4)-\int_{\tilde{\kappa}^4}^{\tilde{\kappa}^5}\left(\frac{\ub{\beta}(z)}{z+\ub{\beta}(z)}\right)^2 \diff z-\frac{\tilde{\kappa}^4\tilde{\kappa}^5}{\tilde{\kappa}^4+\tilde{\kappa}^5}\\
=& \frac{\tilde{\kappa}^5}{2}-\frac{\tilde{\kappa}^4\tilde{\kappa}^5}{\tilde{\kappa}^4+\tilde{\kappa}^5}-\int_{\tilde{\kappa}^4}^{\tilde{\kappa}^5} \left(\frac{\ub{\beta}(z)}{z+\ub{\beta}(z)}\right)^2 \diff z\\
=& \int_{\tilde{\kappa}^4}^{\tilde{\kappa}^5} \left[\left(\frac{\tilde{\kappa}^5}{z+\tilde{\kappa}^5}\right)^2-\left(\frac{\ub{\beta}(z)}{z+\ub{\beta}(z)}\right)^2\right]\diff z\\
\geq & 0\,,
\end{align*}
where the last inequality follows from the fact that $\ub{\beta}(z) \leq \tilde{\kappa}^5$ for all $z \in [\tilde{\kappa}^4,\tilde{\kappa}^5]$. Together, $\tilde{\phi}(v_l)+\tilde{\psi}(v_h) \geq \widetilde{\pi}(v_l,v_h)$, as desired. 

\mbox{}

\noindent \textbf{Case 4:} $v_l \in (\tilde{\kappa}^4,\tilde{\kappa}^5]$.

\mbox{}

\noindent When $v_h \leq \tilde{\kappa}^1$, by \eqref{eq:fact1}, $v_l \geq \tilde{\kappa}^4 \geq \tilde{\kappa}^3 \geq 3\tilde{\kappa}^1 \geq 3v_h$, and hence $\widetilde{\pi}(v_l,v_h)=\nicefrac{v_l}{4}$. Thus, 
\[
\tilde{\phi}(v_l)+\tilde{\psi}(v_h)=\frac{\tilde{\kappa}^4}{4}+\int_{\tilde{\kappa}^4}^{v_l}\left(\frac{\ub{\beta}(z)}{z+\ub{\beta}(z)}\right)^2 \diff z \geq \frac{\tilde{\kappa}^4}{4}+\frac{1}{4}(v_l-\tilde{\kappa}^4)=\frac{v_l}{4}=\widetilde{\pi}(v_l,v_h)\,,
\]
where the inequality follows from the fact that $\ub{\beta}(z) \geq z$ for all $z \in [\tilde{\kappa}^4,\tilde{\kappa}^5]$, which in turn follows from $\tilde{\kappa}^4 \leq v^\star \leq \tilde{\kappa}^5$.  

When $v_h \in (\tilde{\kappa}^1,\tilde{\kappa}^3]$, 
\[
\widetilde{\pi}(v_l,v_h)=\max\left\{\frac{v_lv_h}{v_l+v_h},\frac{v_l}{4}\right\}\,.
\]
Note that, as argued above,
\[
\tilde{\phi}(v_l)+\tilde{\psi}(v_h) \geq \tilde{\phi}(v_l)=\frac{\tilde{\kappa}^4}{4}+\int_{\tilde{\kappa}^4}^{v_l}\left(\frac{\ub{\beta}(z)}{z+\ub{\beta}(z)}\right)^2 \diff z \geq \frac{v_l}{4}\,.
\]
Moreover, since 
\[
(v_l,v_h) \mapsto \tilde{\phi}(v_l)+\tilde{\psi}(v_h)-\frac{v_lv_h}{v_l+v_h}
\]
is increasing in $v_l$ and decreasing in $v_h$, 
\[
\tilde{\phi}(v_l)+\tilde{\psi}(v_h)-\frac{v_lv_h}{v_l+v_h} \geq \tilde{\phi}(\tilde{\kappa}^4)+\tilde{\psi}(\tilde{\kappa}^3)-\frac{\tilde{\kappa}^4\tilde{\kappa}^3}{\tilde{\kappa}^4+\tilde{\kappa}^3}=\frac{1}{4}(\tilde{\kappa}^4+\tilde{\kappa}^3)-\frac{\tilde{\kappa}^4\tilde{\kappa}^3}{\tilde{\kappa}^4+\tilde{\kappa}^3} \geq 0\,,
\]
as desired. 

When $v_h \in (\tilde{\kappa}^3,\tilde{\kappa}^4]$, 
\[
\widetilde{\pi}(v_l,v_h)=\max\left\{\frac{v_lv_h}{v_l+v_h},\frac{v_l}{4}\right\}\,.
\]
Again, as argued above,
\[
\tilde{\phi}(v_l)+\tilde{\psi}(v_h) \geq \tilde{\phi}(v_l)=\frac{\tilde{\kappa}^4}{4}+\int_{\tilde{\kappa}^4}^{v_l}\left(\frac{\ub{\beta}(z)}{z+\ub{\beta}(z)}\right)^2 \diff z \geq \frac{v_l}{4}\,.
\]
Meanwhile, note that the function 
\[
(v_l,v_h) \mapsto \tilde{\phi}(v_l)+\tilde{\psi}(v_h)-\frac{v_lv_h}{v_l+v_h}
\]
is increasing in $v_l$ and decreasing in $v_h$. Thus,
\[
\tilde{\phi}(v_l)+\tilde{\psi}(v_h)-\frac{v_lv_h}{v_l+v_h} \geq \tilde{\phi}(\tilde{\kappa}^4)+\tilde{\psi}(\tilde{\kappa}^4)-\frac{\tilde{\kappa}^4}{2}=0\,.
\]
Together, $\tilde{\phi}(v_l)+\tilde{\psi}(v_h) \geq \widetilde{\pi}(v_l,v_h)$, as desired. 

When $v_h \in (\tilde{\kappa}^4,\tilde{\kappa}^5]$, first note that, since 
\[
(v_l,v_h) \mapsto \frac{v_lv_h}{v_l+v_h}
\]
is supermodular and since $\ub{\beta}$ is increasing, by construction, 
\[
\tilde{\phi}(v_l)=\max_{v_h' \in [\tilde{\kappa}^4,\tilde{\kappa}^5]} \left[\frac{v_lv_h'}{v_l+v_h'}-\psi(v_h')\right] \geq \frac{v_lv_h}{v_l+v_h}-\tilde{\psi}(v_h)\,,
\]
and thus 
\[
\tilde{\phi}(v_l)+\tilde{\psi}(v_h) \geq \widetilde{\pi}(v_l,v_h)\,.
\]
Next, note that since the function 
\[
(v_l,v_h) \mapsto \tilde{\phi}(v_l)+\tilde{\psi}(v_h)-\frac{v_h}{4}
\]
is increasing in $v_l$ and decreasing in $v_h$,
\[
\tilde{\phi}(v_l)+\tilde{\psi}(v_h)-\frac{v_h}{4} \geq \tilde{\phi}(\tilde{\kappa}^4)+\tilde{\psi}(\tilde{\kappa}^5)-\frac{\tilde{\kappa}^5}{4}=\frac{\tilde{\kappa}^4}{4}-\frac{\tilde{\kappa}^2\tilde{\kappa}^1}{\tilde{\kappa}^2+\tilde{\kappa}^1}=\frac{\tilde{\kappa}^4}{4}-\frac{\tilde{\kappa}^3}{4}+\int_{\tilde{\kappa}^2}^{\tilde{\kappa}^3}\left(\frac{\lb{\beta}(z)}{z+\lb{\beta}(z)}\right)^2 \diff z \geq 0\,.
\]
Lastly, note that since the function 
\begin{equation}\label{eq:4.1}
(v_l,v_h) \mapsto \tilde{\phi}(v_l)+\tilde{\psi}(v_h)-\frac{v_l}{4}
\end{equation}
is increasing in both $v_l$ and $v_h$, 
\begin{equation}\label{eq:4.2}
\tilde{\phi}(v_l)+\tilde{\psi}(v_h)-\frac{v_l}{4} \geq \tilde{\phi}(\tilde{\kappa}^4)+\tilde{\psi}(\tilde{\kappa}^4)-\frac{\tilde{\kappa}^4}{4}=\frac{\tilde{\kappa}^4}{4} \geq 0\,.
\end{equation}
Together, we have that 
\[
\tilde{\phi}(v_l)+\tilde{\psi}(v_h) \geq \widetilde{\pi}(v_l,v_h)\,,
\]
as desired. 

When $v_h>\tilde{\kappa}^5$, 
\[
\widetilde{\pi}(v_l,v_h)=\max\left\{\frac{v_lv_h}{v_l+v_h},\frac{v_h}{4}\right\}\,.
\]
Note that 
\[
\tilde{\phi}(v_l)+\tilde{\psi}(v_h) \geq \frac{\tilde{\kappa}^4}{4}+\frac{v_h}{4}-\frac{\tilde{\kappa}^2\tilde{\kappa}^1}{\tilde{\kappa}^2+\tilde{\kappa}^1} \geq \frac{\tilde{\kappa}^3}{4} +\frac{v_h}{4}-\frac{\tilde{\kappa}^2\tilde{\kappa}^1}{\tilde{\kappa}^2+\tilde{\kappa}^1}= \int_{\tilde{\kappa}^2}^{\tilde{\kappa}^3}\left(\frac{\lb{\beta}(z)}{z+\lb{\beta}(z)}\right)^2 \diff z +\frac{v_h}{4} \geq \frac{v_h}{4}\,.
\]
Moreover, since the function 
\[
(v_l,v_h) \mapsto \tilde{\phi}(v_l)+\tilde{\psi}(v_h)-\frac{v_lv_h}{v_l+v_h}
\]
is decreasing in $v_l$ and is increasing in $v_h$, 
\[
\tilde{\phi}(v_l)+\tilde{\psi}(v_h)-\frac{v_lv_h}{v_l+v_h} \geq \tilde{\phi}(\tilde{\kappa}^5)+\tilde{\psi}(\tilde{\kappa}^5)-\frac{\tilde{\kappa}^5}{2}=0\,.
\]
Together, $\tilde{\psi}(v_l)+\tilde{\psi}(v_h) \geq \widetilde{\pi}(v_l,v_h)$, as desired. 

\mbox{}

\noindent \textbf{Case 5:} $v_l>\tilde{\kappa}^5$.

\mbox{}

\noindent When $v_h \leq \tilde{\kappa}^1$, by \eqref{eq:fact1}, $v_l \geq \tilde{\kappa}^5 \tilde{\kappa}^3 \geq 3\tilde{\kappa}^1 \geq 3v_h$, and hence $\widetilde{\pi}(v_l,v_h)=\nicefrac{v_l}{4}$. Thus, 
\[
\tilde{\phi}(v_l)+\tilde{\psi}(v_l) = \frac{v_l}{4}+\frac{\tilde{\kappa}^2\tilde{\kappa}^1}{\tilde{\kappa}^2+\tilde{\kappa}^1} \geq \frac{v_l}{4} = \widetilde{\pi}(v_l,v_h)\,,
\]
as desired. 

When $v_h \in (\tilde{\kappa}^1,\tilde{\kappa}^3]$, 
\[
\widetilde{\pi}(v_l,v_h)=\max\left\{\frac{v_lv_h}{v_l+v_h},\frac{v_l}{4}\right\}\,.
\]
Note that 
\[
\tilde{\phi(v_l)}+\tilde{\psi}(v_h)\geq \tilde{\phi(v_l)} \geq \frac{v_l}{4}\,,
\]
as argued above. Moreover, since 
\[
(v_l,v_h) \mapsto \tilde{\phi}(v_l)+\tilde{\psi}(v_h)-\frac{v_lv_h}{v_l+v_h}
\]
is increasing in $v_l$ and decreasing in $v_h$, 
\[
\tilde{\phi}(v_l)+\tilde{\psi}(v_h)-\frac{v_lv_h}{v_l+v_h} \geq \tilde{\phi}(\tilde{\kappa}^5)+\tilde{\psi}(\tilde{\kappa}^3)-\frac{\tilde{\kappa}^5\tilde{\kappa}^3}{\tilde{\kappa}^5+\tilde{\kappa}^3}
=\frac{\tilde{\kappa}^5}{4}+\frac{\tilde{\kappa}^3}{4}-\frac{\tilde{\kappa}^5\tilde{\kappa}^3}{\tilde{\kappa}^5+\tilde{\kappa}^3}+\frac{\tilde{\kappa}^2\tilde{\kappa}^1}{\tilde{\kappa}^2+\tilde{\kappa}^5} \geq 0\,.
\]
Together, $\tilde{\phi}(v_l)+\tilde{\psi}(v_h) \geq \widetilde{\pi}(v_l,v_h)$, as desired. 

When $v_h \in (\tilde{\kappa}^3,\tilde{\kappa}^4]$, 
\[
\tilde{\phi}(v_l)+\tilde{\psi}(v_h)=\frac{v_l}{4}+\frac{v_h}{4}+\frac{\tilde{\kappa}^2\tilde{\kappa}^1}{\tilde{\kappa}^2+\tilde{\kappa}^1} \geq \max \left\{\frac{v_lv_h}{v_l+v_h},\frac{v_l}{4},\frac{v_h}{4}\right\}=\widetilde{\pi}(v_l,v_h)\,,
\]
as desired. 

When $v_h \in (\tilde{\kappa}^4,\tilde{\kappa}^5]$, 
\[
\widetilde{\pi}(v_l,v_h)=\max\left\{\frac{v_lv_h}{v_l+v_h},\frac{v_l}{4}\right\}\,.
\]
Note that 
\[
\tilde{\phi}(v_l)+\tilde{\psi}(v_h) \geq \tilde{\phi}(v_l)=\frac{v_l}{4}+\frac{\tilde{\kappa}^2\tilde{\kappa}^1}{\tilde{\kappa}^2+\tilde{\kappa}^1} \geq \frac{v_l}{4}\,.
\]
Moreover, since the function 
\[
(v_l,v_h) \mapsto \tilde{\phi}(v_l)+\tilde{\psi}(v_h)-\frac{v_lv_h}{v_l+v_h}
\]
is increasing in $v_l$ and is decreasing in $v_h$, 
\[
\tilde{\phi}(v_l)+\tilde{\psi}(v_h)-\frac{v_lv_h}{v_l+v_h} \geq \tilde{\phi}(\tilde{\kappa}^5)+\tilde{\psi}(\tilde{\kappa}^5)-\frac{\tilde{\kappa}^5}{2}=0\,.
\]
Together, $\tilde{\phi}(v_l)+\tilde{\psi}(v_h) \geq \widetilde{\pi}(v_l,v_h)$, as desired. 

When $v_h>\tilde{\kappa}^5$, 
\[
\tilde{\phi}(v_l)+\tilde{\psi}(v_h)=\frac{1}{4}(v_l+v_h) \geq \max\left\{\frac{v_lv_h}{v_l+v_h},\frac{v_l}{4},\frac{v_h}{4}\right\}\,,
\]
as desired.

Next, let 
\begin{align*}
&\tilde{\gamma}^\star_c(v_l \leq x \mid v_h):=\\
&\begin{cases}
\mathbf{1}\{\lb{\Delta}_c^{-1}(F_{c,h}(v_h+\Delta_c(\tilde{\tilde{\kappa}}_c^3)) \leq x\},& \mbox{if } v_h \leq \tilde{\tilde{\kappa}}_c^1\\
\mathbf{1}\{F_{c,l}^{-1}(F_{c,h}(v_h)+\Delta_c(\tilde{\tilde{\kappa}}_c^3)) \leq x\},&\mbox{if } v_h \in (\tilde{\tilde{\kappa}}_c^1,\tilde{\tilde{\kappa}}^3_c]\\
\mathbf{1}\{v_h \leq x\},& \mbox{if } v_h \in (\tilde{\tilde{\kappa}}_c^3,\tilde{\tilde{\kappa}}_c^4]\\
\mathbf{1}\{F_{c,l}^{-1}(F_{c,h}(v_h)+\Delta_c(\tilde{\tilde{\kappa}}_c^4)) \leq x\}, &\mbox{if } v_h \in (\tilde{\tilde{\kappa}}_c^4,\tilde{\tilde{\kappa}}_c^5]\\
\frac{f_{c,l}(v_h)}{f_{c,h}(v_h)}\cdot \mathbf{1}\{v_h \leq x\}+\frac{f_{c,h}(v_h)-f_{c,l}(v_h)}{f_{c,h}(v_h)} \cdot \mathbf{1}\{F_{c,l}^{-1}(\Delta_c(\tilde{\tilde{\kappa}}_c^5)-\Delta_c(v_h)) \leq x\},&\mbox{if } v_h > \tilde{\tilde{\kappa}}_c^5,
\end{cases}\,,
\end{align*}
for all $x \in V$ and for all $v_h \in V$. 

Then, let $\tilde{\rho}_c^\star \in \Delta(V \times V)$ be defined as 
\begin{equation}\label{eq:sol}
\tilde{\rho}_c^\star(v_l \in A, v_h \in B):=\int_B \gamma_c^\star(A\mid v_h) F_{c,h}(\diff v_h)\,,
\end{equation}
for all measurable sets $A,B \subseteq V$. By construction, the marginals of $\rho_c^\star$ are exactly $F_{c,l}$ and $F_{c,h}$. That is, $\tilde{\rho}_c^\star \in \mathcal{R}_c$.

It remains to show that for any $(v_l,v_h) \in \supp(\tilde{\rho})$, $\tilde{\phi}(v_l)+\tilde{\psi}(v_h)=\widetilde{\pi}(v_l,v_h)$. To see this, consider any $(v_l,v_h) \in \supp(\tilde{\rho})$. If $v_l \leq \tilde{\kappa}^2$, it must be that $v_h \geq \tilde{\kappa}^5$. By \eqref{eq:fact2}, $\widetilde{\pi}(v_l,v_h)=\nicefrac{v_h}{4}$. Therefore, 
\[
\tilde{\phi}(v_l)+\tilde{\psi}(v_h)=\frac{v_h}{4}=\widetilde{\pi}(v_l,v_h)\,,
\]
as desired. If $v_l \in (\tilde{\kappa}^2,\tilde{\kappa}^3]$, then it must be that $v_h=\lb{\beta}(v_l)$. By \eqref{eq:fact1}, it follows that 
\[
\widetilde{\pi}(v_l,v_h)=\frac{v_lv_h}{v_l+v_h}\,.
\]
Therefore, 
\[
\tilde{\phi}(v_l)+\tilde{\psi}(v_h)=\tilde{\phi}(v_l)+\tilde{\psi}(\lb{\beta}(v_l))=\frac{v_l\lb{\beta}(v_l)}{v_l+\lb{\beta}(v_l)}=\widetilde{\pi}(v_l,\lb{\beta}(v_l))=\widetilde{\pi}(v_l,v_h)\,,
\]
as desired. If $v_l \in (\tilde{\kappa}^3,\tilde{\kappa}^4]$, then it must be that $v_h=v_l$. Therefore, 
\[
\tilde{\phi}(v_l)+\tilde{\psi}(v_h)=\frac{v_l}{2}=\widetilde{\pi}(v_l,v_l)=\widetilde{\pi}(v_l,v_h)\,,
\]
as desired. If $v_l \in [\tilde{\kappa}^4,\tilde{\kappa}^5]$, it must be that $v_h=\ub{\beta}(v_l)$, and thus
\[
\tilde{\phi}(v_l)+\tilde{\psi}(v_h)=\tilde{\phi}(v_l)+\tilde{\psi}(\ub{\beta}(v_l))=\frac{v_l\ub{\beta}(v_l)}{v_l+\ub{\beta}(v_l)}\,. 
\]
Moreover, by \eqref{eq:4.1}, 
\[
\frac{v_l\ub{\beta}(v_l)}{v_l+\ub{\beta}(v_l)}=\tilde{\phi}(v_l)+\tilde{\psi}(\ub{\beta}(v_l))\geq \frac{\ub{\beta}(v_l)}{4}\,.
\]
Likewise, by \eqref{eq:4.2}, 
\[
\frac{v_l\ub{\beta}(v_l)}{v_l+\ub{\beta}(v_l)}=\tilde{\phi}(v_l)+\tilde{\psi}(\ub{\beta}(v_l))\geq \frac{v_l}{4}\,.
\]
Together, for any $v_l \in [\tilde{\kappa}^4,\tilde{\kappa}^5]$,
\[
\frac{v_l\ub{\beta}(v_l)}{v_l+\ub{\beta}(v_l)} \geq \max\left\{\frac{v_l}{4},\frac{\ub{\beta}(v_l)}{4}\right\}
\]
and thus 
\[
\widetilde{\pi}(v_l,\ub{\beta}(v_l))=\frac{v_l\ub{\beta}(v_l)}{v_l+\ub{\beta}(v_l)}
\]
for all $v_l \in [\tilde{\kappa}^4,\tilde{\kappa}^5]$. As a result,
\begin{align*}
\tilde{\phi}(v_l)+\tilde{\psi}(v_h)=&\tilde{\phi}(v_l)+\tilde{\psi}(\ub{\beta}(v_l))\\
=&\frac{v_l\ub{\beta}(v_l)}{v_l+\ub{\beta}(v_l)}\\
=& \widetilde{\pi}(v_l,\ub{\beta}(v_l))\\
=& \widetilde{\pi}(v_l,v_h)\,,
\end{align*}
as desired. Lastly, for any $v_l >\tilde{\kappa}^5$, it must be that $v_h=v_l$. Therefore, 
\[
\tilde{\phi}(v_l)+\tilde{\psi}(v_h)=\frac{v_l}{2}=\widetilde{\pi}(v_l,v_l)=\widetilde{\pi}(v_l,v_h)\,,
\]
as desired. Together, this completes the proof. 
\end{proof}

\begin{proof}[Proof of \bref{prop:welfare}]
By \bref{prop:opt-transport}, any optimal non-discriminatory pricing rule $p$ can be identified by a family $\{\rho_c\}_{c \in C}$ of matching schemes, where $\rho_c \in \mathcal{R}_c$ is a solution of 
\[
\max_{\rho \in \mathcal{R}_c} \int_{V^2} \pi_c(v_l,v_h) \diff \rho,.
\]
Thus, it suffices to consider the solutions of the optimal transport problem \eqref{eq:transport-point} for each $c$.   

When $F_{c,l}(c) \geq \|F_{c,l}-F_{c,h}\|$, since $\pi^\star(c)=\EE[(v-c)^+]$, it must be that $CS(c,h;p)=CS(c,l;p)=0$. Now suppose that $F_{c,l}(c)< \|F_{c,l}-F_{c,h}\|$. Fix any such $c \in C$, consider any solution $\rho_c$ of the optimal transport problem \eqref{eq:transport-point}. By \bref{lem:dual}, for any solution $\rho_c$ of \eqref{eq:transport-point}, it must be that 
\[
\phi^\star_c(v_l)+\psi^\star_c(v_h)=\pi_c(v_l,v_h)\,,
\]
for all $(v_l,v_h) \in \supp (\rho_c)$. Therefore, we have 
\begin{align}\label{eq:supp}
\supp(\rho_c) \subseteq [0,\kappa_c^3] \times [\kappa_c^5,\infty) \cup &\{(v_l,v_h) \in [\kappa_c^1,\kappa_c^3] \times [\kappa_c^1,\kappa_c^3]: v_l \geq v_h\} \notag\\
\cup& [\kappa_c^3,\kappa_c^4] \times [0,\kappa_c^1] \notag\\
\cup& \{(v_l,v_h) \in [\kappa_c^3,\kappa_c^4] \times [\kappa_c^3,\kappa_c^4]: v_l=v_h\}\\
\cup& \{(v_l,v_h) \in [\kappa_c^4,\kappa_c^5] \times [\kappa_c^4,\kappa_c^5]: v_l \leq v_h\} \notag\\
\cup& \{(v_l,v_h) \in [\kappa_c^5,\infty) \times [\kappa_c^5,\infty): v_l=v_h\} \notag
\end{align}

Now consider any optimal non-discriminatory pricing rule $p$. By \bref{prop:opt-transport}, there exists $\{\rho_c\}_{c \in C}$ such that $\rho_c$ is a solution of \eqref{eq:transport-point} for all $c \in C$ and that for almost all $c \in C$ and for almost all matched pair $(v_l,v_h) \in \mathrm{supp}(\rho_c)$, these consumers face a price that equals 
\[
\argmax_{x \in \{v_l,v_h\}} (x-c)[\alpha_c \mathbf{1}\{v_h \geq x\}+(1-\alpha_c)\mathbf{1}\{v_l \geq x\}]\,.
\]

Let $\{\rho_c\}_{c \in C}$ be the family of matching schemes associated with the non-discriminatory pricing rule $p$. Note that for any $(v_l, v_h) \in [\kappa_c^3,\kappa_c^4] \times [0,\kappa_c^1]$, $(1-\alpha_c) (v_l-c) \geq (1-\alpha_c) (\kappa_c^3-c)=\kappa_c^1-c \geq v_h-c$. Therefore, the optimal price for these matched pairs equals $v_h$, and hence $h$-consumers purchase at a price equals their values, whereas $l$-consumers do not purchase. In particular, these consumers retain zero surplus, just as under $p^\star$. Likewise, for any $(v_l,v_h) \in [0,\kappa_c^3] \times [\kappa_c^5,\infty)$, $\alpha_c (v_h-c) \geq \alpha_c (\kappa_c^5-c)=\alpha_c (\kappa_c^4-c)+(1-\alpha_c)(\kappa_c^3-c) \geq \alpha_c \kappa_c^3-c+(1-\alpha_c)(\kappa_c^3-c)=\kappa_c^3-c \geq v_l-c$. Thus, the optimal price for these matched pairs $(v_l,v_h)$ must equal $v_h$, and thus $h$-consumers purchase by paying their values, whereas $l$-consumers do not purchase, just as under $p^\star$. Furthermore, for any $(v_l,v_h) \in [\kappa_c^4,\kappa_c^5] \times [\kappa_c^4,\kappa_c^5]$ such that $v_l \leq v_h$, $\alpha_c (v_h-c) \leq \alpha_c( (\kappa_c^5-c)=\alpha_c (\kappa_c^4-c)+(1-\alpha_c)(\kappa_c^3-c) = \alpha_c (\kappa_c^4-c)+(1-\alpha_c)(\kappa_c^3-c) \leq \kappa_c^4-c \leq v_l-c$. The optimal price for these matched pairs $(v_l,v_h)$ must equals $v_l$, and hence both $\theta_l$ and $\theta_h$ consumers purchase by paying the value of $l$-consumers, just as under $p^\star$. Together, the pricing rule $p$ must lead to the same outcomes as $p^\star$ for matched pairs $(v_l,v_h)$ in $[0,\kappa_c^1] \times [\kappa_c^5,\infty)$, $\{(v_l,v_h) \in [\kappa_c^3,\kappa_c^4] \times [0,\kappa_c^1]: v_l \geq v_h\}$, and $\{(v_l,v_h) \in [\kappa_c^4,\kappa_c^5] \times [\kappa_c^4,\kappa_c^5]: v_l \leq v_h\}$.
In the meantime, for any matched pair $(v_l,v_h) \in [\kappa_c^1,\kappa_c^3] \times [\kappa_c^1,\kappa_c^3]$, since $(1-\alpha_c) (v_l-c) \leq (1-\alpha_c)(\kappa_c^3-c)=\kappa_c^1-c \leq v_h-c$, the optimal price for these matched pairs must be $v_h$.

Together, we have
\[
CS(c,h;p)=CS(c,h;p^\star)=\int_{F_{c,h}(\kappa_c^4)}^{F_{c,h}(\kappa_c^5)}(F_{c,h}^{-1}(q)-F_{c,l}^{-1}(q+F_{c,h}(\kappa_c^4)-F_{c,l}(\kappa_c^4)))\diff q\,;
\]
\[
WL(c,h;p)=WL(c,h;p^\star)=\int_0^{\kappa_c^1} v F_{c,h}(\diff v)\,;
\]
\[
CS(c,l;p)=\int_{[\kappa_c^1,\kappa_c^3]^2} (v_l-v_h) \rho(\diff v_l,\diff v_h)\,;
\]
and 
\[
WL(c,l;p)=\int_0^{\kappa_c^1}v F_{c,l}(\diff v)+ \int_{[\kappa_c^1,\kappa_c^3] \times [\kappa_c^5,\infty)} v_l \rho(\diff v_l,\diff v_h) 
\]
By \eqref{eq:supp}, $\rho_c(v_l \notin [\kappa_c^1,\kappa_c^3],v_h \in [\kappa_c^1,x])=0$ and $\rho_c(v_l \in [\kappa_c^1,\kappa_c^3], v_h \notin [\kappa_c^1,\kappa_c^3])=\rho_c(v_l \in[\kappa_c^1,x], v_h > \kappa_c^5)$, for all $x \in [\kappa_c^1,\kappa_c^3]$. Thus,
\begin{align*}
\rho_c(v_l \in [\kappa_c^1,\kappa_c^3], v_h \in [\kappa_c^1,x])=&\rho(v_l \in V, v_h \in [\kappa_c^1,x])-\rho_c(v_l \notin [\kappa_c^1,\kappa_c^3],v_h \in [\kappa_c^1,x])\\
=& F_{c,h}(x)-F_{c,h}(\kappa_c^1)\,,
\end{align*}
and  
\begin{align*}
\rho_c(v_l \in [\kappa_c^1,x], v_h \in [\kappa_c^1,\kappa_c^3])=& \rho_c(v_l \in [\kappa_c^1,x],v_h \in V)-\rho_c(v_l \in [\kappa_c^1,x], v_h \notin [\kappa_c^1,\kappa_c^3])\\
=& F_{c,l}(x)-F_{c,l}(\kappa_c^1)-\rho(v_l \in [\kappa_c^1,x],v_h>\kappa_c^5)\,, 
\end{align*}
for all $x \in [\kappa_c^1,\kappa_c^3]$. Moreover, by \eqref{eq:supp}, since $\rho_c \in \mathcal{R}_c$, it must be that 
\[
\rho(v_l \in [0,\kappa_c^3],v_h>\kappa_c^5)=\Delta_c(\kappa_c^5)\,.
\]
Together with the fact that $F_{c,l}(\kappa_c^2)=\Delta_c(\kappa_c^5)$, it follows that 
\begin{equation}\label{eq:rhobound}
\min\{F_{c,l}(x),F_{c,l}(\kappa_c^2)\} \geq \rho_c(v_l \in [\kappa_c^1,x], v_h>\kappa_c^5)\,.
\end{equation}
As a result, 
\begin{align*}
CS(c,l;p)=&\int_{[\kappa_c^1,\kappa_c^3]^2} (v_l-v_h) \rho_c(\diff v_l, \diff v_h)\\=& \int_{\kappa_c^1}^{\kappa_c^3} v_l F_{c,l}(\diff v_l)-\int_{\kappa_c^1}^{\kappa_c^3} v_l \rho_c(\diff v_l, v_h>\kappa_c^5)-\int_{\kappa_c^1}^{\kappa_c^3} v_h F_{c,h}(\diff v_h)\\
\leq & \int_{\kappa_c^1}^{\kappa_c^3} v F_{c,l}(\diff v)-\int_{\kappa_c^1}^{\kappa_c^2} v F_{c,l}(\diff v)-\int_{\kappa_c^1}^{\kappa_c^3}v F_{c,h}(\diff v)\\
=& \int_{\kappa_c^2}^{\kappa_c^3} v F_{c,l}(\diff v)-\int_{\kappa_c^1}^{\kappa_c^3} v F_{c,h}(\diff v)\\
=& \int_{F_{c,l}(\kappa_c^2)}^{F_{c,l}(\kappa_c^3)} F_{c,l}^{-1}(q) \diff q -\int_{F_{c,h}(\kappa_c^1)}^{F_1(\kappa_c^3)} F_{c,h}^{-1}(q) \diff q\\
=& \int_{F_{c,l}(\kappa_c^2)}^{F_{c,l}(\kappa_c^3)} (F_{c,l}^{-1}(q)-F_{c,h}^{-1}(q+F_{c,h}(\kappa_c^3)-F_{c,l}(\kappa_c^3)))\diff q\\
=&CS(c,l;p^\star)\,,
\end{align*}
where the inequality follows from \eqref{eq:rhobound} and the last equality follows from \eqref{eq:system}. Likewise, 
\begin{align*}
WL(c,l;p)=& \int_0^{\kappa_c^1} v F_{c,l}(\diff v) +\int_{[\kappa_c^1,\kappa_c^3] \times [\kappa_c^5,\infty)} v_l \rho_c(\diff v_l, \diff v_h) \\
\geq & \int_0^{\kappa_c^1} v F_{c,l}(\diff v)+\int_{\kappa_c^1}^{\kappa_c^2} v F_{c,l}(\diff v)\\
=& \int_0^{\kappa_c^2} v F_{c,l}(\diff v)\,,
\end{align*}
where the inequality follows from \eqref{eq:rhobound}. Together, we have that 
\[
CS(c,h;p)=CS(c,h;p^\star)\,; \mbox{ and } WL(c,h;p)=WL(c,h;p^\star)\,,
\]
while 
\[
0 \leq CS(c,l;p) \leq CS(c,l;p^\star)
\]
for all $c \in C$. Since 
\[
\pi^\star(c)+\sum_{\theta \in \{l,h\}} [CS(c,\theta;p)+WL(c,\theta;p)]\,,
\]
it then follows that 
\[
WL(c,l;p^\star) \leq WL(c,l;p) \leq \int_0^{\kappa_c^1} v_l F_{c,l}(\diff v_l)+\int_{\kappa_c^1}^{\kappa_c^3} v (f_{c,l}(v)-f_{c,h}(v))\diff v\,.
\]

It now remains to show that for any $c$ and for any $\sigma_{c,l}$ 
\[
0 \leq \sigma_{c,l} \leq CS(c,l;p^\star)
\]
there exists $\rho_c \in \mathcal{R}_c$ that solves \eqref{eq:transport-point} such that 
\[
\int_{[\kappa_c^1,\kappa_c^3]}(v_l-v_h) \rho_c(\diff v_l,\diff v_h)=\sigma_{c,l}
\]

To this end, for each $c \in C$, let
\begin{align*}
\tilde{\gamma}_c(v_l \leq x \mid v_h):=
\begin{cases}
\mathbf{1}\{\lb{\Delta}_c^{-1}(F_{c,h}(v_h)+\Delta_c(\kappa_c^3)) \leq x\},& \mbox{if } v_h \leq \kappa_c^1\\
\mathbf{1}\{v_h \leq x\},&\mbox{if } v_h \in (\kappa_c^1,\kappa_c^4]\\
\mathbf{1}\{F_{c,l}^{-1}(F_{c,h}(v_h)+\Delta_c(\kappa_c^4)) \leq x\}, &\mbox{if } v_h \in (\kappa_c^4,\kappa_c^5]\\
\frac{f_{c,l}(v_h)}{f_{c,h}(v_h)}\cdot \mathbf{1}\{v_h \leq x\}+\frac{f_{c,h}(v_h)-f_{c,l}(v_h)}{f_{c,h}(v_h)} \cdot \mathbf{1}\{J_c^{-1}(\Delta_c(\kappa_c^5)-\Delta_c(v_h)) \leq x\},&\mbox{if } v_h > \kappa_c^5,
\end{cases}\,,
\end{align*}
where 
\[
J_c(v):=\begin{cases}
\min\{F_{c,l}(v),\Delta_c(v)+F_{c,h}(\kappa_c^1)\},&\mbox{if } v \leq \kappa_c^3\\
\Delta_c(\kappa_c^3)+F_{c,h}(\kappa_c^3),&\mbox{if } v>\kappa_c^3
\end{cases}
\]
for all $v \in V$. By the same argument as the proof of \bref{lem:rfeas}, $\tilde{\gamma}_c$ is indeed a transition probability. Then, let
\[
\tilde{\rho}_c(v_l \in A, v_h \in B):=\int_{B} \tilde{\gamma}_c(A\mid v_h)F_{c,h}(\diff v_h)\,.
\]
Since $\rho_c^\star \in \mathcal{R}_c$ and since $F_{c,l}(\kappa_c^2)=\Delta_c(\kappa_c^5)$, it follows that $\tilde{\rho}_c \in \mathcal{R}_c$ as well. Moreover, for all $(v_l,v_h) \in \supp(\tilde{\rho}_c)$, $\phi_c^\star(v_l)+\psi_c^\star(v_h)=\pi_c(v_l,v_h)$. Thus, by \bref{lem:dual}, $\tilde{\rho}_c$ solves \eqref{eq:transport-point}. In the meantime, by construction, $v_l=v_h$ for all $(v_l,v_h) \in \supp(\tilde{\rho}_c) \cap [\kappa_c^1,\kappa_c^3] \times [\kappa_c^1,\kappa_c^3]$. Therefore, 
\[
\int_{[\kappa_c^1,\kappa_c^3]^2} (v_l-v_h) \diff \tilde{\rho}_c=0\,.
\]
Therefore, for any $c$ and for any $\sigma_c,l \in [0, CS(c,l;p^\star)]$, 
\[
\rho_c:=\frac{\sigma_{c,l}}{CS(c,l;p^\star)} \rho^\star_c+\left(1-\frac{\sigma_{c,l}}{CS(c,l;p^\star)}\right) \tilde{\rho}_c\,.
\]
Since $\mathcal{R}_c$ is convex and since both $\rho_c^\star$ and $\tilde{\rho}_c$ are solutions of \eqref{eq:transport-point}, $\rho_c$ is in $\mathcal{R}_c$ solves \eqref{eq:transport-point} as well. Moreover,
\begin{align*}
\int_{[\kappa_c^1,\kappa_c^3]^2} (v_l-v_h) \diff \rho_c=&\frac{\sigma_{c,l}}{CS(c,l;p^\star)} \int_{[\kappa_c^1,\kappa_c^3]^2} (v_l-v_h) \diff \rho_c^\star+\left(1-\frac{\sigma_{c,l}}{CS(c,l;p^\star)}\right) \int_{[\kappa_c^1,\kappa_c^3]^2} (v_l-v_h) \tilde{\rho}_c\\
=&CS(c,l;p^\star) \cdot \frac{\sigma_{c,l}}{CS(c,l;p^\star)}+0 \cdot \left(1-\frac{\sigma_{c,l}}{CS(c,l;p^\star)}\right) \\
=&\sigma_{c,l}\,,
\end{align*}
as desired. This completes the proof. 
\end{proof}
\subsection{More on the Partly Anti-Assortative Pricing Rule}
Consider any partly anti-assortative pricing rule $p^{anti}$ with quantiles $\{q_c\}_{c \in C}$. Note that all consumers with $\theta=h$ and $v \geq c$ would purchase and pay their values. For $l$-consumers, if $v < F_{c,l}^{-1}(q_c)$, then since 
\[
F_{c,l}(x) \geq F_{c,h}(x) > F_{c,h}(x)-(1-q_c) 
\]
for all $x$, 
\[
p^{anti}(v,c,l)=F_{c,h}^{-1}(F_{c,l}(v)+(1-q_c)) > v\,,
\]
and thus they would not purchase. In the meantime, if $v \geq F_{c,l}^{-1}(q_c)$, note that 
\[
p^{anti}(v,c,l)=F_{c,h}^{-1}(F_{c,l}(v)-q_c) \leq v
\]
if and only if 
\[
F_{c,l}(v)-F_{c,h}(v)=\Delta_c(v) \leq q_c\,.
\]
Therefore, such a consumer would purchase if and only if $\Delta_c(v) \leq q_c$, and will purchase at a price $F_{c,h}^{-1}(F_{c,l}(v)-q_c)$. As a result, for any $c \in C$, the smallest $q_c$ such that all consumers with $\theta=l$ and $v \geq F_{c,l}^{-1}(q_c)$ would purchase is $q_c=\Delta_c(v^\star_c)$. In this case, the seller's profit is given by 
\[
\E{(p^{anti}-c)\mathbf{1}\{v \geq p^{anti}\}}=\alpha_c \int_V (v-c)^+ F_{c,h}(\diff v)+(1-\alpha_c)\left[\int_0^{F_{c,h}^{-1}(1-q_c)}(v-c)^+F_{c,h}(\diff v)\right]\,.
\]

\subsection{Assumption \ref{ass:no-full-surplus-extraction} under a Scaled Family}
Suppose that there exists $F_h,F_l$ such that $F_{c,h}(x)=F_h(\nicefrac{x}{c})$ and $F_{c,l}(x)=F_{l}(\nicefrac{x}{c})$ for all $c$ and for all $x$. Then 
\begin{align*}
\|F_{c,l}-F_{c,h}\|=&\max_{v} \Delta_c(v)=\max_v [F_{c,l}(v)-F_{c,h}(v)]\\
=&\max_{v} \left[F_{l}\left(\frac{v}{c}\right)-F_{h}\left(\frac{v}{c}\right)\right]\\
=&\max_{\tilde{v}} [F_{l}(\tilde{v})-F_{h}(\tilde{v})] \tag{$\tilde{v}=\frac{v}{c}$}\\
=&\|F_l-F_h\|\,.
\end{align*}
Therefore, \bref{ass:no-full-surplus-extraction} is equivalent to 
\[
F_{c,l}(c)=F_l(1)<\|F_l-F_h\|=\|F_{c,l}-F_{c,h}\|\,.
\]

Furthermore, suppose that $\alpha_c=\alpha$ for all $c \in C$. Let $\kappa_1$ be the solution to \eqref{eq:system} when $c=1$, and let $\kappa_c:=c\cdot \kappa_1$. Then, for all $c \in C$, $F_{c,l}(\kappa_c^j)=F_{1,l}(\kappa_1^j)$ and $F_{c,h}(\kappa_c^j)=F_{1,h}(\kappa_1^j)$ for all $j \in \{1,2,3,4,5\}$, and 
\[
\kappa_c^1-c=c(\kappa_c^1-1)=c\alpha(\kappa_1^3-1)=(1-\alpha)(\kappa_c^3-c)
\]
while 
\[
(1-\alpha)(\kappa_c^3-c)=c(1-\alpha)(\kappa_1^3-1)=c\alpha(\kappa_1^5-\kappa_1^4)=\alpha(\kappa_c^5-\kappa_c^4)\,.
\]
Therefore, $\kappa_c=c\cdot \kappa_1$ must solve \eqref{eq:system}.

Now suppose that $F_{c,l}(v)=1-e^{-v/\lambda_l c}$ and $F_{c,h}(v)=1-e^{-v/\lambda_h c}$ for all $v \geq 0$ and for some $0<\lambda_l<\lambda_h$. Let $\gamma:=\nicefrac{\lambda_h}{\lambda_l}$. Then, $\Delta_c(v)=e^{-v/\lambda_h c}-e^{-v/\lambda_lc}$ for all $v \geq 0$. Moreover, since $v^\star_c$ is the unique maximize of $\Delta_c$, by the first order condition, $\Delta_c'(v^\star_c)=f_{c,l}(v^\star_c)-f_{c,h}(v^\star_c)=0$. Therefore, 
\[
\gamma=\frac{\lambda_h}{\lambda_l}=\exp\left(\frac{-v^\star_c}{\lambda_h c}+\frac{v^\star_c}{\lambda_l c}\right)=\exp\left(-\frac{v^\star_c}{\lambda_h c}(1-\gamma)\right)\,.
\]
Therefore, 
\[
\exp\left(-\frac{v^\star_c}{\lambda_h c}\right)=\gamma^{-\frac{1}{\gamma-1}}
\]
and 
\[
\exp\left(-\frac{v^\star_c}{\lambda_l c}\right)=\left(\exp\left(-\frac{v^\star_c}{\lambda_h c}\right)\right)^{\frac{\lambda_h}{\lambda_l}}=\gamma^{-\frac{\gamma}{\gamma-1}}\,.
\]
As a result, 
\[
\Delta_c(v^\star_c)=\exp\left(-\frac{v^\star_c}{\lambda_h c}\right)-\exp\left(-\frac{v^\star_c}{\lambda_l c}\right)=\gamma^{-\frac{1}{\gamma-1}}-\gamma^{-\frac{\gamma}{\gamma-1}}=\gamma^{-\frac{\gamma}{\gamma-1}}(\gamma-1)\,,
\]
and hence \bref{ass:no-full-surplus-extraction} simplifies to 
\[
F_{c,l}(c)=1-e^{\frac{-1}{\lambda_l}}<\gamma^{-\frac{\gamma}{\gamma-1}}(\gamma-1)=\Delta_c(v^\star_c)\,.
\]
\end{document}